\documentclass{emulateapj}
\usepackage{graphicx}

\slugcomment{Accepted for publication in ApJ}

\newcommand{\for}[1]{\begin{equation} #1 \end{equation}}
\newcommand{\st}[1]{_{\rm{#1}}}

\newcommand{\gl}[1]{(\ref{#1})}
\def\gsim{\mathrel{\raise0.35ex\hbox{$\scriptstyle >$}\kern-0.6em\lower0.40ex\hbox{{$\scriptstyle \sim$}}}}
\def\lsim{\mathrel{\raise0.35ex\hbox{$\scriptstyle <$}\kern-0.6em\lower0.40ex\hbox{{$\scriptstyle \sim$}}}}

\begin{document}

\title{An ALMA survey of submillimeter galaxies in the Extended \textit{Chandra} Deep Field South: Source Catalog and multiplicity}

\shorttitle{An ALMA survey of SMGs in ECDFS}
\shortauthors{Hodge et al.}

\author{J. A. Hodge\altaffilmark{1}}
\altaffiltext{1}{Max--Planck Institute for Astronomy, K\"{o}nigstuhl 17, 69117 Heidelberg, Germany}
\email{hodge@mpia.de}

\author{A. Karim\altaffilmark{2}}
\altaffiltext{2}{Institute for Computational Cosmology, Durham University, South Road, Durham, DH1 3LE, UK}

\author{I. Smail\altaffilmark{2}}

\author{A. M. Swinbank\altaffilmark{2}}

\author{F. Walter\altaffilmark{1}} 

\author{A. D. Biggs\altaffilmark{3}}
\altaffiltext{3}{European Southern Observatory, Karl--Schwarzschild Strasse 2, D--85748 Garching, Germany}

\author{R. J. Ivison\altaffilmark{4,5}}
\altaffiltext{4}{UK Astronomy Technology Center, Science and Technology Facilities Council, Royal Observatory, Blackford Hill, Edinburgh EH9 3HJ, UK}
\altaffiltext{5}{Institute for Astronomy, University of Edinburgh, Blackford Hill, Edinburgh EH9 3HJ, UK}

\author{A. Weiss\altaffilmark{6}}
\altaffiltext{6}{Max--Planck Institut f\"ur Radioastronomie, Auf dem H\"ugel 69, D--53121 Bonn, Germany}

\author{D. M. Alexander\altaffilmark{2}}

\author{F. Bertoldi\altaffilmark{7}}
\altaffiltext{7}{Argelander--Institute of Astronomy, Bonn University, Auf dem H\"ugel 71, D--53121 Bonn, Germany}

\author{W. N. Brandt\altaffilmark{8,9}}
\altaffiltext{8}{Department of Astronomy \& Astrophysics, 525 Davey Lab, Pennsylvania State University, University Park, Pennsylvania, 16802, USA}
\altaffiltext{9}{Institute for Gravitation and the Cosmos, The Pennsylvania State University, University Park, PA 16802, USA}  

\author{S. C. Chapman\altaffilmark{10,11}}
\altaffiltext{10}{Institute of Astronomy, University of Cambridge, Madingley Road, Cambridge CB3 0HA, UK}
\altaffiltext{11}{Department of Physics and Atmospheric Science, Dalhousie University, Coburg Road Halifax, B3H 4R2, UK}

\author{K. E. K. Coppin\altaffilmark{12}}
\altaffiltext{12}{Department of Physics, McGill University, 3600 Rue University, Montreal, QC H3A 2T8, Canada}

\author{P. Cox\altaffilmark{13}}
\altaffiltext{13}{IRAM, 300 rue de la piscine, F--38406 Saint--Martin d'H\'eres, France}

\author{A. L. R. Danielson\altaffilmark{2}}

\author{H. Dannerbauer\altaffilmark{14}}
\altaffiltext{14}{Universit\"at Wien, Institut f\"ur Astrophysik, T\"urkenschanzstrasse 17, 1180 Wien, Austria}

\author{C. De Breuck\altaffilmark{3}}

\author{R. Decarli\altaffilmark{1}}

\author{A. C. Edge\altaffilmark{2}}

\author{T. R. Greve\altaffilmark{15}}
\altaffiltext{15}{University College London, Department of Physics \& Astronomy, Gower Street, London, WC1E 6BT, UK}

\author{K. K. Knudsen\altaffilmark{16}}
\altaffiltext{16}{Department of Earth and Space Sciences, Chalmers University of Technology, Onsala Space Observatory, SE--43992 Onsala, Sweden}

\author{K. M. Menten\altaffilmark{6}}

\author{H.--W. Rix\altaffilmark{1}}

\author{E. Schinnerer\altaffilmark{1}}

\author{J. M. Simpson\altaffilmark{2}}

\author{J. L. Wardlow\altaffilmark{17}}
\altaffiltext{17}{Department of Physics \& Astronomy, University of California, Irvine, CA 92697, USA}

\author{P. van der Werf\altaffilmark{18}}
\altaffiltext{18}{Leiden Observatory, Leiden University, PO Box 9513, 2300 RA Leiden, Netherlands}

\begin{abstract}
\noindent We present an Atacama Large Millimeter/submillimeter Array (ALMA) Cycle 0 survey of 126 submillimeter sources from the LABOCA ECDFS Submillimeter Survey (LESS). 
Our 870$\mu$m survey with ALMA (ALESS) has produced maps 
$\sim$3$\times$ deeper and with a beam area $\sim$200$\times$ smaller than the original LESS observations, doubling the current number of interferometrically--observed submillimeter sources. 
The high resolution of these maps allows us to resolve sources that were previously blended and accurately identify the origin of the submillimeter emission. 
We discuss the creation of the ALESS submillimeter galaxy (SMG) catalog, including the main sample of 99 SMGs and a supplementary sample of 32 SMGs. 
We find that at least 35\% (possibly up to 50\%) of the detected LABOCA sources have been resolved into multiple SMGs, and that the average number of SMGs per LESS source increases with LESS flux density. 
Using the (now precisely known) SMG positions, we empirically test the theoretical expectation for the uncertainty in the single--dish source positions. 
We also compare our catalog to the previously predicted radio/mid--infrared counterparts, finding that 45\% of the ALESS SMGs were missed by this method. 
Our $\sim$1.6$^{\prime\prime}$ resolution allows us to measure a size of $\sim$9 kpc$\times$5 kpc for the rest--frame $\sim$300$\mu$m emission region in one resolved SMG, implying a star formation rate surface density of 80 M$_{\odot}$ yr$^{-1}$ kpc$^{-2}$, and we constrain the emission regions in the remaining SMGs to be $<$10 kpc. 
As the first statistically reliable survey of SMGs, this will provide the basis for an unbiased multiwavelength study of SMG properties.

\noindent\textit{Key words:} galaxies: starburst -- galaxies: high-redshift -- submillimeter -- catalogs

\end{abstract}

\section{INTRODUCTION}
\label{Intro}

Since their discovery over a decade ago, it has been known that submillimeter--luminous galaxies \citep[SMGs;][]{2002PhR...369..111B} are undergoing massive bursts of star formation at rates unheard of in the local universe ($\sim$1000 M$_{\odot}$ yr$^{-1}$). 
One thousand times more numerous than local ultra--luminous infrared galaxies \citep[ULIRGs;][]{1996ARA&A..34..749S}, they could host up to half of the star formation rate density at $z\sim2$ \citep[e.g.][]{2005ApJ...622..772C}. 
They are thought to be linked to both QSO activity and the formation of massive ellipticals in the local universe, making them key players in models of galaxy formation and evolution.

\begin{deluxetable*}{ l l c l l }
\tabletypesize{\small}
\tablewidth{0pt}
\tablecaption{Summary of ALESS Observations \label{tab-1}}
\tablehead{
\colhead{SB\tablenotemark{a}} & \colhead{Date} & \colhead{Antennas\tablenotemark{b}} & \colhead{Fields\tablenotemark{c}} & \colhead{Notes}
}
\startdata
SB1 & 18 Oct 2011 & 15 & 16 & No flux calibrator; used flux scale solutions from SB3\\
SB2 & 18 Oct 2011 & 15 & 15 & Science fields observed at low elevation (20$^{\circ}$--40$^{\circ}$)\\
SB3 & 20 Oct 2011 & 15 & 16 & --\\
SB4 & 20 Oct 2011 & 15 & 15 & No flux calibrator; used flux scale solutions from SB3\\
SB5 & 20 Oct 2011 & 15 & 16 & Science fields observed at low elevation (20$^{\circ}$--40$^{\circ}$)\\
SB6 & 21 Oct 2011 & 13 & 16 & Flux calibrator unusable (20$^{\circ}$ elevation); used flux scale solutions from SB5\\
SB7 & 21 Oct 2011 & 12 & 13 & Science fields observed at low elevation (20$^{\circ}$--30$^{\circ}$)\\
SB8 & 03 Nov 2011 & 14 & 15 & -- 
\enddata
\tablenotetext{a}{Scheduling block}
\tablenotetext{b}{Number of antennas in that SB; 12--m only}
\tablenotetext{c}{Number of LESS fields observed in that SB}
\end{deluxetable*}

Previous surveys identifying submillimeter sources have used telescopes such as the JCMT, IRAM 30m, and APEX single dishes equipped with the SCUBA, MAMBO, and LABOCA bolometer arrays \citep{1997ApJ...490L...5S, 1998Natur.394..248B, 1998Natur.394..241H, 1999ApJ...515..518E, 2000A&A...360...92B, 2004MNRAS.354..779G, 2006MNRAS.372.1621C, 2009ApJ...707.1201W}.
The main limitation of single--dish submillimeter surveys is their angular resolution ($\sim$15$^{\prime\prime}$--20$^{\prime\prime}$ FWHM), leading to multiple issues with the interpretation of the data.
In particular, one of the most challenging issues is the identification of counterparts at other wavelengths.
Because of the large uncertainties on the submillimeter source position, and the presence of multiple possible counterparts within the large beam, these studies rely on statistical associations.
Most studies attempt to identify SMGs by comparing the corrected Poissonian probabilities \citep[$P$--statistic;][]{1978MNRAS.182..181B, 1986MNRAS.218...31D} for all possible radio/mid--infrared counterparts within a given search radius. 
However, the underlying correlation between submillimeter and radio emission from SMGs is poorer than expected \citep[from local studies -- e.g.][]{2000MNRAS.319..813D}, 
possibly due to the presence of radio--loud AGN and cold dust which is not as strongly associated with massive star formation. 
A recent study has suggested that only $\sim$50\% of the single--dish detected SMGs have correctly--identified counterparts assigned with this method \citep[although the results are complicated by the use of both millimeter/submillimeter--selected sources;][]{2012A&A...548A...4S}.
Moreover, the radio/mid--IR do not benefit from the negative K--correction like the submillimeter does, meaning that these studies are biased against the faintest/highest redshift SMGs.

Another related issue is blending in fields with multiple SMGs within the beam.
For example, \citet{2007MNRAS.380..199I} found that many single--dish submillimeter sources have multiple ``robust" radio counterparts, suggesting a significant fraction are interacting pairs on scales of a few arcseconds. 
The existing millimeter/submillimeter interferometry also suggests some sources are resolved into multiple SMGs at $\sim$arcsecond resolution \citep[e.g.,][]{2011ApJ...726L..18W},
though it is not always clear whether the SMGs are interacting, companions, or entirely unrelated.
Finally, multiplicity in the single--dish beam is also expected from evidence of strong clustering among SMGs \citep[e.g.;][]{2004ApJ...611..725B, 2006MNRAS.370.1057S, 2009ApJ...707.1201W, 2012MNRAS.421..284H}.
Whatever their relation, multiple SMGs blended into one source can further complicate
the identification of counterparts as mentioned above.
They can also contribute additional scatter/bias to the far--IR/radio correlation at high redshift.
Most crucially for galaxy formation modeling, it can confuse the observed number counts, 
mimicking a population of brighter sources and affecting the slope
of, e.g., flux versus redshift diagrams.

A number of interferometric submillimeter observations of SMGs have been carried out, 
achieving resolutions of a few arcseconds or less, but the sensitivity of existing interferometers has generally been too poor to observe more than a handful of sources in a reasonable amount of time \citep[e.g.,][]{2000MNRAS.316L..51G, 2000AJ....120.1668F, 2001A&A...378...70L, 2002ApJ...573..473D, 2004ApJ...613..655W, 2007ApJ...670L..89W, 2006ApJ...640L...1I, 2007ApJ...671.1531Y, 2008ApJ...673L.127D, 2008MNRAS.387..707Y, 2009ApJ...704..803Y, 2010ApJ...719L..15A, 2011ApJ...726L..18W, 2012ApJS..200...10S, 2012arXiv1209.1626B}.
Two notable exceptions are the recent PdBI/SMA surveys 
by \citet{2012A&A...548A...4S} and \citet{2012arXiv1209.1626B}.
These studies, which show the state--of--the--art prior to ALMA, emphasize the importance of interferometric observations for a complete and unbiased view of SMGs.
However, 
the interpretation of their results is complicated by the mix of millimeter-- and submillimeter--selected sources and small sample sizes.

With the Atacama Large Millimeter/submillimeter Array (ALMA) now online, the situation is fundamentally changed.
Even 
with the limited capabilities offered in Cycle 0, it
has the resolution and sensitivity necessary to 
double the total number of interferometrically observed submillimeter sources in a matter of hours.
We therefore 
used ALMA in Cycle 0 to observe a large sample of submillimeter galaxies in the Extended \textit{Chandra} Deep Field South (ECDFS),
a 30$^{\prime}$ $\times$ 30$^{\prime}$ field with deep, multi--wavelength coverage from the radio to the X--ray \citep{2001ApJ...551..624G, 2004ApJ...600L..93G, 2005ApJS..161...21L, 2006AJ....132.1729B, 2008ApJS..179...19L, 2008ApJS..179..114M, 2009ApJ...707.1201W, 2009Natur.458..737D, 2010MNRAS.405.2260S, 2010MNRAS.402..245I, 2013arXiv1301.7004M}. 
The 126 submillimeter sources we target were previously detected in the LABOCA ECDFS Submillimeter Survey \citep[LESS;][]{2009ApJ...707.1201W},
the largest, most homogenous, and most sensitive blind 870$\mu$m survey to date. 
We call our 870$\mu$m ALMA survey of these 126 LESS sources `ALESS'.

The ALESS data have already been used in part in a number of papers \citep{2012MNRAS.427.1066S, 2012arXiv1208.4846C},
including a new study constraining the 870$\mu$m number counts \citep{Karim_aless}.
Here, we present 
the overarching results of the survey and 
the full catalog of SMGs. 
We begin in \S\ref{data} by describing the observations, 
our data reduction strategy, 
and presenting the final maps. 
\S\ref{ALESSsources} describes the creation of the ALESS SMG sample,
including source extraction and characterization, 
the associated completeness and reliability, 
and checks on the absolute flux scale 
and astrometry. 
The ALESS SMG catalog is described in \S\ref{catalog}, 
including the definition of the samples 
and some notes on using the catalog. 
\S\ref{results} contains our results,
including details of the SMG sizes, 
a discussion of LESS sources which have been resolved into multiple SMGs 
and those which have no detected ALMA sources, 
an empirical calibration of the LABOCA source positional offsets, 
and a comparison with previously--identified radio and mid--infrared counterparts. 
We end with a summary in \S\ref{summary}.

\section{THE ALMA DATA}
\label{data}

\subsection{Observations}
\label{obs}
The ALMA observations were taken 
between 18 Oct 2011--03 Nov 2011 as part of Cycle 0 Project \#2011.1.00294.S.
The targets were the 126 submillimeter sources originally detected in LESS, which had an angular resolution of $\sim$19$^{\prime\prime}$ FWHM
and an rms sensitivity of $\sigma_{\rm 870\mu m}$ $=$ 1.2\,mJy\,beam$^{-1}$.
The 126 LESS sources were selected above a significance level of 3.7$\sigma$, 
with the 
estimate 
that $\sim$5 of the sources are false detections.

We observed the LESS sources with ALMA's Band 7 centered at 344 GHz (870$\mu$m) -- the same frequency as LESS for direct comparison of the measured flux densities.
These ALMA LESS observations (ALESS) utilized the ``single continuum'' spectral mode, 
with 4 $\times$ 128 dual polarization channels over the full 8 GHz bandwidth, 7.5 GHz of which was usable after flagging edge channels.
The observations were taken in ALMA's Cycle 0 compact configuration, which had a maximum baseline of 125m (corresponding to an angular resolution FWHM of $\sim$1.5$^{\prime\prime}$ FWHM at 344 GHz).
The 126 sources were split into eight scheduling blocks (SBs), each of which was observed once (Table~\ref{tab-1}).
To ensure our survey was unbiased if it was not fully completed, targets were 
assigned to these SBs in an alternating fashion.
Table~\ref{tab-1} also includes details for each SB on how many 12--m antennas were present, how many fields (i.e., LESS sources) were observed, and whether that SB was taken at low elevation and/or is missing the flux calibrator observation.

\begin{figure*}[]
\centering
\includegraphics[scale=0.78]{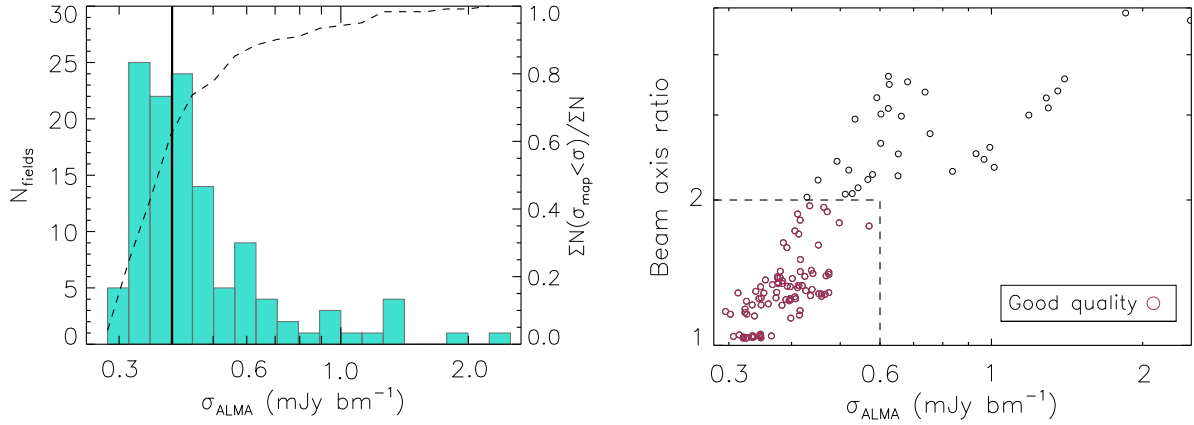}
\caption{Plots showing the properties of the final maps. 
\textit{Left:} Histogram showing the rms noise achieved (logarithmic x--axis) in all 122 fields observed. The dashed line shows the cumulative distribution function $\Sigma$(N). The median rms noise of the maps is 0.4\,mJy\,beam$^{-1}$ (vertical line), equivalent to the requested sensitivity.
\textit{Right:} RMS noise versus beam axis ratio for all 122 fields observed. 
Fields with elongated beam shapes were observed at low elevation and tend to have high values of rms noise.
The boundaries defining good quality maps (i.e.\ rms $<$ 0.6\,mJy\,beam$^{-1}$, axis ratio $<$ 2) are indicated with the dashed box and were chosen to include as many fields as possible with relatively round beam shapes and low rms noise.}
\label{fig:rmsfig}
\end{figure*}

At the frequency of our observations, ALMA's primary beam is 17.3$^{\prime\prime}$ (FWHM).
The beam was centered on the catalogued positions of the LESS sources \citep{2009ApJ...707.1201W},
and each field was observed for 120 seconds.
The beam size matches that of the LESS beam, making it possible to detect all SMGs contributing to the submillimeter source.
The phase stability/weather conditions were good, with a PWV $\lsim$ 0.5 mm.
Three of the scheduling blocks were observed 
at low elevation
($<$30$^{\circ}$), 
affecting the rms and resolution achieved.
In particular, the rms of the QA2--passed delivered data products for $\sim$20 sources are substantially worse than the requested 0.4\,mJy bm$^{-1}$ (Figure~\ref{fig:rmsfig}).
Four fields of the 126 were never observed (LESS 52, 56, 64, and 125).


Depending on the SB, either Mars or Uranus served 
to set the absolute flux density scale.
The quasar B0537-441 was used for bandpass calibration.
In three SBs, the flux calibrator observation was missing or unusable, in which case we used the flux solutions from the next available SB with a reliable planetary observation.
The primary phase calibrator (the quasar B0402-362) was observed for 25 seconds before and after every target field,
and a secondary phase calibrator (the quasar B0327-241) was observed after every other target field.
The secondary phase calibrator was closer to our target field but six times weaker
and was used 
to check the phase referencing and astrometric accuracy of the observations.

\subsection{Data Reduction}
\label{reduction}
The data were reduced and imaged using the Common Astronomy Software Application ({\sc casa}) version 3.4.0\footnote{http://casa.nrao.edu}.
The initial part of data reduction involved, for each SB, converting the native ALMA data format into a measurement set (MS), calibrating the system temperature (Tsys calibration), and applying these corrections along with the phase corrections as measured by the water vapor radiometers on each antenna.
We then visually inspected the $uv$--data, flagging shadowed antennas, the autocorrelation data, and any other obvious problems.
A clean model was used to set the flux density of the flux calibrator.
Prior to bandpass calibration, 
we determined phase--only gain solutions for the bandpass calibrator over a small range of channels at the center of the bandpass.
This step prevents decorrelation of the vector--averaged bandpass solutions.
We then bandpass--calibrated the data and inspected the solutions for problems, flagging data as needed.

We applied the bandpass solutions on--the--fly during the phase calibration. 
Phase--only solutions were determined for each integration time and applied on--the--fly to derive amplitude solutions for each scan. 
A separate phase--only calibration was also run over the scan time for application to the targets. 
All phase and amplitude solutions were examined, and any phase jumps or regions of poor phase stability were flagged. 
The calibration solutions were then tied to the common flux scale and applied to the data.
If the calibration was deemed inadequate (based on inspection of the calibrated calibrators) then more data was flagged and the process was repeated.
The flux density of the primary phase calibrator B0402-362 varied from $\sim$1.23--1.58\,Jy, indicating variability.
The secondary phase calibrator ranged from $\sim$0.20--0.23\,Jy, and the bandpass calibrator ranged from $\sim$2.0--2.2\,Jy.
The absolute flux calibration has an uncertainty of $\sim$15\%, and this uncertainty is not included in the error bars for individual SMG flux densities.

The $uv$--data were Fourier--transformed and the resulting ``dirty" image was deconvolved from the point spread function (i.e.\ the ``dirty beam") using the {\sc clean} algorithm and natural weighting.
The images are 25.6$^{\prime\prime}$ (128 pixels) per side and have a pixel scale of 0.2$^{\prime\prime}$.
The depth of the {\sc clean} process was determined iteratively and depends on the presence of strong sources in the field.
To begin with, we created a dirty image of each target field using the entire 7.5 GHz bandpass. 
These images were used to calculate the initial rms noise for each field.
All images were then cleaned to a depth of 3$\sigma$ over the entire image,
and a new rms noise value was calculated.
We then used the task BOXIT to automatically identify significant ($>$5$\sigma$) sources.
The average number of $>$5$\sigma$ sources per field was 0.6.
If a field did not contain any $>$5$\sigma$ sources, the image produced from the previous clean was considered to be the final one.
If a field did contain any sources $>$5$\sigma$, tight clean boxes were placed around these sources and the image was cleaned down to 1.5$\sigma$ using these clean boxes.
Note that the 5$\sigma$ threshold was chosen to ensure we only cleaned real sources, and the 1.5$\sigma$ clean threshold was chosen to thoroughly clean those sources.
The image produced from this cleaning was then considered to be the final one.
No sources were deemed bright enough to reliably self--calibrate.


\subsection{Final Maps}
\label{maps}

The final {\sc clean}ed ALESS images are shown in order of their LESS field number in Figure~\ref{fig:contourplots} of the Appendix.
These maps have not been corrected for the response of the primary beam, which increases the flux density scale away from field center.
The field of view of the primary beam (17.3$^{\prime\prime}$ FWHM) is indicated by the large circles,
and not only matches the LABOCA beam, but is sufficient to encompass the error--circles of the SMGs from the LESS maps, $\lsim $\,5$''$ \citep{2009ApJ...707.1201W} even in confused situations.
The ellipses in the bottom left--hand corner of each map indicate the angular resolution achieved.
Note that every fourth LABOCA source on the first few pages is noisier because of the distribution of sources into scheduling blocks.

The properties of the final cleaned maps, including rms noise and beam shape, are listed by LESS ID in Table~\ref{tab-2}. 
The rms noise measurements were derived from the non--primary--beam--corrected maps by averaging over several rectangular apertures and will describe the rms at field center in the primary--beam--corrected maps.
A histogram of these rms noise values is shown in Figure~\ref{fig:rmsfig}, 
where we also show the cumulative distribution function.
The median rms noise (at field center) of the maps is $\sigma = 0.4$\,mJy\,beam$^{-1}$, or $\sim$3$\times$ deeper than the original LABOCA data.
We find that 85\% of the maps achieve an rms noise of $\sigma < 0.6$\,mJy\,beam$^{-1}$.
Also plotted in Figure~\ref{fig:rmsfig} is rms noise versus beam axis ratio, defined as the ratio of major to minor axis of the synthesized beam.
The median axis ratio of all of the maps is 1.4, corresponding to a median angular resolution of 1.6$^{\prime\prime}$$\times$1.15$^{\prime\prime}$.
This resolution corresponds to a physical scale of $\sim$13 kpc $\times$ 9 kpc at $z\sim2.5$
and is $>$10$\times$ better than the LABOCA maps.
In terms of beam areas, the improvement is $\sim$200$\times$.
Figure~\ref{fig:rmsfig} also demonstrates that
fields with more elongated beam shapes tend to be noisier.
These fields were observed at very low ($<$20--30$^{\circ}$) elevation, 
and many are of poor quality.
We therefore define `good quality' maps as the subset of maps having relatively round beam shapes (corresponding to axis ratios $\leq$ 2) and an 
rms noise at field center within 50\% of the requested rms (i.e.\ an observed rms noise of $<$0.6\,mJy\,beam$^{-1}$, and note that this condition naturally follows from the first).
While 
we 
will concentrate on these maps for the quality checks in Sections~\ref{sims}--\ref{astrometry}, 
all of the maps are used for the creation of the final SMG catalog, though with the necessary caveats.
We will discuss the catalog samples further in \S\ref{catalog}.

\section{The ALESS SMG SAMPLE}
\label{ALESSsources}

\subsection{Source Extraction and Characterization}
\label{srcextract}
We used custom--written {\sc idl}--based source extraction software to identify and extract sources from the final, cleaned ALMA maps.
We started by identifying individual pixels with flux densities above 2.5$\sigma$ in descending order of significance.
At each position found, and within a box of $2^{\prime\prime} \times 2^{\prime\prime}$, we determined the elliptical Gaussian that best described the underlying signal distribution using an {\sc idl}-implemented Metropolis-Hastings Markov chain Monte Carlo (MH-MCMC) algorithm. 
In the simplest case, and as most ALMA sources appeared to be unresolved, each Gaussian was described by a simple point--source model with only three free parameters: its peak flux density, and its position. 
The values of the major axis, minor axis, and position angle were held fixed to the clean beam values for the given map.
To account for any extended sources, we also repeated the fitting process using a six--parameter fit  
(including axial ratio, major axis size, and orientation).
We discuss which fits are preferred in \S\ref{sizes}.


\begin{figure*}[]
\centering
\includegraphics[scale=0.5]{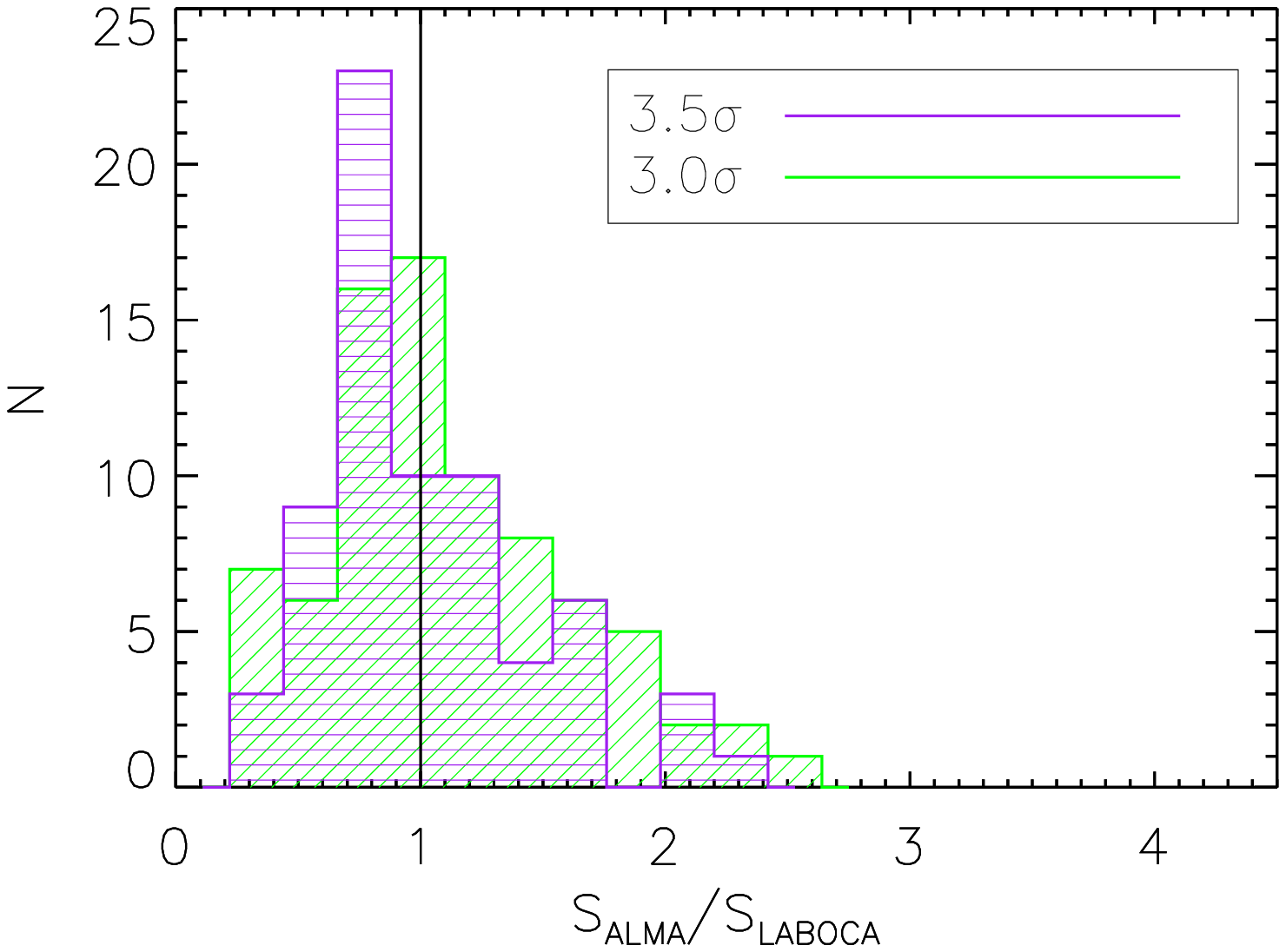}
\hfil
\includegraphics[scale=0.5]{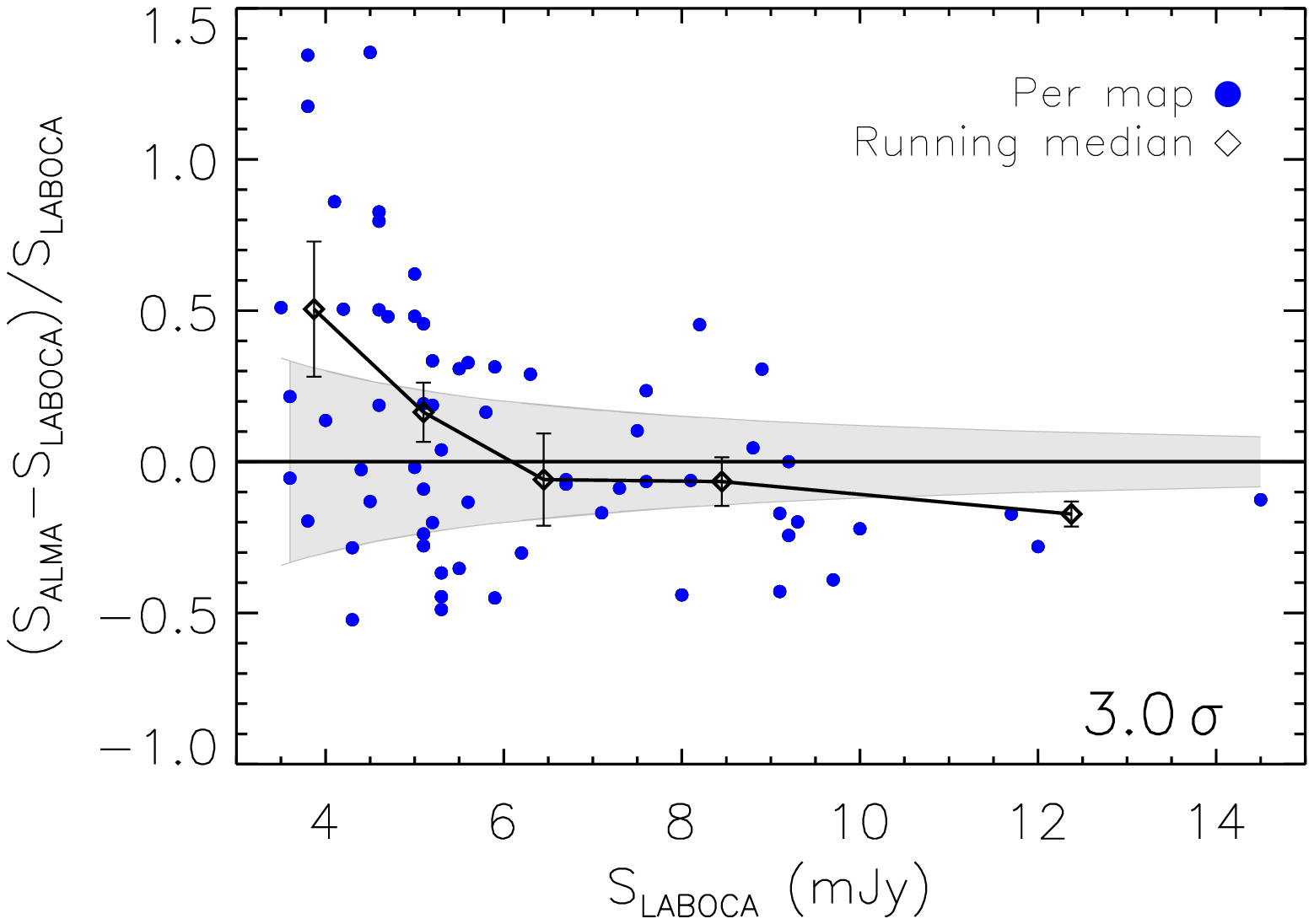}
\caption{Plots showing the results of a flux comparison between the integrated emission from SMGs in ALMA maps and the (deboosted) LABOCA flux density. 
\textit{Left:} 
Histograms of the flux density ratio S$_{\rm ALMA}$/S$_{\rm LABOCA}$ calculated by including SMGs in the ALMA maps down to a source threshold of 3.5$\sigma$ and 3$\sigma$. 
\textit{Right:} 
Plot of (S$_{\rm ALMA}$-S$_{\rm LABOCA}$)/S$_{\rm LABOCA}$ versus S$_{\rm LABOCA}$ for ALMA maps including all SMGs down to a S/N threshold of 3$\sigma$ and convolved with the LABOCA primary beam. We also show the running median, 
and the shaded region indicates the 1$\sigma$ uncertainty expected from the error in the LABOCA flux densities.
The ALMA and LABOCA flux scales are overall in good agreement for a 3$\sigma$ threshold. }
\label{fig:fluxcomp}
\end{figure*}

For each parameter, the mean of the posterior distribution determines its best fit value.
The full set of parameters obtained in this way is used as a flux density model to be subtracted off the initial map before proceeding to the next signal peak within the map.
This process is repeated until the 2.5$\sigma$ threshold is reached, 
with an average of 12 such sources per map. 
The end result is a combined model as well as a residual map. 
The flux densities resulting from both the 3-- and 6--parameter fits are listed in Table~\ref{tab-3} -- See \S\ref{sizes} for more details.

As a check on our source extraction routine, we also used {\sc casa}'s {\sc imfit} task to fit all bright ($>$4$\sigma$) sources.
Overall, we find good agreement between our best--fit parameters and those returned by {\sc imfit}, confirming our results. 
A comparison of the integrated flux densities derived by both methods yields (S$_{\rm IMFIT}$ $-$ S$_{\rm SOURCE}$)/S$_{\rm SOURCE}$ $=$ 0.001 $\pm$ 0.09,
and all flux density estimates are in agreement within the error bars. 
Our quoted errors are, however, slightly larger, as they take into account the correlated nature of the noise.
We will therefore use the parameters derived from our software for the remainder of the paper.
For further information on the source extraction and characterization, error determination, integrated flux densities, and beam deconvolution, see the Appendix.

\subsection{Completeness and Reliability}
\label{sims}

To determine the completeness and reliability of extracted sources above a given S/N threshold, we carried out two different tests.
In the first test, we extracted all sources down to 2.5$\sigma$ in a given map,
producing a residual map.
We then inserted five fake sources per map, a number chosen to build up a significant sample of false sources with separations typical of the final science sources without overcrowding the field.
The peak flux densities follow a steeply declining flux density distribution and with S/N $\sim$ 2--20. 
We repeated this process 16 times per map, re--running our source extraction algorithm each time and deriving the recovered fraction and spurious fraction as a function of S/N.
The result, presented in a companion paper (Karim et al.\ 2012), is that the source extraction recovers $\sim$99\% of all sources above 3.5$\sigma$, with a spurious fraction of only 1.6\% (Karim et al.\ 2012).

As a second test, we ran the source extraction algorithm on the regular and inverted ALMA maps. 
In the simplified case of uncorrelated, Gaussian noise, comparing these results at a given threshold would allow us to estimate the reliability for a source above that threshold. 
Using a source threshold of 3.5$\sigma$, we determined a reliability of $\sim$75\%. This estimate rises to $\sim$90\% for a 4$\sigma$ detection threshold and nearly $\sim$100 for 5$\sigma$ (Karim et al.\ 2012).
Since the noise in the maps is 
more complex in nature,
these reliability estimates are likely lower limits.
The choice of threshold is, as always, a compromise between excluding real sources and including noise.
We will therefore take 3.5$\sigma$ as our source detection threshold in the good quality maps -- see \S\ref{samples}).

\subsection{Absolute Flux Scale}
\label{fluxscale}

To test the absolute flux scale, we compared our results to those of the original LABOCA survey 
taken at the same frequency \citep{2009ApJ...707.1201W}.
Here (and in the rest of the paper) we refer to the deboosted LABOCA flux densities, as these are the best estimates of the `true' LABOCA flux density.
There is a significant difference between the ALMA and LABOCA bandwidths (2$\times$4 GHz versus 60 GHz), so we should expect to see a small systematic difference for SMGs with a steep Rayleigh--Jeans tail.
A more immediate complication is the vastly different angular resolutions of the two surveys, which may cause fainter and/or more extended emission to be resolved out in the ALMA maps.
We therefore modeled the ALMA maps as idealized distributions of perfect point sources 
using only good quality ALMA maps with at least one bright ($>$4$\sigma$) source
and including all sources in such maps down to our source threshold of 3.5$\sigma$.
To accurately represent the true sky distribution, we used the primary beam--corrected `best' flux density values for the sources (see \S\ref{columns}),
and we also included the negative peaks exceeding our source threshold.
We then determined what 
flux would be measured by a telescope with a 19$^{\prime\prime}$ beam, the FWHM of LABOCA,
by convolving the maps to the LABOCA resolution.
This method 
allows for a fairer flux comparison while ensuring that the large--scale noise properties of the ALMA maps do not dominate the convolved images.

The method outlined above results in a median ALMA--to--LABOCA flux density ratio of S$_{\rm ALMA}$/S$_{\rm LABOCA}$ $=$ 0.83$\pm^{0.09}_{0.04}$.
A histogram of the values for different fields is shown in Figure~\ref{fig:fluxcomp}, exhibiting a strong peak below one.
The obvious implication is that a 3.5$\sigma$ threshold is not low enough, in general, to capture all of the true flux in the maps.
The specific threshold chosen for including ALMA SMGs in the model maps obviously affects the peak flux densities measured in the final, convolved maps.
Using a lower threshold ensures that fainter SMGs are accounted for, but the chance of including random noise in the model ALMA maps increases. 
We have attempted to counter this effect by including both the positive and negative peaks in the maps.
We therefore decreased the threshold to 3.0$\sigma$, deriving a median flux density ratio of
S$_{\rm ALMA}$/S$_{\rm LABOCA}$ $=$ 0.97$\pm^{0.07}_{0.04}$, consistent with 
equality of the flux scales.

To determine if these results are biased by extended emission in the SMGs, we performed two further tests.
Although the majority of the SMGs appear unresolved (see \S\ref{columns}), leading us to use the peak flux density as the best flux density estimate,
we tried modeling the SMGs using their (primary beam corrected) integrated flux density estimates.
We also tried tapering all of the ALMA maps to a lower resolution (i.e. increasing the beam area by a factor of a $\sim$few), re--running the source extraction algorithm, and using the (primary beam corrected) peak flux density values from these maps in our model maps. 
Neither test changed the results of the flux comparison significantly, indicating that we are not `missing' flux by taking the majority of the SMGs to be point sources.

Figure~\ref{fig:fluxcomp} compares the ALMA and LABOCA flux densities using a 3$\sigma$ threshold for individual fields as a function of LABOCA S/N.
Also shown are the running median and
the expected 1$\sigma$ uncertainty based on the typical error in LACOBA flux density estimates (1.2\,mJy).
The running median shows a preference for higher ALMA flux densities/lower LABOCA flux densities for the faintest LABOCA sources, 
and lower ALMA flux densities/higher LABOCA flux densities for the brightest LABOCA sources.
This may indicate that there are additional, faint ($<$3$\sigma$) SMGs that are being missed from our models of the brightest LABOCA sources, a theory which we will come back to in Section~\ref{multiples}.
Nevertheless, Figure~\ref{fig:fluxcomp} demonstrates that there is no overall systematic bias between the ALMA and LABOCA flux density scales.  

\begin{figure}[]
\centering
\includegraphics[scale=0.6]{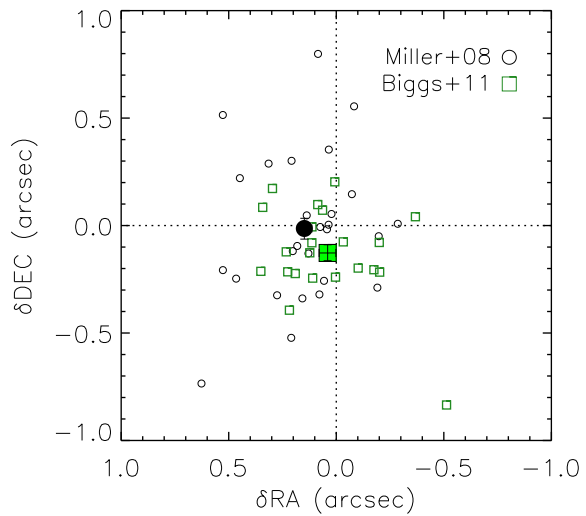}
\caption{Plot showing the astrometric offset between the VLA 1.4 GHz data and the ALMA data. The Miller et al.\ (2008) reduction of the VLA data shows a significant offset in RA, while the \citet{2011MNRAS.413.2314B} re--reduction of the same data shows a significant offset in Declination. The mean offsets for both reductions are indicated with the large, solid symbols.}
\label{fig:astrometry}
\end{figure}

\subsection{Astrometry}
\label{astrometry}

To confirm our astrometry, we looked at all $>$3.5$\sigma$ ALESS SMGs in good quality maps with VLA 1.4 GHz counterparts (see \S\ref{radioIDs}). 
The original 1.4 GHz map of the ECDFS that was used for counterpart identification of LESS sources by \citet{2011MNRAS.413.2314B} was presented in \citet{2008ApJS..179..114M}. 
Matching this data to the ALMA data, we measure a scatter of 0.3$^{\prime\prime}$ in both RA and Declination,
and mean offsets of 0.15$^{\prime\prime}$$\pm$0.03$^{\prime\prime}$ in RA and 0.01$^{\prime\prime}$$\pm$0.04$^{\prime\prime}$ in Dec (in the sense VLA -- ALMA).
A scatter plot of the offsets for individual SMGs in shown in Figure~\ref{fig:astrometry}, where the significant systematic offset in RA is visible.

The same radio data were also re--reduced by \citet{2011MNRAS.413.2314B},
including slight changes to the modeling of the phase calibrator field to account for its resolved structure as well as an additional source in the field. 
The Biggs et al.\ reduction achieved an rms just below 7$\mu$Jy at its deepest point (versus 6.5$\mu$Jy for the Miller et al.\ reduction) and the flux density scale of the two reductions differed by $<$1\% \citep{2011MNRAS.413.2314B}.
In Figure~\ref{fig:astrometry}, we overplot the offsets measured between the ALMA SMGs and the radio counterparts extracted from this map.
Using the Biggs et al.\ reduction, the significant systematic offset seen in the RA coordinate with the Miller et al.\ reduction is no longer present.
However, it has been replaced by a significant systematic offset in Dec.
We measure mean offsets of 0.04$^{\prime\prime}$$\pm$0.04$^{\prime\prime}$ in RA and -0.13$^{\prime\prime}$$\pm$0.04$^{\prime\prime}$ in Dec (in the sense VLA -- ALMA).
We also measure a (smaller) scatter of 0.2$^{\prime\prime}$ in both RA and Declination, though we caution that fewer ALMA SMGs have radio matches.

Recently, \citet{2013arXiv1301.7004M} also released another re--reduction of the \citet{2008ApJS..179..114M} data, which they refer to as the second data release (DR2). Using this new re--reduction, we again recover a significant systematic offset in RA (0.12$^{\prime\prime}$$\pm$0.04$^{\prime\prime}$) with no significant offset in Declination (-0.05$^{\prime\prime}$$\pm$0.06$^{\prime\prime}$). 
Thus comparison to \citet{2008ApJS..179..114M} and \citet{2013arXiv1301.7004M} yields similar results, and both are contrary to the re--reduction of \citet{2011MNRAS.413.2314B}.

Since it appears that the astrometry of the 1.4 GHz radio data is extremely sensitive to the details of the calibration, we take the systematic offsets measured as being entirely due to the radio data.
We conclude that the ALMA SMG positions are accurate to within 0.2$^{\prime\prime}$--0.3$^{\prime\prime}$.
For comparison, the expected astrometric accuracy is usually estimated as $\sim\Theta$/(S/N), or 0.17$^{\prime\prime}$ using the median resolution and S/N of the matched sample.

In addition to the VLA data, we also compared the ALMA SMG positions to the positions of the (confirmed) 24$\mu$m MIPS and 3.6$\mu$m IRAC counterparts presented in \citet{2011MNRAS.413.2314B} and discussed further in \S\ref{radioIDs}.
For the 24$\mu$m data, we measure mean offsets of $-$0.16$^{\prime\prime}$$\pm$$^{0.08}_{0.11}$$^{\prime\prime}$ in RA and 0.15$^{\prime\prime}$$\pm$$^{0.06}_{0.05}$$^{\prime\prime}$ in Dec (in the sense 24$\mu$m -- ALMA).
Similarly, the 3.6$\mu$m data show offsets of $-$0.10$^{\prime\prime}$$\pm$0.05$^{\prime\prime}$ in RA and 0.42$^{\prime\prime}$$\pm$$^{0.05}_{0.04}$$^{\prime\prime}$ in Dec (in the same sense).
A systematic offset between the radio and 24$\mu$m data was previously noted by \citet{2011MNRAS.413.2314B}, who measured mean offsets of $-$0.25$^{\prime\prime}$ in RA and +0.29$^{\prime\prime}$ in Dec (in the sense MIPS -- radio),
in agreement with what we measure here.


\section{The Catalog}
\label{catalog}

\subsection{Sample Definitions}
\label{samples}

We define the MAIN ALESS SMG sample as
consisting of all SMGs satisfying the following criteria: the rms of the ALMA map is less than 0.6\,mJy\,beam$^{-1}$; the ratio of the major and minor axes of the synthesized beam is less than two; 
the SMG lies inside the ALMA primary beam FWHM; and
the S/N ratio of the SMG (defined as the ratio of the best fit peak flux density from the 3--parameter point source model\footnote{Note that since the S/N ratio from the \textit{model fit} is used to identify SMGs in the catalog, there may be SMGs in Figure~\ref{fig:contourplots} which appear to be fainter than 3.5$\sigma$ (based on the number of contours) but which which are identified as MAIN SMGs (and vice versa).} to the background rms) is greater than 3.5. 
The SMGs in the MAIN sample are indicated in the catalog with the flag ALESS$\_$SAMPLE $=$ 1.
These 99 SMGs are the most reliable SMGs, coming from within the primary beam FWHM of the good--quality maps (Figure~\ref{fig:rmsfig}).

In addition to the MAIN sample, we define a supplementary sample comprised of two different selections (Table~\ref{tab-4}).
The first component of the supplementary sample consists of SMGs satisfying the following criteria: the rms of the ALMA map is less than 0.6\,mJy\,beam$^{-1}$; the ratio of the major and minor axes of the synthesized beam is less than two;
the SMG lies outside the ALMA primary beam FWHM; and the S/N ratio of the SMG is greater than 4.
This selection consists of SMGs that -- like the MAIN sample -- come from the maps designated as good--quality (Figure~\ref{fig:rmsfig}), but they lie \textit{outside} the primary beam FWHM. 
Because the telescope sensitivity in this region is $<$50\% of the maximum, leading to a higher fraction of spurious sources, 
we have raised the S/N threshold slightly.
The 18 SMGs which satisfy these criteria are indicated in the catalog with the flag ALESS$\_$SAMPLE $=$ 2 and should be used with care.

The second component of the supplementary sample consists of SMGs satisfying the criteria:
the rms of the ALMA map is greater than 0.6\,mJy\,beam$^{-1}$ \textbf{OR} the ratio of the major and minor axes of the synthesized beam is greater than two; the SMG lies inside the ALMA primary beam FWHM; and the S/N ratio of the SMG is greater than 4.
These SMGs are found in maps which range in quality from just slightly worse than the good quality maps to significantly worse.
Because of this, we have (again) raised the S/N threshold to 4.
SMGs which come from maps of just slightly worse quality than the `good quality' maps are likely reliable, while those from the noisiest maps (Figure~\ref{fig:contourplots}) should be used with extreme care, as even SMGs near the phase center may be spurious.
The 14 SMGs which satisfy these criteria are indicated in the catalog with the flag ALESS$\_$SAMPLE $=$ 3.


\subsection{Using the Catalog}
\label{columns}

The ALESS catalog can be found accompanying this paper or from the ALESS website\footnote{http://www.astro.dur.ac.uk/LESS}.
For a detailed description of the data columns in the catalog, see the README file accompanying the catalog. 
Some of the relevant columns for the MAIN and Supplementary samples are also listed in Tables~\ref{tab-3} and \ref{tab-4}
and described below.
Note that while we list both the observed and primary beam corrected flux densities for reference, the primary beam corrected flux densities should always be used for science applications.

\begin{itemize}
\item LESS ID: LESS source ID in order of appearance in the S/N--sorted \citet{2009ApJ...707.1201W} catalog.

\item ALESS ID: Official IAU short ID for ALESS SMGs (ALESS XXX.X), based on LESS ID and ranking in S/N of any subcomponents. Note that higher S/N subcomponents will not make it into the MAIN catalog if they are outside the primary beam FWHM.

\item ALMA Position: Right Ascension and Declination (J2000) of the SMG, based on the 3--parameter point source model fit.

\item $\delta$RA/$\delta$Dec: The 1$\sigma$ uncertainty on the ALMA position in arcseconds.  Please see the Appendix for a detailed discussion of its calculation.

\item S$_{\rm pk}$: The non--primary--beam--corrected best fit peak flux density in mJy\,beam$^{-1}$ based on 3--parameter point source model fit. For details of the error estimation, see the Appendix.

\item S$_{\rm int}$: The non--primary--beam--corrected best fit integrated flux density in mJy from the 6--parameter model fit.

\item S/N$_{\rm pk}$: Signal--to--noise ratio calculated using S$_{\rm pk}$.


\item S$_{\rm BEST,pbcorr}$: The primary--beam--corrected best flux determination in mJy. This is the flux density estimate that we recommend for use in any analysis, and it is equal to the primary--beam--corrected peak flux density from the 3--parameter point source model fit for all SMGs except ALESS 007.1 
(see \S\ref{sizes}).

\item Sample: Corresponding to ALESS$\_$SAMPLE in the online catalog, this column is only shown for the Supplementary sources (Table~\ref{tab-4}), as all MAIN sample sources have ALESS$\_$SAMPLE $=$ 1 by definition. As described in \S\ref{samples}, ALESS$\_$SAMPLE $=$ 2 sources come from outside the primary beam in good quality maps, and ALESS$\_$SAMPLE $=$ 3 sources come from poor quality maps.

\item Biggs et al.\ ID: A flag indicating whether the SMG confirms the position of a robust (\textbf{r}) or tentative (t) counterpart in the catalog of radio/mid--infrared counterparts of \citet{2011MNRAS.413.2314B}. Sources that do not correspond to either a robust or tentative counterpart are indicated with~`--'. See \S\ref{radioIDs} for further details.
\end{itemize}

\begin{figure*}[]
\centering
\includegraphics[scale=0.7]{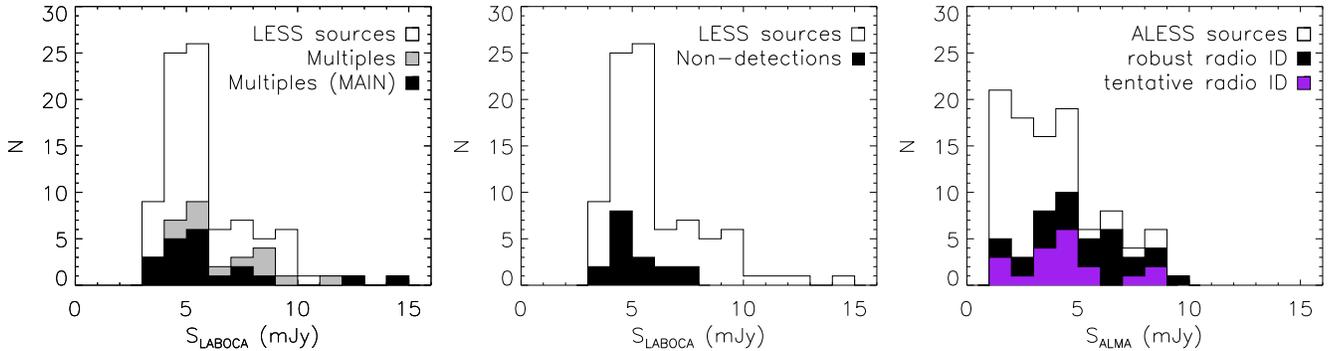}
\caption{Histograms of LABOCA/ALMA flux density values in mJy. 
\textit{Left:} The LABOCA flux density values for all LESS sources with good quality (i.e.\ rms $<$ 0.6\,mJy\,beam$^{-1}$, axis ratio $<$ 2) ALMA maps. The black filled histogram shows the subset of fields which have been resolved by ALMA into multiple SMGs within the primary beam (i.e., in the MAIN sample). The gray filled histogram shows the additional LESS sources that are resolved into multiples if we include supplementary SMGs outside the primary beam FWHM in these maps. The brightest LABOCA sources have a high fraction of multiples.
\textit{Center:} Here we show the LABOCA flux densities for the subset of LESS sources which are non--detections with ALMA. 
These sources tend to lie on the fainter end of the LABOCA flux density distribution.
\textit{Right:} ALMA flux density values for all SMGs in the MAIN sample. All LESS sources above 9\,mJy have been resolved into multiple SMGs or are fainter in the ALMA maps. We also show the subset of SMGs which confirm previously identified radio/mid--infrared robust counterparts proposed by \citet{2011MNRAS.413.2314B}, and (independently) the subset of SMGs which confirm tentative counterparts. The radio/mid--infrared counterparts correctly predict a large percentage of the bright sources but miss 55\% of the SMGs overall.}
\label{fig:blanks/blends}
\end{figure*}

\section{RESULTS}
\label{results}

\subsection{Source sizes}
\label{sizes}

Based on the source catalog, we have first tried to estimate which (if any) of the SMGs may be extended.
While there is some evidence that a number of sources may be marginally--extended, 
the error bars produced by the deconvolution algorithm are generally too large to make any conclusive statements.
In addition, many sources are not bright enough to reliably measure a significant source extension in the current data.
We therefore conclude that all of the SMGs except one 
are best described as point sources,
and we have set the ``best" flux determination (S$_{\rm BEST}$) equal to the peak flux density from the 3--parameter point--source fit for all SMGs in the catalog (except one -- see below).

The fact that the majority of the ALESS SMGs are unresolved suggests
that their rest--frame $\sim$300$\mu$m emission is arising in a region with a size $<$10 kpc.
This upper limit agrees with observations of high--$J$ ($J>2$) CO transitions, which 
typically report sizes in the range 4--6 kpc \citep[FWHM;][]{2006ApJ...640..228T, 2008ApJ...680..246T, 2010MNRAS.405..219B, 2010ApJ...724..233E, 2012arXiv1205.1511B}. 
Some observations of lower--$J$ CO transitions and radio continuum emission, on the other hand, 
have found extended gas reservoirs of $>$10 kpc \citep[e.g.,][]{2004ApJ...611..732C, 2008MNRAS.385..893B, 2010MNRAS.404..198I, 2011MNRAS.412.1913I, 2011ApJ...733L..11R,  2011ApJ...739L..31R, 2012ApJ...760...11H}.

Assuming a $\sim$10 kpc upper limit on the (median) size of the ALESS SMGs
corresponds to a lower limit on the (median) SFR surface density of $>$14 M$_{\odot}$ yr$^{-1}$ kpc$^{-2}$.
Here, we have assumed the median SFR of the LESS SMGs \citep[1100 M$_{\odot}$ yr$^{-1}$;][]{2011MNRAS.415.1479W}.
Using the interquartile range on SFR of 300--1900 M$_{\odot}$ yr$^{-1}$ results in SFR surface densities of $>$4--24 M$_{\odot}$ yr$^{-1}$ kpc$^{-2}$, 
still well below the limit for Eddington--limited star formation in a radiation pressure supported starburst \citep[1000 M$_{\odot}$ yr$^{-1}$ kpc$^{-2}$;][]{2005ApJ...630..167T}.

Only a single SMG (ALESS 007.1) is bright enough to reliably resolve \textit{and} appears to be significantly extended along both axes (i.e.\ even within the error margins, the intrinsic major and minor axes are inconsistent with a point source model).
For ALESS 007.1, the best flux density estimate (S$_{\rm BEST}$) is set to the integrated flux density from the 6--parameter model fit.
This fit produced an intrinsic source size of (1.1$^{\prime\prime}$$\pm$0.3$^{\prime\prime}$)$\times$(0.7$^{\prime\prime}$$\pm$0.2$^{\prime\prime}$).
Taking this SMG's photometric redshift estimate of $z_{\rm phot}=2.81\pm^{0.18}_{0.07}$ \citep{2011MNRAS.415.1479W}, 
this corresponds to a physical size of $\sim$9 kpc $\times$ 5 kpc, implying that the star formation is spread out over many kpc.
Its SFR of 2800$\pm$$^{400}_{700}$ \citep{2011MNRAS.415.1479W} implies an SFR surface density of $\sim$80 M$_{\odot}$ yr$^{-1}$ kpc$^{-2}$,
consistent with previous measurements for SMGs via indirect means like high--$J$ CO \citep[$\sim$80 M$_{\odot}$ yr$^{-1}$ kpc$^{-2}$; e.g.,][]{2006ApJ...640..228T}.

\subsection{Multiplicity}
\label{multiples}

One of the main results from this survey is that a large fraction of the LESS sources have been resolved into multiple SMGs.
Considering LESS sources with at least one detection in the MAIN ALESS SMG sample, we find that 24 of 69 LESS sources split into two or more MAIN ALESS SMGs.
If we also consider as reliable the SMGs which lie outside the primary beam in these maps (i.e.\ ALESS$\_$SAMPLE $=$ 2), 
then we find that the majority of these SMGs lie in maps which had at least one MAIN SMG, while only two SMGs lie in maps with no MAIN sources.
This brings the total number of LESS sources considered to 71, of which we find that 32 split into multiple SMGs.
We therefore find that for the good quality maps with at least one 
SMG somewhere in the map, $\sim$35\%--45\% consist of multiple SMGs.

Despite the large fraction of detected multiples, it is important to note that the median number of detected SMGs per good quality map is only one. 
This is true whether or not we also include supplementary sources outside the primary beam FWHM in these fields. 
For this calculation, we have also included those maps (discussed in the next section) which are of good quality but devoid of any detected SMGs.
If we consider that a portion of the 
non--detections are also due to LESS sources resolved into multiple SMGs, then the true fraction of single--dish sources consisting of multiple SMGs may be even higher.
In the extreme case, 
the fraction of multiple SMGs may increase to 50\% (see \S\ref{blanks}).

That some of the single--dish sources are multiples is not unexpected theoretically and has also been reported in some of the first interferometric submillimeter studies.
For example, \citet{2012A&A...548A...4S} presented interferometric millimeter observations of submillimeter--selected sources in COSMOS.
In their 870$\mu$m--selected sample of 27 LABOCA sources observed interferometrically, 22\% turned out to be blended.
However, their results are complicated by the fact that their interferometric observations were taken at a different waveband from the original submillimeter source selection.
A more reliable analysis by \citet{2012arXiv1209.1626B} of 16 SCUBA--2 identified sources with the SMA at the same wavelength suggested that 40\% of bright sources are actual blends, though they cautioned that their results relied on small number statistics.
Both results are in 
agreement with what we find with our larger sample.

Figure~\ref{fig:blanks/blends} shows a histogram of LABOCA flux density values for the fields with multiple ALESS SMGs compared to that of all good quality 
maps.
If we only treat as reliable those SMGs detected within the primary beam FWHM (i.e., MAIN SMGs), then the brightest two LESS sources are resolved into multiples. 
If we also include 
the supplementary SMGs from outside the primary beam FWHM, then we find that the brightest three LESS sources are resolved into multiples. 
The fluxes for all distinct SMGs in the ALESS MAIN sample are shown in Figure~\ref{fig:blanks/blends},
where there are now no SMGs with flux densities above 9.0\,mJy (the brightest ALESS SMG).
All LESS sources which previously had S $>$ 9\,mJy have either been resolved into multiple, fainter SMGs, or appear to be single SMGs but with slightly lower (S $<$ 9\,mJy) flux densities in the ALMA observations.
This is also reflected in Figure~\ref{fig:fluxcomp} (right) and may indicate that these LESS sources consist of multiple SMGs as well, but that the additional components are below our detection threshold.

\begin{figure}
\centering
\includegraphics[scale=0.6]{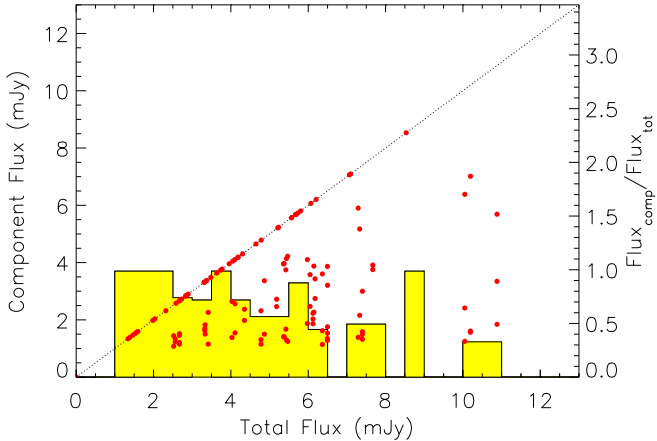}
\caption{Component flux for MAIN sample SMGs versus total flux, where total flux is defined as the sum of the ALESS SMGs. The points show the component flux of individual SMGs. When a point falls on the diagonal line, it means it was the only SMG in that field (and therefore the component flux equals the total flux). The histogram shows the average ratio of component to total flux for a given total flux bin, with values indicated on the right--hand axis. As the total flux increases, the average fractional contribution decreases, indicating a larger average number of SMGs.}
\label{fig:mult}
\end{figure}

It is interesting to speculate on whether Figure~\ref{fig:blanks/blends} implies that bright sources are more likely to be multiples.
The brightest LESS source, LESS 1, is actually a triple, and if you count the faint submillimeter emission in LESS 2 that is coincident with a 24$\mu$m source but just below the 3.5$\sigma$ detection threshold (see Figure~\ref{fig:contourplots} in the Appendix), then LESS 2 may be a triple as well.
Another way to visualize the sample multiplicity is shown in Figure~\ref{fig:mult}, which plots component flux for MAIN sample SMGs against total flux (defined, in this case, as the sum of the SMGs).
Also indicated is the average fractional contribution per component in a given total flux bin, which can be inverted to give the average number of SMGs in that bin (e.g., a fractional contribution of 0.33 implies three SMGs).
From this figure, it appears that as the total flux increases, the average fractional contribution decreases, indicating an increasing number of SMGs per field.
However, because of our sensitivity limit, we can only detect individual SMGs above 3.5$\sigma$, so it is less likely that we would detect multiple SMGs from a fainter LESS source.
Moreover, our observations specifically target sources which were bright in the single--dish (i.e., low--resolution) map.
This strategy would preferentially target regions with multiple SMGs within the single--dish beam, even if SMGs were not thought to be strongly--clustered \citep{2011ApJ...733...92W}.

A separate but related question is whether the multiple SMGs are physically associated. 
A number of studies have concluded that SMGs are strongly--clustered \citep[e.g.,][]{2004ApJ...611..725B,2009ApJ...707.1201W, 2012MNRAS.421..284H}, although other studies have questioned whether this can be explained entirely by the low resolution of single--dish surveys \citep{2011ApJ...733...92W}.
An interferometric study by \citet{2011ApJ...726L..18W} presented two examples of submillimeter sources that were resolved into multiple, physically--unrelated galaxies.
Based on the ALESS number counts alone \citep{Karim_aless}, such a coincidence would seem unlikely, although it is difficult to quantify given that a chance coincidence would also be brighter and thus more easily detectable for a single--dish telescope.

Assuming that the nearby SMGs are physically--related, we can measure their projected separations. 
Do the multiple SMGs tend to be closely grouped together, or do they span the full range of possible separations within the beam?
The predominant theory for SMG formation is that the majority of such sources are starbursting major mergers \citep[e.g.,][]{2003ApJ...599...92C, 2010ApJ...724..233E,2010MNRAS.401.1613N,2011ApJ...743..159H,2012arXiv1203.1318H}. 
In the merger scenario, the gravitational torques induced are thought to be efficient at funneling cold gas to the galaxy's center \citep{1996ApJ...471..115B}, thereby inducing a nuclear starburst and boosting the submillimeter flux observed. 
If these gravitational torques 
are effective at boosting the submillimeter flux on the scales probed here,
we might expect to see an 
increasing number of SMG pairs with smaller separations.

\begin{figure}
\centering
\includegraphics[scale=0.51]{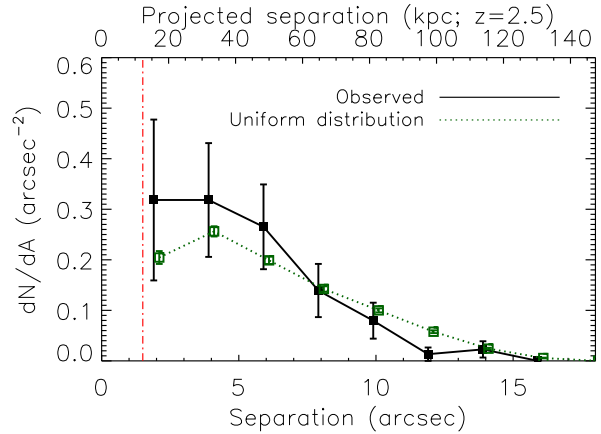}
\caption{Number of MAIN SMG pairs with a given separation and normalized by annular area. 
The separations seen in the MAIN sample are limited to within the primary beam FWHM by definition. 
We also show the simulated result for SMGs with the same multiplicity and flux density distribution as the MAIN sources and placed randomly within the primary beam FWHM. 
The dot--dashed line indicates the typical ALESS resolution. 
Both samples are affected by the decreasing sensitivity of the telescope for large separations, and the ALESS SMGs 
show no evidence for an excess of sources at small separations.} 
\label{fig:sep}
\end{figure}

We have plotted the measured separations for the MAIN ALESS SMGs in Figure~\ref{fig:sep}.
The upper axis converts these separations to projected distance in kpc assuming a typical SMG redshift of $z=2.5$.
The upper limit on separation is set (by definition) by the FWHM of the primary beam.
For comparison, we have also calculated the separations observed for a simulated catalog of SMGs with the same flux density and multiplicity as the MAIN ALESS SMGs, but randomly placed within the primary beam FWHM. 
As with the actual data, simulated SMGs could not be closer than 1.5$^{\prime\prime}$ apart.
We repeated this simulation 100 times to increase statistics, and the results are shown along with the MAIN sample. 
The drop in simulated number density at larger separations is due to the decreasing sensitivity of the telescope with distance from the phase center. 
This affects our ability to detect large separations, as (for example) two MAIN sample SMGs spaced 17$^{\prime\prime}$ apart will both be at the 50\% sensitivity contour, and will thus be harder to detect.

Figure~\ref{fig:sep} shows that, for the scales we are sensitive to, the number density of ALESS SMGs as a function of separation is consistent with what we would expect from a uniformly--distributed population.
This holds true for projected separations from $\sim$13 kpc (our resolution limit at $z\sim2.5$) out to $>$ 140 kpc. 
If we bin all points within 7$^{\prime\prime}$ (60 kpc), where there appears to be a trend toward higher number densities than in the simulation, we find that the slight excess is not statistically significant.
Therefore, the ALESS SMGs 
show no evidence for 
an excess of sources at small separations.



\subsection{Non--detections}
\label{blanks}

There are 88 ALMA maps in total which meet the first two criteria for the MAIN sample (rms $<$ 0.6\,mJy\,beam$^{-1}$, axis ratio $<$ 2) and thus are considered to be of good quality.
However, only 69 
of these fields contain SMGs 
from the MAIN ALESS sample.
If we relax our criteria slightly and also include SMGs from the good quality maps in the Supplementary sample (\S\ref{catalog}), then we find an additional two maps (LESS 91 and 106) which contain a $>$4$\sigma$ SMG just outside the primary beam FWHM.
We therefore find that 17/88 ($\sim$19\%$\pm$4\%) of the LESS sources with high--quality ALMA coverage are non--detections.
A histogram of the LESS flux densities for these fields is shown in Figure~\ref{fig:blanks/blends} compared to the distribution for all good quality maps.
The blank maps may be slightly fainter, in general, than the rest of the LESS sources. 
The average flux of the undetected LESS sources is 5.0$\pm$0.2\,mJy versus 5.9$\pm$0.2\,mJy for all LESS sources, 
and a Komolgorov--Smirnov 
test returns a probability of 80\% that the blank maps are drawn from a different parent distribution.
In particular, there are no sources with S$_{\rm LABOCA}$ $>$\,7.6mJy that are undetected with ALMA.

\begin{figure*}[]
\centering
\includegraphics[scale=0.75]{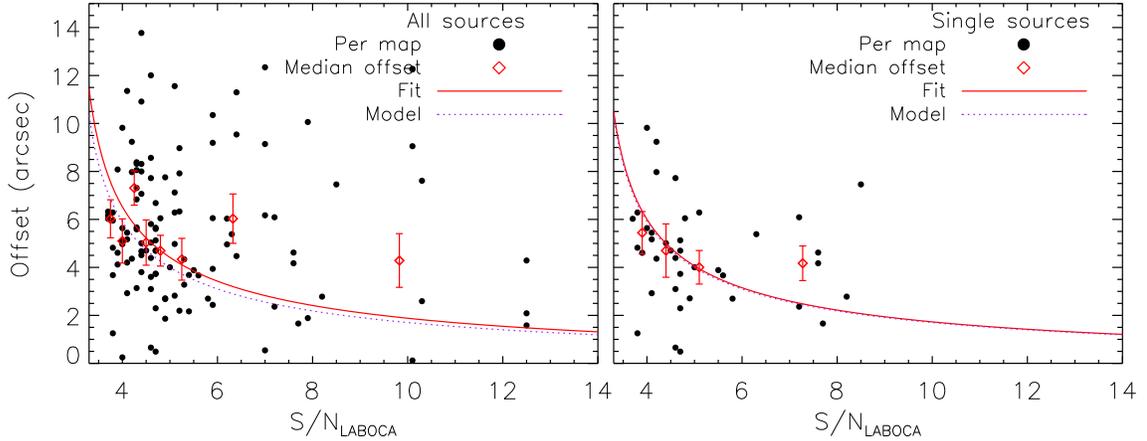}
\caption{Radial offsets between ALESS SMGs and LESS sources as a function of observed (smoothed) LABOCA S/N.
\textit{Left:}
Individual offsets are shown for all ALESS SMGs in good quality maps.
\textit{Right:} Offsets are shown only for ALESS SMGs that constitute the lone source in the field.
For both panels, we also show the median offsets in bins of S/N$_{\rm LABOCA}$. 
The dotted line shows the theoretical expectation for the positional uncertainty \citep{2007MNRAS.380..199I}
assuming the LESS resolution of 19.2$^{\prime\prime}$. 
The solid line shows a fit of the same form. 
The fitted function is consistent with the theoretical prediction,
indicating that there are no additional sources of astrometric error in the LABOCA source positions. 
The median offsets, however, show a systematic bias toward larger offsets than predicted at high S/N, bringing into question the use of S/N--derived search radii in counterpart searches where blending and confusion are significant.}
\label{fig:SNRvsOffset}
\end{figure*}

There are several possible reasons why we may not have detected an SMG with ALMA.
The first possibility is that the LESS source was spurious.
According to \citet{2009ApJ...707.1201W}, only $\sim$5 of the 126 LESS sources are expected be false detections.
In our sample of 88 good quality maps, we might therefore expect $5\times(88/126)=3.5$ of the ALMA maps to target spurious LESS sources.
The large majority of the apparent non--detections therefore cover sources which we expect to be real. 

Alternately, very extended/diffuse SMGs could fall below the detection threshold of the ALMA observations. 
While the ALMA observations are $\sim$3 times deeper than the LABOCA data, the angular resolution is also $>$10 times higher. 
If a 4\,mJy LESS source were a single SMG extended over three or more ALMA beams ($\geq$15--20 kpc in diameter) as is observed in some SMGs in low--$J$ CO transitions \citep{2010MNRAS.404..198I, 2011MNRAS.412.1913I, 2011ApJ...733L..11R, 2011ApJ...739L..31R, 2012ApJ...760...11H}, 
it would be too extended to detect in a map with our median rms sensitivity of 0.4\,mJy\,beam$^{-1}$.
Of the SMGs we do detect, however, the majority are consistent with point sources (\S\ref{sizes}), making this scenario difficult to believe unless the SMG population is a heterogeneous mix of submillimeter morphologies \citep[but see][]{2012arXiv1203.1318H}.

Finally, the LESS source may have been resolved into multiple distinct SMGs which are too faint to be detected individually. 
If all 17 non--detections are actually undetected multiples, then the fraction of multiple sources rises to $\sim$50\%.
The number of components would not need to be very large for a faint LESS source to become undetectable in these ALMA data.
For example, if a 4\,mJy LESS source were composed of three 1.3 mJy SMGs, these SMGs would be below the detection threshold for the maps achieving our target sensitivity.
As an experiment, we have taken the LESS source flux densities and rms noise values of the maps with non--detections and calculated 
how many SMGs would be required for the S/N of each SMG to fall right at the 3.5$\sigma$ detection threshold for that map.
We have simply rounded to the nearest integer number of SMGs, as the difference between this estimate and the LESS flux density is always $\leq$50\% of the 1.2\,mJy uncertainty in LESS flux density.
We have then used the number of SMGs per map (for which we find a median of 4) and the flux densities of these hypothetical sources
to put a lower limit on the faint--end number counts of SMGs,
finding that at S$_{\rm 870\mu m}\sim$1.3\,mJy, $>$500 SMGs mJy$^{-1}$ deg$^{-2}$ would be required to explain our non--detections.
This number is consistent with the faint--end extrapolation of the ALESS number counts presented in \citet{Karim_aless}. 
There is even some tentative observational evidence that this may be the case in at least one of the ALESS SMGs. 
For example,
in LESS 27 (a blank map) there are three 3$\sigma$ sources lying exactly on top of three radio/mid--IR--predicted counterparts, suggesting they may be real despite the fact that they are all below our adopted detection threshold.
Future, deeper observations with ALMA will 
reveal whether 
all of the undetected LESS sources are indeed multiple, faint SMGs.


\subsection{LABOCA Positional Offsets}
\label{offsets}

Prior to the existence of large, interferometrically--observed SMG samples such as that presented here, the identification of counterparts at other wavelengths relied on statistical methods to estimate the likelihood of particular associations \citep[e.g.,][]{2011MNRAS.413.2314B}.
In such cases, the positional offsets between the submillimeter source and wavelength of interest are dominated by the uncertainty in the (single--dish) submillimeter positions, and this uncertainty, in turn, is thought to be primarily a function of S/N of the single--dish submillimeter source \citep[e.g.,][]{2007MNRAS.380..199I}.
Specifically, in the limit where centroiding uncertainty dominates over systematic astrometry errors, and for uncorrelated Gaussian noise, 
the theoretical expectation for the positional uncertainty in single--dish submillimeter sources is 
\begin{equation}
\Delta\alpha = \Delta\delta = k\,\theta[(S/N_{\rm app})^2-(2\beta+4)]^{-1/2}
\label{eqn:offset}
\end{equation}
where $\Delta\alpha$ and $\Delta\delta$ are the rms positional errors in right ascension and declination, respectively, 
the constant $k = 0.6$,
$\theta$ is the FWHM of the single--dish primary beam (19.2$^{\prime\prime}$), S/N$_{\rm app}$ is the apparent, smoothed signal--to--noise ratio before correcting for flux boosting, and the $\beta$ term provides the correction to intrinsic S/N \citep{2007MNRAS.380..199I}.
Some counterpart searches, such as that done by \citet{2011MNRAS.413.2314B} for LESS, have used this fact to vary the search radius based on the S/N of the submillimeter source.

Now that we have both single--dish LABOCA and ALMA observations of a large number of submillimeter sources, we are in a position to empirically calibrate the positional uncertainties of the LABOCA sources, and thus the search radius needed to recover SMGs from multiwavelength comparisons.
Figure~\ref{fig:SNRvsOffset} (left) shows radial offset between the ALMA and LABOCA positions (where the latter correspond to phase center in the ALMA maps) as a function of apparent (smoothed) S/N for all ALESS SMGs in the good quality maps (i.e., MAIN and Supplementary sample 2). 
Also shown are the median offsets in bins of S/N$_{\rm app}$.
Fitting a function to the median data points of the form given in Equation~\ref{eqn:offset}
(with a $\sqrt{2}$ correction to radial offset),
we find $k = 0.66\pm0.03$, or $\sim$10\% worse than the theoretical expectation using this FWHM.

Note that many maps have $\geq$2 SMGs over the S/N threshold which are blended into one brighter source in the LABOCA maps.
The theoretical prediction, on the other hand, describes the positional offset expected for a single source of the given S/N.
If we limit the comparison to maps with only a single ALESS SMG (Figure~\ref{fig:SNRvsOffset}, right), 
then the fitted function shifts down slightly,
giving
$k = 0.61\pm0.06$, consistent with the theoretical prediction.
We therefore find that there are no additional sources of astrometric error for the LESS sources.

Although the fitted function is consistent with theory, the normalization of the function is largely constrained by the data points at the low--S/N end.
At higher S/N, we see in Figure~\ref{fig:SNRvsOffset} that the median offsets measured are systematically higher than the predicted offsets.
This is likely to be a consequence of our finding earlier that a larger fraction of bright submillimeter sources are multiples.
Hence, although recent searches for counterparts at other wavelengths have employed a S/N--dependent search radius, Figure~\ref{fig:SNRvsOffset} implies that a fixed search radius based on the typical S/N of the data may be just as good or even better than a S/N--dependent radius.




\subsection{Radio/MIR IDs}
\label{radioIDs}

Previously, \citet{2011MNRAS.413.2314B} used radio and mid--infrared data from the Very Large Array and \textit{Spitzer} to identify potential counterparts to the LESS sources. 
Using their radio and/or 24$\mu$m emission along with the corrected Poissonian probability \citep[$p$--statistic;][]{1978MNRAS.182..181B,1986MNRAS.218...31D} 
and a S/N--dependent search radius (discussed in the previous section),
they identified statistically robust counterparts to 62 of the 126 submillimeter sources in the LESS survey. 
(Note that they did not correct for the small systematic offset between the radio and MIPS data discussed in Section~\ref{astrometry}, as they concluded that the effect was miniscule.)
\citeauthor{2011MNRAS.413.2314B} then used a color--flux cut on IRAC 3.6$\mu$m and 5.8$\mu$m sources (chosen to maximize the number of secure radio and 24$\mu$m counterparts) to identify 17 additional sources 
whose IRAC colors suggest they are likely counterparts.
In total, they identified robust counterparts to 79 LESS sources, some of which have multiple radio/mid--infrared identifications (Figure~\ref{fig:falsecolor} in the Appendix).

With our interferometric submillimeter imaging, we are now in a position to test how many of the ALESS SMGs were correctly predicted by these radio/mid--infrared identifications.
Taking the 99 ALESS SMGs in the MAIN sample, we find that 45 have robust radio/mid--infrared counterparts.
These SMGs are indicated with a boldface `r' in Tables~\ref{tab-3} and \ref{tab-4}.
Note that a counterpart was considered `robust' by \citeauthor{2011MNRAS.413.2314B} if it had a corrected Poissonian probability $p \leq 0.05$ in either the radio, 
MIPS, or IRAC data,
or a value $0.05 < p < 0.1$ in two separate wavelengths.
If we also include tentative counterparts (defined as having $0.05 < p < 0.1$ in just one wavelength), then we find an additional 9 ALESS SMGs.
These SMGs are indicated with a `t' in Tables~\ref{tab-3} and \ref{tab-4}.
The remaining 45 ALESS SMGs have neither robust nor tentative counterparts.
The radio/mid--infrared counterpart identification is therefore $\sim$55\% complete if we include both robust and tentative counterparts.

From Figure~\ref{fig:falsecolor}, it is immediately obvious that a significant number of maps have ALESS SMGs outside the search radius used by \citet{2011MNRAS.413.2314B} to predict these counterparts.
Of the four brightest LESS sources with good quality ALMA maps (LESS 1, 2, 3, and 5), all four have at least one SMG outside the search radius, and in two of these cases, it is the brightest source in the map.
In total, 21 out of 88 good--quality fields have at least one ALESS SMG outside the search radius, 
whereas \citet{2011MNRAS.413.2314B} estimated that only 1\% of all the LESS SMGs (i.e., 1.3 of 126 sources) would have counterparts missed for this reason. 
The LESS sources with counterparts outside the search radius tend to be the brighter sources, meaning that they had a smaller adopted search radius.
This implies, as we already indicated in \S\ref{offsets}, that the S/N--dependent search radius may not be the most effective means to identify counterparts.

\begin{figure}[]
\centering
\includegraphics[scale=0.5]{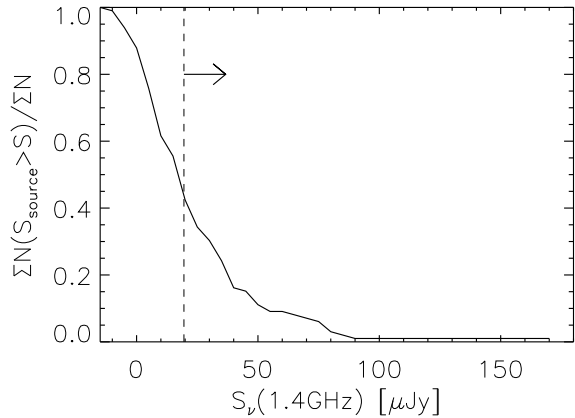}
\caption{Cumulative distribution function of 1.4 GHz flux density values extracted from the map of \citet{2013arXiv1301.7004M} at the positions of the MAIN ALESS SMGs. The dashed vertical line shows the 3$\sigma$ detection limit at the deepest part of the map, which is therefore a lower limit.
Despite reaching an rms of 6$\mu$Jy at its deepest point, at least 55\% of the ALESS SMGs are still undetected in the radio data. Our stacking analysis indicates that the median 1.4 GHz flux density of the ALESS SMGs is 15$\pm$1 $\mu$Jy.}
\label{fig:VLAfluxhist}
\end{figure}

Another possible issue with the counterpart identification is the depth of the multiwavelength data used.
This could result in missed counterparts no matter what the search radius.
The 1.4 GHz data used for the LESS source counterpart identification had an rms of $\sim$6.5$\mu$Jy at its deepest point \citep{2008ApJS..179..114M}, resulting in a 3$\sigma$ detection threshold of 19.5$\mu$Jy or higher.
We used the map from the second data release \citep[DR2;][]{2013arXiv1301.7004M} to extract cutouts at the positions of all the ALESS SMGs in the MAIN sample.
Stacking these cutouts gives a median flux density of 15$\pm$1 $\mu$Jy (a 15$\sigma$ detection), with a statistical error of 2.7$\mu$Jy.
The cumulative distribution function of the peak flux densities is shown in Figure~\ref{fig:VLAfluxhist},
where we see that at least 55\% of the ALESS SMGs are still undetected.
In comparison, \citet{2012arXiv1209.1626B} recently took 16 interferometrically--observed SMGs and looked for counterparts in a deeper 1.4 GHz image (rms of 2.5$\mu$Jy), detecting all 16 SMGs at $>$5$\sigma$.
This difference may be explained largely by the relative brightness of their detected SMGs (S$_{860\mu m}>3.2\,mJy$),
though at least two of their SMGs fall below our nominal 1.4 GHz detection limit.
\citet{2011ApJ...737...83L} also report a very high radio--detection rate for submillimeter sources in a deep (rms$=$ 2.7 $\mu$Jy) radio map.
Despite our high rate of non--detections in the radio, the radio data still help to identify a larger fraction of ALESS counterparts overall than the MIPS or IRAC data.
If we categorize the results based on wavelength, we find that 28\% of the correctly predicted counterparts were based on the radio data alone, 13\% were based on the MIPS data alone, and 31\% were based on IRAC data alone,
with 
an additional 28\% based on either radio$+$MIPS (26\%) or radio$+$IRAC (2\%).

The 55\% overall completeness quoted above refers to the percentage of all ALESS SMGs predicted, regardless of whether some of the SMGs correspond to the same LESS source (i.e.\ are multiples). 
While there may also be multiple radio/mid--infrared IDs per field, it may be interesting to look at what fraction of LESS sources have \textit{at least} one correct ID. 
In total, of the 69 LESS sources covered by the MAIN ALESS sample, 52 (75\%) have at least one correct robust or tentative radio/mid--infrared ID.
We find that for those LESS sources with multiple ALESS SMGs, 80\% have at least one SMG that was correctly predicted.
The brightest ALESS SMG is among the predicted SMGs for the majority (80\%) of those cases. 
Therefore, while the radio/mid--infrared ID process only predicts 55\% of SMGs in total, it has a higher success rate if we consider only the brightest SMG in each field.

The flux density distributions of the confirmed robust/tentative counterparts are shown in Figure~\ref{fig:blanks/blends}. 
The robust IDs clearly favor the brighter ALESS SMGs, with 75\% of the SMGs above 5\,mJy matching previously predicted radio/mid--infrared counterparts (versus only 35\% of the SMGs below 5\,mJy).
If we include tentative counterparts, then the flux density above which the predictions are 75\% complete drops to only 3\,mJy,
below which only 25\% of the SMGs were predicted with robust/tentative radio/mid--infrared counterparts.

We can also test how many of the total number of predicted counterparts are now confirmed (i.e.\ the reliability).
Considering only robust IDs, there are a total of 57 distinct proposed radio/mid--infrared counterparts in maps covered by the ALESS MAIN sample. 
Of these counterparts, 45 are confirmed by the ALMA maps. 
Therefore, 80\% of the robust radio/mid--infrared IDs have been confirmed by the ALMA maps, and 20\% have been shown to be incorrect.
Although formally the robust IDs should have a $>$95\% chance of being correct, a reliability of 80\% is still encouraging given the uncertainties.
Moreover, our results really only give a lower limit on the reliability, since there may be some counterparts with submillimeter emission just below our detection threshold.
For example, in LESS 2, there is a predicted counterpart at the position of a $\sim$3$\sigma$ submillimeter peak, which is just below our detection threshold but (given this alignment) likely real.
If we consider both robust and tentative IDs, then there are a total of 85 radio/mid--infrared counterparts (57 robust $+$ 28 tentative) falling in MAIN sample maps. 
Of these, 54 are confirmed by the ALMA maps 
($\sim$65\%).

Note that we did not include the blank maps in the above analysis, even though they are also of high quality, since the positions of the SMG(s) in those maps remain unconstrained.
Of the 17 blank maps, 11 have at least one predicted radio/mid--IR counterpart.
In several cases -- e.g., LESS 27, LESS 47, LESS 95 (slightly offset), and LESS 120 -- these counterparts coincide with a $\sim$3$\sigma$ submillimeter peak, indicating that the predicted SMG may be correct but just below our detection threshold.
In the case of LESS 27, there are actually three predicted counterparts, and all three coincide with $\sim$3$\sigma$ sources in the ALMA map. 
This case in particular supports the proposal put forward in \S\ref{blanks} that these maps are blank because they have been resolved into multiple SMGs with flux densities too faint to detect.

To summarize, if we consider only robust counterparts, then the radio/mid--infrared identification process has a completeness of only $\sim$45\%, but a reliability of $\sim$80\%.
If we include tentative counterparts as well, then the completeness rises to $\sim$55\% and the reliability drops to $\sim$65\%.
This process therefore still misses $\sim$45\% of SMGs, and of those it claims to find, approximately one--third are incorrect.
We conclude that submillimeter interferometry is really the best and only way to obtain accurate identifications for SMGs.


\section{SUMMARY}
\label{summary}

We have presented an ALMA Cycle 0 survey of 
SMGs in the Extended \textit{Chandra} Deep Field South. 
These SMGs were originally detected with the (single--dish) LESS survey on the APEX telescope,
the single largest, most homogenous 
870$\mu$m survey to date.
Our ALMA observations utilize the Cycle 0 compact array in Band 7 to map 122 of the 126 SMGs at the same central frequency as LESS.
In just two minutes per source, we reach a median rms noise of 0.4\,mJy\,beam$^{-1}$, three times as deep as the LESS observations.
Our median angular resolution (1.6$^{\prime\prime}$ $\times$ 1.15$^{\prime\prime}$) represents an improvement in beam area of $\sim$200 times, allowing us to precisely locate the origin of the submillimeter emission from the SMGs.

We 
identify and extract sources from the final maps.
recovering 99\% of all SMGs above 3.5$\sigma$ with a spurious fraction of 1.6\%.
From an analysis of the inverted maps, we estimate that 75\%/90\% of SMGs above 3.5$\sigma$/4$\sigma$ may be regarded as reliable, though these estimates should be regarded 
as lower limits due to the complex and correlated nature of the noise in the maps.

To test the absolute flux scale, we model the true sky distribution using both positive and negative sources down to 3$\sigma$ and convolving with the resolution of LABOCA for a more meaningful comparison.
We find S$_{\rm ALMA}$/S$_{\rm LABOCA}$ $=$ 0.97$\pm^{0.07}_{0.04}$,
confirming that the ALMA and LABOCA flux density scales are in reasonable agreement.
An analysis of the astrometry indicates that the ALMA SMG positions are accurate to within 0.2$^{\prime\prime}$--0.3$^{\prime\prime}$.

We discuss the creation of the catalog, including a MAIN sample of 99 ALESS SMGs and a supplementary sample of 32 SMGs.
The MAIN sample SMGs are the most reliable, 
while the supplementary sample SMGs should be used with care.

Using this catalog, we put constraints on the sizes of the SMGs in submillimeter continuum. 
We find that all but one SMG are consistent with point sources, suggesting sizes $<$10 kpc and a median SFR surface density $>$14 M$_{\odot}$ yr$^{-1}$ kpc$^{-2}$.
The resolved SMG (ALESS 007.1) has an intrinsic source size of (1.1$^{\prime\prime}$$\pm$0.3$^{\prime\prime}$) $\times$ (0.7$^{\prime\prime}$$\pm$0.2$^{\prime\prime}$),
corresponding to $\sim$9 kpc $\times$ 5 kpc at its estimated redshift of $z_{\rm phot} = 2.8$.
Its SFR surface density is $\sim$80 M$_{\odot}$ yr$^{-1}$ kpc$^{-2}$, consistent with previous measurements for SMGs from high--$J$ CO lines and well below the limit for Eddington--limited star formation.

With our $\sim$1.6$^{\prime\prime}$ resolution, we find that many of the LESS sources are resolved into multiple distinct SMGs.
Of the good quality maps with at least one detection, $\sim$35\%--45\% contain multiple SMGs.
This is likely due to the low resolution of the original LESS survey and/or the clustering of SMGs.
We find that the brighter sources are preferentially affected, though this may be an artifact of the single--dish selection and sensitivity limitations.
An analysis of the separation between multiple SMGs yields no evidence for a significant excess of SMGs with small ($<$60 kpc) projected separations.

We report that 17 out of 88 good--quality maps ($\sim$20\%) have no detected SMGs, 
even though only $\sim$3.5 of the LESS sources are expected to be spurious.
It may be that the emission from these LESS sources has been spread out 
into diffuse or multi--component morphologies below our detection threshold.
Assuming 
it is the latter,
we calculate 
the number of SMGs necessary in each map for the flux densities of the SMGs to be just below the detection threshold. 
We find that at S$_{\rm 870\mu m} \sim 1.3$ mJy, $>$500 SMGs mJy$^{-1}$ deg$^{-2}$ would be necessary in order to explain our non--detections, a number which is consistent with the faint end extrapolation of the ALESS number counts \citep{Karim_aless}.
If all 17 blank--map sources are actually multiple SMGs, the total fraction of multiple SMGs in the survey rises to $\sim$50\%.

We use the ALESS catalog to
empirically calibrate the uncertainty in the LABOCA positions as a function of signal--to--noise.
We find that the observed positional offsets are consistent with the theoretical prediction,
indicating no additional sources of astrometric error for the LESS sources.
However, the median offsets measured are systematically higher than predicted offsets at large values of S/N, 
implying that counterpart searches may be better off using a fixed search radius than a S/N--dependent search radius. 

Finally, we use the precise submillimeter positions for our SMGs to test the reliability/completeness of 
radio/mid--IR counterparts previously predicted using a probabilistic method.
We find that overall, the radio/mid--infrared ID process only correctly predicts $\sim$55\% of ALESS SMGs.
It has a higher success rate for the brighter LESS sources, or if we only consider the brightest ALESS SMG in fields with multiple SMGs.
The depth of the radio data may be one limiting factor, although there are also a significant number of cases where the SMG falls outside the (S/N--dependent) search radius,
again bringing its effectiveness into question.
Of the counterparts predicted by the radio/mid--IR method, we also find that one--third are incorrect, further highlighting the importance of submillimeter interferometry for obtaining accurate identifications for SMGs.


In conclusion, we have completed the first statistically--reliable survey of SMGs by combining a wide--field single--dish 870$\mu$m survey using LABOCA with targeted interferometric 870$\mu$m observations of detected sources with ALMA.
This study provides the basis for the unbiased multiwavelength study of the properties of this population, including the high--resolution number counts of SMGs \citep{Karim_aless},
constraints on the number density of $z\sim4$ SMGs \citep{2012MNRAS.427.1066S},
the properties of AGN in SMGs \citep[][Wang et al.\ in prep]{2012arXiv1208.4846C},
and their redshift distribution (Simpson et al.\ in prep).

\acknowledgements
The authors wish to thank George Bendo and the Manchester ALMA ARC node for their support.
AK acknowledges support from STFC, and AMS and TRG gratefully acknowledge STFC Advanced Fellowships.
IRS acknowledges support from STFC and a Leverhume Fellowship. 
KEKC acknowledges support from the endowment of the Lorne Trottier Chair in Astrophysics and Cosmology at McGill, the Natural Science and Engineering Research Council of Canada (NSERC), and a L'Oreal Canada for Women in Science Research Excellence Fellowship, with the support of the Canadian Commission for UNESCO.
TRG acknowledges the IDA and DARK.
This paper makes use of the following ALMA data: ADS/JAO.ALMA\#2011.1.00294.S. 
ALMA is a partnership of ESO (representing its member states), NSF (USA) and NINS (Japan), together with NRC (Canada) and NSC and ASIAA (Taiwan), in cooperation with the Republic of Chile. The Joint ALMA Observatory is operated by ESO, AUI/NRAO and NAOJ.
This publication also makes use of data acquired with the APEX under program IDs 078.F-9028(A), 079.F-9500(A), 080.A-3023(A) and 081.F-9500(A). APEX is a collaboration between the Max--Planck--Institut f\"ur Radioastronomie, the European Southern Observatory and the Onsala Space Observatory.


\appendix

\section{Details of the source extraction and characterization}

As discussed in \S\ref{srcextract}, we used custom--written {\sc idl}--based source extraction software to identify and fit SMGs in the ALMA maps.
We already demonstrated that {\sc casa}'s {\sc imfit} task returns comparable values for the fits, verifying our results.
Nevertheless, for completeness, we now elaborate further on our source extraction and characterization process.

For each of the fits (three-- and six--parameters),
we determined the step size for each parameter that best balanced fast convergence and good mixing of the Markov chain by looking at individual maps.
On average, we reached an acceptance probability of 30\% after an initial and sufficiently long thermalization phase.
This value is within the optimal range for a multi-dimensional MH-MCMC algorithm. 
The box size was optimized to ensure that all visually apparent double sources in the maps were individually recovered while minimizing the number of clear fitting artifacts.

There are physical prior constraints for a valid model description of the flux density distribution of an elliptical source model in an interferometry map. 
In order not to violate the detailed balance condition and to ensure ergodicity of the Markov chain, however, it is important not to constrain the random walk through the six-dimensional parameter space which is consequently chosen to follow a symmetric attempt distribution about the previous value for a given parameter.
We do not violate the MH-MCMC principles if we constrain the actual move acceptance which therefore underlies the following constraints:
\begin{enumerate}
\item The minor extent of a given source cannot be smaller than the clean beam minor axis.
\item The ratio of major to minor axis must not exceed twice the corresponding axis ratio of the clean beam.
\item The peak flux density must be a positive value.
\end{enumerate}

For each parameter, the mean of the posterior distribution determined its best fit value.
The full set of parameters obtained in this way was used as a flux density model to be subtracted off the initial map before proceeding to the next signal peak within the map.
This process was repeated until the threshold was reached, and the end result was a combined model as well as a residual map.

\section{Error determination, integrated flux densities, and beam deconvolution}

In the presence of correlations in the Markov chain, effectively only $N/2\theta_A$ statistically independent measurements of the parameter $A$ have been performed, where $\theta_A$ denotes the data autocorrelation length and $N$ is the number of MCMC steps after the equilibration phase. 
In practice, we therefore monitor the autocorrelation of the Markov chain and bin the data in $N_B=N/N_L$ blocks of length $N_L > \theta_A$. We then calculate for each bin the mean value:

\for{\label{binmean}\overline{A}_{i} = \frac{1}{N_L} \sum_{k=1}^{N_L}A_{(i-1)N_L+k}, \qquad i=1,\ldots,N_B,}

\noindent where the index denotes that we are dealing with consecutive members of the Markov chain. 
The set of bin mean values $\big\{\overline{A}_{i}\big\}$ will then be statistically independent so that the best--fit parameter solution is simply given by the mean of all bin means, while its corrected standard error is given by:

\for{\label{estimerr} \Delta A  \approx  \frac{1}{\sqrt{N_B-1}}  \left(\frac{1}{N_B} \sum_{i=1}^{N_B}  \Big(\overline{A}_{i}\Big)^2-  \left(\frac{1}{N_B}\sum_{i=1}^{N_B} \overline{A}_{i}\right)^2\right)^{-1/2}}

The integrated flux density is derived as the product of the best--fit peak flux density and the ratio of the fitted source area to the clean beam area.
The appropriate determination of measurement errors for elliptical source fits in the presence of correlated noise in radio maps has been extensively worked out by \citet{1997PASP..109..166C}. \citep[See also][]{1984A&AS...58....1W, 2003AJ....125..465H, 2004AJ....128.1974S, 2010ApJS..188..384S, 2011ApJ...730...61K}.
We adopt those findings for 
error estimates for our measured source parameters, taking into account also the pure autocorrelation--corrected fitting errors based on the posterior distribution function for each parameter.


For this study the main quantity of interest is the total flux density and its full error is given by:
\for{\label{eq:errftot}\frac{\sigma_{\rm{Total}}}{\langle F_{\rm{Total}} \rangle}=\sqrt{\left( \frac{\sigma_{\rm{data}}}{\langle F_{\rm{Total}} \rangle} \right)^2 + \left( \frac{\sigma_{\rm{fit}}}{\langle F_{\rm{Total}}\rangle} \right)^2 },}
where 
\begin{eqnarray} \label{eq:errftotd}
\frac{\sigma_{\rm{data}}}{\langle F_{\rm{Total}} \rangle} &=& \sqrt{\left(\frac{S}{N}\right)^{-2} + \left( \frac{1}{100} \right)^2}\\ \label{eq:errftotf}
\frac{\sigma_{\rm{fit}}}{\langle F_{\rm{Total}} \rangle} &=& \sqrt{\frac{2}{\rho_S} + \left( \frac{\theta\st{B} \theta\st{b}}{\theta\st{M} \, \theta\st{m}} \right) \left( \frac{2}{\rho^2_{\psi}} + \frac{2}{\rho^2_{\phi}} \right)}.
\end{eqnarray}
$\theta\st{B}$ is the major axis and $\theta\st{b}$ the minor axis of a given clean beam while $\theta\st{M}$ and $\theta\st{m}$ are the major and minor axis of the measured (hence convolved) source ellipse. The signal-to-noise ratio $S/N= \overline{F}_{\rm{peak}} /\sigma_{\rm{map}}$ denotes the ratio of the best-fit peak flux density and RMS noise for a given map. The same applies to the parameter-dependent estimators of the fit entering equation \gl{eq:errftotf} that are given by:
\for{\rho^2_X = \frac{\theta\st{M} \, \theta\st{m}}{4 \, \theta\st{B} \theta\st{b}} \left( 1 + \frac{\theta\st{B}}{\theta\st{M}} \right)^{a} \left( 1 + \frac{\theta\st{b}}{\theta\st{m}} \right)^{b} \left(\frac{S}{N} \right)^2}
and $a = b = 1.5$ for $ \rho_F$, $a = 2.5$ and $b = 0.5$ for $ \rho\st{M}$ as well as $a =0.5$ and $b = 2.5$ for $ \rho\st{m}$.

For each source found we attempt to trigonometrically deconvolve the clean beam of the corresponding ALMA map
in order to determine its intrinsic extent, shape and orientation, assuming the underlying source can be described by an elliptical Gaussian using our own implementation of the well-tested {\sc AIPS}-equivalent.
We call a source unresolved in case both its deconvolved minor and major axis result in negative values while we allow sources to be extended only in one dimension. Given the typical resolution achieved in this survey sources are mostly barely resolved and the (on average) low reliability of deconvolved quantities for (partially) resolved sources is reflected in their typically large errors.


\bibliographystyle{apj}		
\bibliography{aLESScat}

\begin{thebibliography}{75}
\expandafter\ifx\csname natexlab\endcsname\relax\def\natexlab#1{#1}\fi

\bibitem[{{Aravena} {et~al.}(2010){Aravena}, {Younger}, {Fazio}, {Gurwell},
  {Espada}, {Bertoldi}, {Capak}, \& {Wilner}}]{2010ApJ...719L..15A}
{Aravena}, M., {Younger}, J.~D., {Fazio}, G.~G., {Gurwell}, M., {Espada}, D.,
  {Bertoldi}, F., {Capak}, P., \& {Wilner}, D. 2010, \apjl, 719, L15

\bibitem[{{Barger} {et~al.}(1998){Barger}, {Cowie}, {Sanders}, {Fulton},
  {Taniguchi}, {Sato}, {Kawara}, \& {Okuda}}]{1998Natur.394..248B}
{Barger}, A.~J., {Cowie}, L.~L., {Sanders}, D.~B., {Fulton}, E., {Taniguchi},
  Y., {Sato}, Y., {Kawara}, K., \& {Okuda}, H. 1998, \nat, 394, 248

\bibitem[{{Barger} {et~al.}(2012){Barger}, {Wang}, {Cowie}, {Owen}, {Chen}, \&
  {Williams}}]{2012arXiv1209.1626B}
{Barger}, A.~J., {Wang}, W.-H., {Cowie}, L.~L., {Owen}, F.~N., {Chen}, C.-C.,
  \& {Williams}, J.~P. 2012, ArXiv e-prints

\bibitem[{{Barnes} \& {Hernquist}(1996)}]{1996ApJ...471..115B}
{Barnes}, J.~E. \& {Hernquist}, L. 1996, \apj, 471, 115

\bibitem[{{Beckwith} {et~al.}(2006){Beckwith}, {Stiavelli}, {Koekemoer},
  {Caldwell}, {Ferguson}, {Hook}, {Lucas}, {Bergeron}, {Corbin}, {Jogee},
  {Panagia}, {Robberto}, {Royle}, {Somerville}, \&
  {Sosey}}]{2006AJ....132.1729B}
{Beckwith}, S.~V.~W., {Stiavelli}, M., {Koekemoer}, A.~M., {Caldwell},
  J.~A.~R., {Ferguson}, H.~C., {Hook}, R., {Lucas}, R.~A., {Bergeron}, L.~E.,
  {Corbin}, M., {Jogee}, S., {Panagia}, N., {Robberto}, M., {Royle}, P.,
  {Somerville}, R.~S., \& {Sosey}, M. 2006, \aj, 132, 1729

\bibitem[{{Bertoldi} {et~al.}(2000)}]{2000A&A...360...92B}
{Bertoldi}, F. {et~al.} 2000, \aap, 360, 92

\bibitem[{{Biggs} \& {Ivison}(2008)}]{2008MNRAS.385..893B}
{Biggs}, A.~D. \& {Ivison}, R.~J. 2008, \mnras, 385, 893

\bibitem[{{Biggs} {et~al.}(2011)}]{2011MNRAS.413.2314B}
{Biggs}, A.~D. {et~al.} 2011, \mnras, 413, 2314

\bibitem[{{Blain} {et~al.}(2004){Blain}, {Chapman}, {Smail}, \&
  {Ivison}}]{2004ApJ...611..725B}
{Blain}, A.~W., {Chapman}, S.~C., {Smail}, I., \& {Ivison}, R. 2004, \apj, 611,
  725

\bibitem[{{Blain} {et~al.}(2002){Blain}, {Smail}, {Ivison}, {Kneib}, \&
  {Frayer}}]{2002PhR...369..111B}
{Blain}, A.~W., {Smail}, I., {Ivison}, R.~J., {Kneib}, J.-P., \& {Frayer},
  D.~T. 2002, \physrep, 369, 111

\bibitem[{{Bothwell} {et~al.}(2010)}]{2010MNRAS.405..219B}
{Bothwell}, M.~S. {et~al.} 2010, \mnras, 405, 219

\bibitem[{{Bothwell} {et~al.}(2012)}]{2012arXiv1205.1511B}
---. 2012, MNRAS, in press

\bibitem[{{Browne} \& {Cohen}(1978)}]{1978MNRAS.182..181B}
{Browne}, I.~W.~A. \& {Cohen}, A.~M. 1978, \mnras, 182, 181

\bibitem[{{Chapman} {et~al.}(2005){Chapman}, {Blain}, {Smail}, \&
  {Ivison}}]{2005ApJ...622..772C}
{Chapman}, S.~C., {Blain}, A.~W., {Smail}, I., \& {Ivison}, R.~J. 2005, \apj,
  622, 772

\bibitem[{{Chapman} {et~al.}(2004){Chapman}, {Smail}, {Windhorst}, {Muxlow}, \&
  {Ivison}}]{2004ApJ...611..732C}
{Chapman}, S.~C., {Smail}, I., {Windhorst}, R., {Muxlow}, T., \& {Ivison},
  R.~J. 2004, \apj, 611, 732

\bibitem[{{Chapman} {et~al.}(2003){Chapman}, {Windhorst}, {Odewahn}, {Yan}, \&
  {Conselice}}]{2003ApJ...599...92C}
{Chapman}, S.~C., {Windhorst}, R., {Odewahn}, S., {Yan}, H., \& {Conselice}, C.
  2003, \apj, 599, 92

\bibitem[{{Condon}(1997)}]{1997PASP..109..166C}
{Condon}, J.~J. 1997, \pasp, 109, 166

\bibitem[{{Coppin} {et~al.}(2006)}]{2006MNRAS.372.1621C}
{Coppin}, K. {et~al.} 2006, \mnras, 372, 1621

\bibitem[{{Coppin} {et~al.}(2012)}]{2012arXiv1208.4846C}
---. 2012, MNRAS, in press

\bibitem[{{Dannerbauer} {et~al.}(2002){Dannerbauer}, {Lehnert}, {Lutz},
  {Tacconi}, {Bertoldi}, {Carilli}, {Genzel}, \&
  {Menten}}]{2002ApJ...573..473D}
{Dannerbauer}, H., {Lehnert}, M.~D., {Lutz}, D., {Tacconi}, L., {Bertoldi}, F.,
  {Carilli}, C., {Genzel}, R., \& {Menten}, K. 2002, \apj, 573, 473

\bibitem[{{Dannerbauer} {et~al.}(2008){Dannerbauer}, {Walter}, \&
  {Morrison}}]{2008ApJ...673L.127D}
{Dannerbauer}, H., {Walter}, F., \& {Morrison}, G. 2008, \apjl, 673, L127

\bibitem[{{Devlin} {et~al.}(2009){Devlin}, {Ade}, {Aretxaga}, {Bock}, {Chapin},
  {Griffin}, {Gundersen}, {Halpern}, {Hargrave}, {Hughes}, {Klein}, {Marsden},
  {Martin}, {Mauskopf}, {Moncelsi}, {Netterfield}, {Ngo}, {Olmi}, {Pascale},
  {Patanchon}, {Rex}, {Scott}, {Semisch}, {Thomas}, {Truch}, {Tucker},
  {Tucker}, {Viero}, \& {Wiebe}}]{2009Natur.458..737D}
{Devlin}, M.~J., {Ade}, P.~A.~R., {Aretxaga}, I., {Bock}, J.~J., {Chapin},
  E.~L., {Griffin}, M., {Gundersen}, J.~O., {Halpern}, M., {Hargrave}, P.~C.,
  {Hughes}, D.~H., {Klein}, J., {Marsden}, G., {Martin}, P.~G., {Mauskopf}, P.,
  {Moncelsi}, L., {Netterfield}, C.~B., {Ngo}, H., {Olmi}, L., {Pascale}, E.,
  {Patanchon}, G., {Rex}, M., {Scott}, D., {Semisch}, C., {Thomas}, N.,
  {Truch}, M.~D.~P., {Tucker}, C., {Tucker}, G.~S., {Viero}, M.~P., \& {Wiebe},
  D.~V. 2009, \nat, 458, 737

\bibitem[{{Downes} {et~al.}(1986){Downes}, {Peacock}, {Savage}, \&
  {Carrie}}]{1986MNRAS.218...31D}
{Downes}, A.~J.~B., {Peacock}, J.~A., {Savage}, A., \& {Carrie}, D.~R. 1986,
  \mnras, 218, 31

\bibitem[{{Dunne} {et~al.}(2000){Dunne}, {Clements}, \&
  {Eales}}]{2000MNRAS.319..813D}
{Dunne}, L., {Clements}, D.~L., \& {Eales}, S.~A. 2000, \mnras, 319, 813

\bibitem[{{Eales} {et~al.}(1999){Eales}, {Lilly}, {Gear}, {Dunne}, {Bond},
  {Hammer}, {Le F{\`e}vre}, \& {Crampton}}]{1999ApJ...515..518E}
{Eales}, S., {Lilly}, S., {Gear}, W., {Dunne}, L., {Bond}, J.~R., {Hammer}, F.,
  {Le F{\`e}vre}, O., \& {Crampton}, D. 1999, \apj, 515, 518

\bibitem[{{Engel} {et~al.}(2010)}]{2010ApJ...724..233E}
{Engel}, H. {et~al.} 2010, \apj, 724, 233

\bibitem[{{Frayer} {et~al.}(2000){Frayer}, {Smail}, {Ivison}, \&
  {Scoville}}]{2000AJ....120.1668F}
{Frayer}, D.~T., {Smail}, I., {Ivison}, R.~J., \& {Scoville}, N.~Z. 2000, \aj,
  120, 1668

\bibitem[{{Gear} {et~al.}(2000){Gear}, {Lilly}, {Stevens}, {Clements}, {Webb},
  {Eales}, \& {Dunne}}]{2000MNRAS.316L..51G}
{Gear}, W.~K., {Lilly}, S.~J., {Stevens}, J.~A., {Clements}, D.~L., {Webb},
  T.~M., {Eales}, S.~A., \& {Dunne}, L. 2000, \mnras, 316, L51

\bibitem[{{Giacconi} {et~al.}(2001){Giacconi}, {Rosati}, {Tozzi}, {Nonino},
  {Hasinger}, {Norman}, {Bergeron}, {Borgani}, {Gilli}, {Gilmozzi}, \&
  {Zheng}}]{2001ApJ...551..624G}
{Giacconi}, R., {Rosati}, P., {Tozzi}, P., {Nonino}, M., {Hasinger}, G.,
  {Norman}, C., {Bergeron}, J., {Borgani}, S., {Gilli}, R., {Gilmozzi}, R., \&
  {Zheng}, W. 2001, \apj, 551, 624

\bibitem[{{Giavalisco} {et~al.}(2004)}]{2004ApJ...600L..93G}
{Giavalisco}, M. {et~al.} 2004, \apjl, 600, L93

\bibitem[{{Greve} {et~al.}(2004){Greve}, {Ivison}, {Bertoldi}, {Stevens},
  {Dunlop}, {Lutz}, \& {Carilli}}]{2004MNRAS.354..779G}
{Greve}, T.~R., {Ivison}, R.~J., {Bertoldi}, F., {Stevens}, J.~A., {Dunlop},
  J.~S., {Lutz}, D., \& {Carilli}, C.~L. 2004, \mnras, 354, 779

\bibitem[{{Hayward} {et~al.}(2012){Hayward}, {Jonsson}, {Kere{\v s}},
  {Magnelli}, {Hernquist}, \& {Cox}}]{2012arXiv1203.1318H}
{Hayward}, C.~C., {Jonsson}, P., {Kere{\v s}}, D., {Magnelli}, B., {Hernquist},
  L., \& {Cox}, T.~J. 2012, ArXiv e-prints

\bibitem[{{Hayward} {et~al.}(2011){Hayward}, {Kere{\v s}}, {Jonsson},
  {Narayanan}, {Cox}, \& {Hernquist}}]{2011ApJ...743..159H}
{Hayward}, C.~C., {Kere{\v s}}, D., {Jonsson}, P., {Narayanan}, D., {Cox},
  T.~J., \& {Hernquist}, L. 2011, \apj, 743, 159

\bibitem[{{Hickox} {et~al.}(2012){Hickox}, {Wardlow}, {Smail}, {Myers},
  {Alexander}, {Swinbank}, {Danielson}, {Stott}, {Chapman}, {Coppin}, {Dunlop},
  {Gawiser}, {Lutz}, {van der Werf}, \& {Wei{\ss}}}]{2012MNRAS.421..284H}
{Hickox}, R.~C., {Wardlow}, J.~L., {Smail}, I., {Myers}, A.~D., {Alexander},
  D.~M., {Swinbank}, A.~M., {Danielson}, A.~L.~R., {Stott}, J.~P., {Chapman},
  S.~C., {Coppin}, K.~E.~K., {Dunlop}, J.~S., {Gawiser}, E., {Lutz}, D., {van
  der Werf}, P., \& {Wei{\ss}}, A. 2012, \mnras, 421, 284

\bibitem[{{Hodge} {et~al.}(2012){Hodge}, {Carilli}, {Walter}, {de Blok},
  {Riechers}, {Daddi}, \& {Lentati}}]{2012ApJ...760...11H}
{Hodge}, J.~A., {Carilli}, C.~L., {Walter}, F., {de Blok}, W.~J.~G.,
  {Riechers}, D., {Daddi}, E., \& {Lentati}, L. 2012, \apj, 760, 11

\bibitem[{{Hopkins} {et~al.}(2003){Hopkins}, {Afonso}, {Chan}, {Cram},
  {Georgakakis}, \& {Mobasher}}]{2003AJ....125..465H}
{Hopkins}, A.~M., {Afonso}, J., {Chan}, B., {Cram}, L.~E., {Georgakakis}, A.,
  \& {Mobasher}, B. 2003, \aj, 125, 465

\bibitem[{{Hughes} {et~al.}(1998)}]{1998Natur.394..241H}
{Hughes}, D.~H. {et~al.} 1998, \nat, 394, 241

\bibitem[{{Iono} {et~al.}(2006)}]{2006ApJ...640L...1I}
{Iono}, D. {et~al.} 2006, \apjl, 640, L1

\bibitem[{{Ivison} {et~al.}(2010{\natexlab{a}}){Ivison}, {Alexander}, {Biggs},
  {Brandt}, {Chapin}, {Coppin}, {Devlin}, {Dickinson}, {Dunlop}, {Dye},
  {Eales}, {Frayer}, {Halpern}, {Hughes}, {Ibar}, {Kov{\'a}cs}, {Marsden},
  {Moncelsi}, {Netterfield}, {Pascale}, {Patanchon}, {Rafferty}, {Rex},
  {Schinnerer}, {Scott}, {Semisch}, {Smail}, {Swinbank}, {Truch}, {Tucker},
  {Viero}, {Walter}, {Wei{\ss}}, {Wiebe}, \& {Xue}}]{2010MNRAS.402..245I}
{Ivison}, R.~J., {Alexander}, D.~M., {Biggs}, A.~D., {Brandt}, W.~N., {Chapin},
  E.~L., {Coppin}, K.~E.~K., {Devlin}, M.~J., {Dickinson}, M., {Dunlop}, J.,
  {Dye}, S., {Eales}, S.~A., {Frayer}, D.~T., {Halpern}, M., {Hughes}, D.~H.,
  {Ibar}, E., {Kov{\'a}cs}, A., {Marsden}, G., {Moncelsi}, L., {Netterfield},
  C.~B., {Pascale}, E., {Patanchon}, G., {Rafferty}, D.~A., {Rex}, M.,
  {Schinnerer}, E., {Scott}, D., {Semisch}, C., {Smail}, I., {Swinbank}, A.~M.,
  {Truch}, M.~D.~P., {Tucker}, G.~S., {Viero}, M.~P., {Walter}, F., {Wei{\ss}},
  A., {Wiebe}, D.~V., \& {Xue}, Y.~Q. 2010{\natexlab{a}}, \mnras, 402, 245

\bibitem[{{Ivison} {et~al.}(2011){Ivison}, {Papadopoulos}, {Smail}, {Greve},
  {Thomson}, {Xilouris}, \& {Chapman}}]{2011MNRAS.412.1913I}
{Ivison}, R.~J., {Papadopoulos}, P.~P., {Smail}, I., {Greve}, T.~R., {Thomson},
  A.~P., {Xilouris}, E.~M., \& {Chapman}, S.~C. 2011, \mnras, 412, 1913

\bibitem[{{Ivison} {et~al.}(2010{\natexlab{b}}){Ivison}, {Smail},
  {Papadopoulos}, {Wold}, {Richard}, {Swinbank}, {Kneib}, \&
  {Owen}}]{2010MNRAS.404..198I}
{Ivison}, R.~J., {Smail}, I., {Papadopoulos}, P.~P., {Wold}, I., {Richard}, J.,
  {Swinbank}, A.~M., {Kneib}, J.-P., \& {Owen}, F.~N. 2010{\natexlab{b}},
  \mnras, 404, 198

\bibitem[{{Ivison} {et~al.}(2007)}]{2007MNRAS.380..199I}
{Ivison}, R.~J. {et~al.} 2007, \mnras, 380, 199

\bibitem[{{Karim} {et~al.}(2011)}]{2011ApJ...730...61K}
{Karim}, A. {et~al.} 2011, \apj, 730, 61

\bibitem[{{Karim} {et~al.}(2012)}]{Karim_aless}
---. 2012, submitted

\bibitem[{{Lehmer} {et~al.}(2005){Lehmer}, {Brandt}, {Alexander}, {Bauer},
  {Schneider}, {Tozzi}, {Bergeron}, {Garmire}, {Giacconi}, {Gilli}, {Hasinger},
  {Hornschemeier}, {Koekemoer}, {Mainieri}, {Miyaji}, {Nonino}, {Rosati},
  {Silverman}, {Szokoly}, \& {Vignali}}]{2005ApJS..161...21L}
{Lehmer}, B.~D., {Brandt}, W.~N., {Alexander}, D.~M., {Bauer}, F.~E.,
  {Schneider}, D.~P., {Tozzi}, P., {Bergeron}, J., {Garmire}, G.~P.,
  {Giacconi}, R., {Gilli}, R., {Hasinger}, G., {Hornschemeier}, A.~E.,
  {Koekemoer}, A.~M., {Mainieri}, V., {Miyaji}, T., {Nonino}, M., {Rosati}, P.,
  {Silverman}, J.~D., {Szokoly}, G., \& {Vignali}, C. 2005, \apjs, 161, 21

\bibitem[{{Lindner} {et~al.}(2011){Lindner}, {Baker}, {Omont}, {Beelen},
  {Owen}, {Bertoldi}, {Dole}, {Fiolet}, {Harris}, {Ivison}, {Lonsdale}, {Lutz},
  \& {Polletta}}]{2011ApJ...737...83L}
{Lindner}, R.~R., {Baker}, A.~J., {Omont}, A., {Beelen}, A., {Owen}, F.~N.,
  {Bertoldi}, F., {Dole}, H., {Fiolet}, N., {Harris}, A.~I., {Ivison}, R.~J.,
  {Lonsdale}, C.~J., {Lutz}, D., \& {Polletta}, M. 2011, \apj, 737, 83

\bibitem[{{Luo} {et~al.}(2008){Luo}, {Bauer}, {Brandt}, {Alexander}, {Lehmer},
  {Schneider}, {Brusa}, {Comastri}, {Fabian}, {Finoguenov}, {Gilli},
  {Hasinger}, {Hornschemeier}, {Koekemoer}, {Mainieri}, {Paolillo}, {Rosati},
  {Shemmer}, {Silverman}, {Smail}, {Steffen}, \&
  {Vignali}}]{2008ApJS..179...19L}
{Luo}, B., {Bauer}, F.~E., {Brandt}, W.~N., {Alexander}, D.~M., {Lehmer},
  B.~D., {Schneider}, D.~P., {Brusa}, M., {Comastri}, A., {Fabian}, A.~C.,
  {Finoguenov}, A., {Gilli}, R., {Hasinger}, G., {Hornschemeier}, A.~E.,
  {Koekemoer}, A., {Mainieri}, V., {Paolillo}, M., {Rosati}, P., {Shemmer}, O.,
  {Silverman}, J.~D., {Smail}, I., {Steffen}, A.~T., \& {Vignali}, C. 2008,
  \apjs, 179, 19

\bibitem[{{Lutz} {et~al.}(2001)}]{2001A&A...378...70L}
{Lutz}, D. {et~al.} 2001, \aap, 378, 70

\bibitem[{{Miller} {et~al.}(2013){Miller}, {Bonzini}, {Fomalont}, {Kellermann},
  {Mainieri}, {Padovani}, {Rosati}, {Tozzi}, \&
  {Vattakunnel}}]{2013arXiv1301.7004M}
{Miller}, N.~A., {Bonzini}, M., {Fomalont}, E.~B., {Kellermann}, K.~I.,
  {Mainieri}, V., {Padovani}, P., {Rosati}, P., {Tozzi}, P., \& {Vattakunnel},
  S. 2013, ArXiv e-prints

\bibitem[{{Miller} {et~al.}(2008){Miller}, {Fomalont}, {Kellermann},
  {Mainieri}, {Norman}, {Padovani}, {Rosati}, \& {Tozzi}}]{2008ApJS..179..114M}
{Miller}, N.~A., {Fomalont}, E.~B., {Kellermann}, K.~I., {Mainieri}, V.,
  {Norman}, C., {Padovani}, P., {Rosati}, P., \& {Tozzi}, P. 2008, \apjs, 179,
  114

\bibitem[{{Narayanan} {et~al.}(2010){Narayanan}, {Hayward}, {Cox}, {Hernquist},
  {Jonsson}, {Younger}, \& {Groves}}]{2010MNRAS.401.1613N}
{Narayanan}, D., {Hayward}, C.~C., {Cox}, T.~J., {Hernquist}, L., {Jonsson},
  P., {Younger}, J.~D., \& {Groves}, B. 2010, \mnras, 401, 1613

\bibitem[{{Riechers} {et~al.}(2011{\natexlab{a}}){Riechers}, {Carilli},
  {Walter}, {Weiss}, {Wagg}, {Bertoldi}, {Downes}, {Henkel}, \&
  {Hodge}}]{2011ApJ...733L..11R}
{Riechers}, D.~A., {Carilli}, C.~L., {Walter}, F., {Weiss}, A., {Wagg}, J.,
  {Bertoldi}, F., {Downes}, D., {Henkel}, C., \& {Hodge}, J.
  2011{\natexlab{a}}, \apjl, 733, L11

\bibitem[{{Riechers} {et~al.}(2011{\natexlab{b}}){Riechers}, {Hodge}, {Walter},
  {Carilli}, \& {Bertoldi}}]{2011ApJ...739L..31R}
{Riechers}, D.~A., {Hodge}, J., {Walter}, F., {Carilli}, C.~L., \& {Bertoldi},
  F. 2011{\natexlab{b}}, \apjl, 739, L31

\bibitem[{{Sanders} \& {Mirabel}(1996)}]{1996ARA&A..34..749S}
{Sanders}, D.~B. \& {Mirabel}, I.~F. 1996, \araa, 34, 749

\bibitem[{{Schinnerer} {et~al.}(2004){Schinnerer}, {Carilli}, {Scoville},
  {Bondi}, {Ciliegi}, {Vettolani}, {Le F{\`e}vre}, {Koekemoer}, {Bertoldi}, \&
  {Impey}}]{2004AJ....128.1974S}
{Schinnerer}, E., {Carilli}, C.~L., {Scoville}, N.~Z., {Bondi}, M., {Ciliegi},
  P., {Vettolani}, P., {Le F{\`e}vre}, O., {Koekemoer}, A.~M., {Bertoldi}, F.,
  \& {Impey}, C.~D. 2004, \aj, 128, 1974

\bibitem[{{Schinnerer} {et~al.}(2010){Schinnerer}, {Sargent}, {Bondi}, {Smol{\v
  c}i{\'c}}, {Datta}, {Carilli}, {Bertoldi}, {Blain}, {Ciliegi}, {Koekemoer},
  \& {Scoville}}]{2010ApJS..188..384S}
{Schinnerer}, E., {Sargent}, M.~T., {Bondi}, M., {Smol{\v c}i{\'c}}, V.,
  {Datta}, A., {Carilli}, C.~L., {Bertoldi}, F., {Blain}, A., {Ciliegi}, P.,
  {Koekemoer}, A., \& {Scoville}, N.~Z. 2010, \apjs, 188, 384

\bibitem[{{Scott} {et~al.}(2010){Scott}, {Yun}, {Wilson}, {Austermann},
  {Aguilar}, {Aretxaga}, {Ezawa}, {Ferrusca}, {Hatsukade}, {Hughes}, {Iono},
  {Giavalisco}, {Kawabe}, {Kohno}, {Mauskopf}, {Oshima}, {Perera}, {Rand},
  {Tamura}, {Tosaki}, {Velazquez}, {Williams}, \&
  {Zeballos}}]{2010MNRAS.405.2260S}
{Scott}, K.~S., {Yun}, M.~S., {Wilson}, G.~W., {Austermann}, J.~E., {Aguilar},
  E., {Aretxaga}, I., {Ezawa}, H., {Ferrusca}, D., {Hatsukade}, B., {Hughes},
  D.~H., {Iono}, D., {Giavalisco}, M., {Kawabe}, R., {Kohno}, K., {Mauskopf},
  P.~D., {Oshima}, T., {Perera}, T.~A., {Rand}, J., {Tamura}, Y., {Tosaki}, T.,
  {Velazquez}, M., {Williams}, C.~C., \& {Zeballos}, M. 2010, \mnras, 405, 2260

\bibitem[{{Scott} {et~al.}(2006){Scott}, {Dunlop}, \&
  {Serjeant}}]{2006MNRAS.370.1057S}
{Scott}, S.~E., {Dunlop}, J.~S., \& {Serjeant}, S. 2006, \mnras, 370, 1057

\bibitem[{{Smail} {et~al.}(1997){Smail}, {Ivison}, \&
  {Blain}}]{1997ApJ...490L...5S}
{Smail}, I., {Ivison}, R.~J., \& {Blain}, A.~W. 1997, \apjl, 490, L5

\bibitem[{{Smol{\v c}i{\'c}} {et~al.}(2012{\natexlab{a}}){Smol{\v c}i{\'c}},
  {Aravena}, {Navarrete}, {Schinnerer}, {Riechers}, {Bertoldi}, {Feruglio},
  {Finoguenov}, {Salvato}, {Sargent}, {McCracken}, {Albrecht}, {Karim},
  {Capak}, {Carilli}, {Cappelluti}, {Elvis}, {Ilbert}, {Kartaltepe}, {Lilly},
  {Sanders}, {Sheth}, {Scoville}, \& {Taniguchi}}]{2012A&A...548A...4S}
{Smol{\v c}i{\'c}}, V., {Aravena}, M., {Navarrete}, F., {Schinnerer}, E.,
  {Riechers}, D.~A., {Bertoldi}, F., {Feruglio}, C., {Finoguenov}, A.,
  {Salvato}, M., {Sargent}, M., {McCracken}, H.~J., {Albrecht}, M., {Karim},
  A., {Capak}, P., {Carilli}, C.~L., {Cappelluti}, N., {Elvis}, M., {Ilbert},
  O., {Kartaltepe}, J., {Lilly}, S., {Sanders}, D., {Sheth}, K., {Scoville},
  N.~Z., \& {Taniguchi}, Y. 2012{\natexlab{a}}, \aap, 548, A4

\bibitem[{{Smol{\v c}i{\'c}}
  {et~al.}(2012{\natexlab{b}})}]{2012ApJS..200...10S}
{Smol{\v c}i{\'c}}, V. {et~al.} 2012{\natexlab{b}}, \apjs, 200, 10

\bibitem[{{Swinbank} {et~al.}(2012){Swinbank}, {Karim}, {Smail}, {Hodge},
  {Walter}, {Bertoldi}, {Biggs}, {de Breuck}, {Chapman}, {Coppin}, {Cox},
  {Danielson}, {Dannerbauer}, {Ivison}, {Greve}, {Knudsen}, {Menten},
  {Simpson}, {Schinnerer}, {Wardlow}, {Wei{\ss}}, \& {van der
  Werf}}]{2012MNRAS.427.1066S}
{Swinbank}, A.~M., {Karim}, A., {Smail}, I., {Hodge}, J., {Walter}, F.,
  {Bertoldi}, F., {Biggs}, A.~D., {de Breuck}, C., {Chapman}, S.~C., {Coppin},
  K.~E.~K., {Cox}, P., {Danielson}, A.~L.~R., {Dannerbauer}, H., {Ivison},
  R.~J., {Greve}, T.~R., {Knudsen}, K.~K., {Menten}, K.~M., {Simpson}, J.~M.,
  {Schinnerer}, E., {Wardlow}, J.~L., {Wei{\ss}}, A., \& {van der Werf}, P.
  2012, \mnras, 427, 1066

\bibitem[{{Tacconi} {et~al.}(2006){Tacconi}, {Neri}, {Chapman}, {Genzel},
  {Smail}, {Ivison}, {Bertoldi}, {Blain}, {Cox}, {Greve}, \&
  {Omont}}]{2006ApJ...640..228T}
{Tacconi}, L.~J., {Neri}, R., {Chapman}, S.~C., {Genzel}, R., {Smail}, I.,
  {Ivison}, R.~J., {Bertoldi}, F., {Blain}, A., {Cox}, P., {Greve}, T., \&
  {Omont}, A. 2006, \apj, 640, 228

\bibitem[{{Tacconi} {et~al.}(2008)}]{2008ApJ...680..246T}
{Tacconi}, L.~J. {et~al.} 2008, \apj, 680, 246

\bibitem[{{Thompson} {et~al.}(2005){Thompson}, {Quataert}, \&
  {Murray}}]{2005ApJ...630..167T}
{Thompson}, T.~A., {Quataert}, E., \& {Murray}, N. 2005, \apj, 630, 167

\bibitem[{{Wang} {et~al.}(2004){Wang}, {Cowie}, \&
  {Barger}}]{2004ApJ...613..655W}
{Wang}, W.-H., {Cowie}, L.~L., \& {Barger}, A.~J. 2004, \apj, 613, 655

\bibitem[{{Wang} {et~al.}(2011){Wang}, {Cowie}, {Barger}, \&
  {Williams}}]{2011ApJ...726L..18W}
{Wang}, W.-H., {Cowie}, L.~L., {Barger}, A.~J., \& {Williams}, J.~P. 2011,
  \apjl, 726, L18

\bibitem[{{Wang} {et~al.}(2007){Wang}, {Cowie}, {van Saders}, {Barger}, \&
  {Williams}}]{2007ApJ...670L..89W}
{Wang}, W.-H., {Cowie}, L.~L., {van Saders}, J., {Barger}, A.~J., \&
  {Williams}, J.~P. 2007, \apjl, 670, L89

\bibitem[{{Wardlow} {et~al.}(2011)}]{2011MNRAS.415.1479W}
{Wardlow}, J.~L. {et~al.} 2011, \mnras, 415, 1479

\bibitem[{{Wei{\ss}} {et~al.}(2009)}]{2009ApJ...707.1201W}
{Wei{\ss}}, A. {et~al.} 2009, \apj, 707, 1201

\bibitem[{{Williams} {et~al.}(2011){Williams}, {Giavalisco}, {Porciani}, {Yun},
  {Pope}, {Scott}, {Austermann}, {Aretxaga}, {Hatsukade}, {Lee}, {Wilson},
  {Cybulski}, {Hughes}, {Kawabe}, {Kohno}, {Perera}, \&
  {Schloerb}}]{2011ApJ...733...92W}
{Williams}, C.~C., {Giavalisco}, M., {Porciani}, C., {Yun}, M.~S., {Pope}, A.,
  {Scott}, K.~S., {Austermann}, J.~E., {Aretxaga}, I., {Hatsukade}, B., {Lee},
  K.-S., {Wilson}, G.~W., {Cybulski}, R., {Hughes}, D.~H., {Kawabe}, R.,
  {Kohno}, K., {Perera}, T., \& {Schloerb}, F.~P. 2011, \apj, 733, 92

\bibitem[{{Windhorst} {et~al.}(1984){Windhorst}, {van Heerde}, \&
  {Katgert}}]{1984A&AS...58....1W}
{Windhorst}, R.~A., {van Heerde}, G.~M., \& {Katgert}, P. 1984, \aaps, 58, 1

\bibitem[{{Younger} {et~al.}(2007)}]{2007ApJ...671.1531Y}
{Younger}, J.~D. {et~al.} 2007, \apj, 671, 1531

\bibitem[{{Younger} {et~al.}(2008)}]{2008MNRAS.387..707Y}
---. 2008, \mnras, 387, 707

\bibitem[{{Younger} {et~al.}(2009)}]{2009ApJ...704..803Y}
---. 2009, \apj, 704, 803

\end{thebibliography}


\begin{figure*}
\centering
\vspace{-20mm}
\includegraphics[scale=0.9]{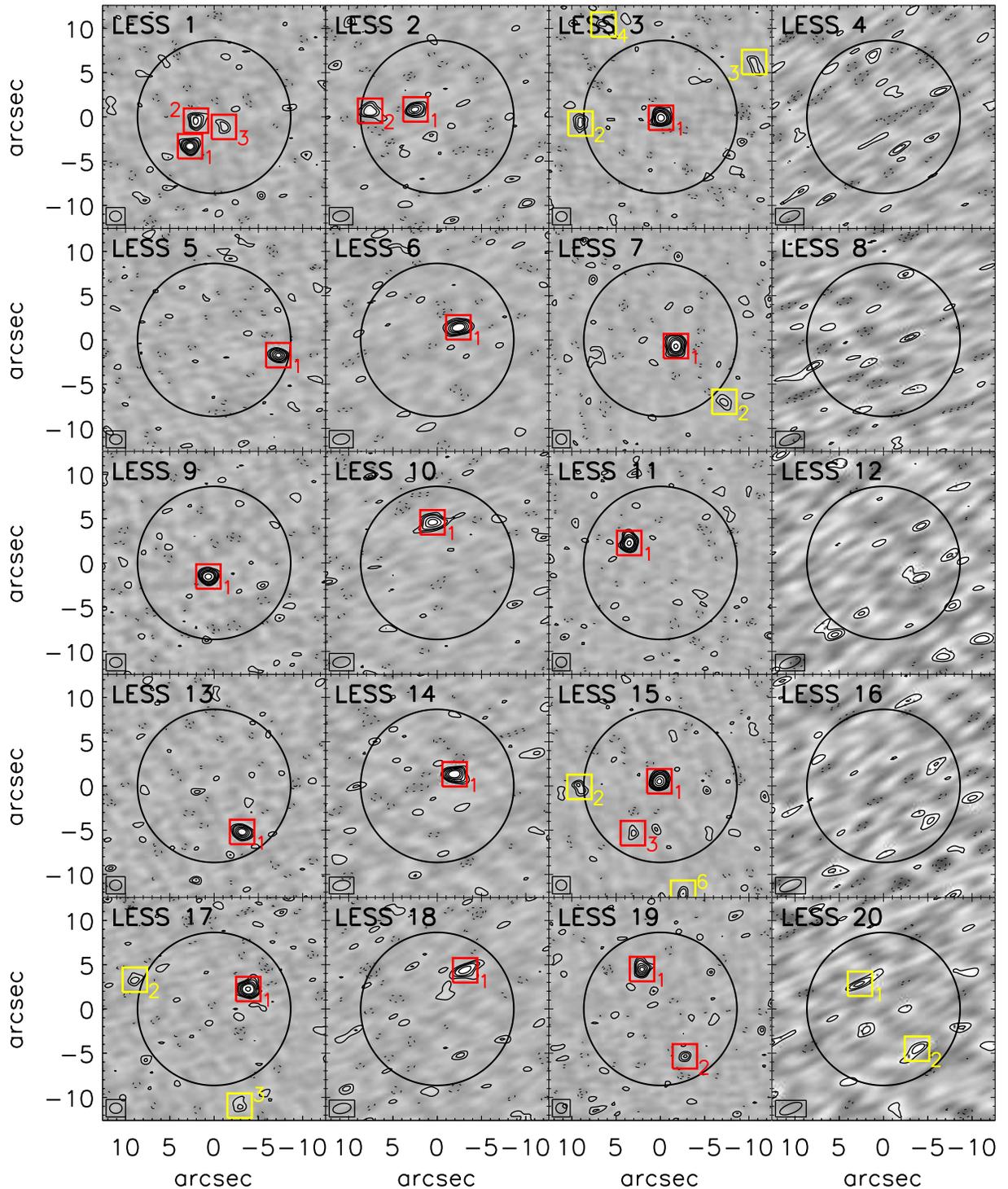}
\vspace{-20mm}
\caption{ALMA 870$\mu$m maps in order of LESS source number. 
The maps are $\sim$26$^{\prime\prime}$ per side and have pixels of 0.2$^{\prime\prime}$. 
Contours start at $\pm$2$\sigma$ and increase in steps of 1$\sigma$, where $\sigma$ is the rms noise measured in that map (Table~\ref{tab-2}). 
SMGs that appear in the MAIN/supplementary catalog are indicated with red/yellow squares and labeled with their ALESS sub--ID. For example, the source labeled `1' in the map for LESS 1 corresponds to source ALESS 001.1 in Table~\ref{tab-3}.
The synthesized beam is shown in the bottom left corner of each map, and the large circle indicates the primary beam FWHM. 
The images show a range in quality, with fields observed at low elevation appearing noisier and having more elongated synthesized beams. 
Note that LESS 52, 56, 64, and 125 were not observed with ALMA, and the quality of the ALMA maps for LESS 48 and 60 is so poor that we do not show them here.}
\label{fig:contourplots}
\end{figure*}

\begin{figure*}
\centering
\vspace{-20mm}
\includegraphics[scale=0.9]{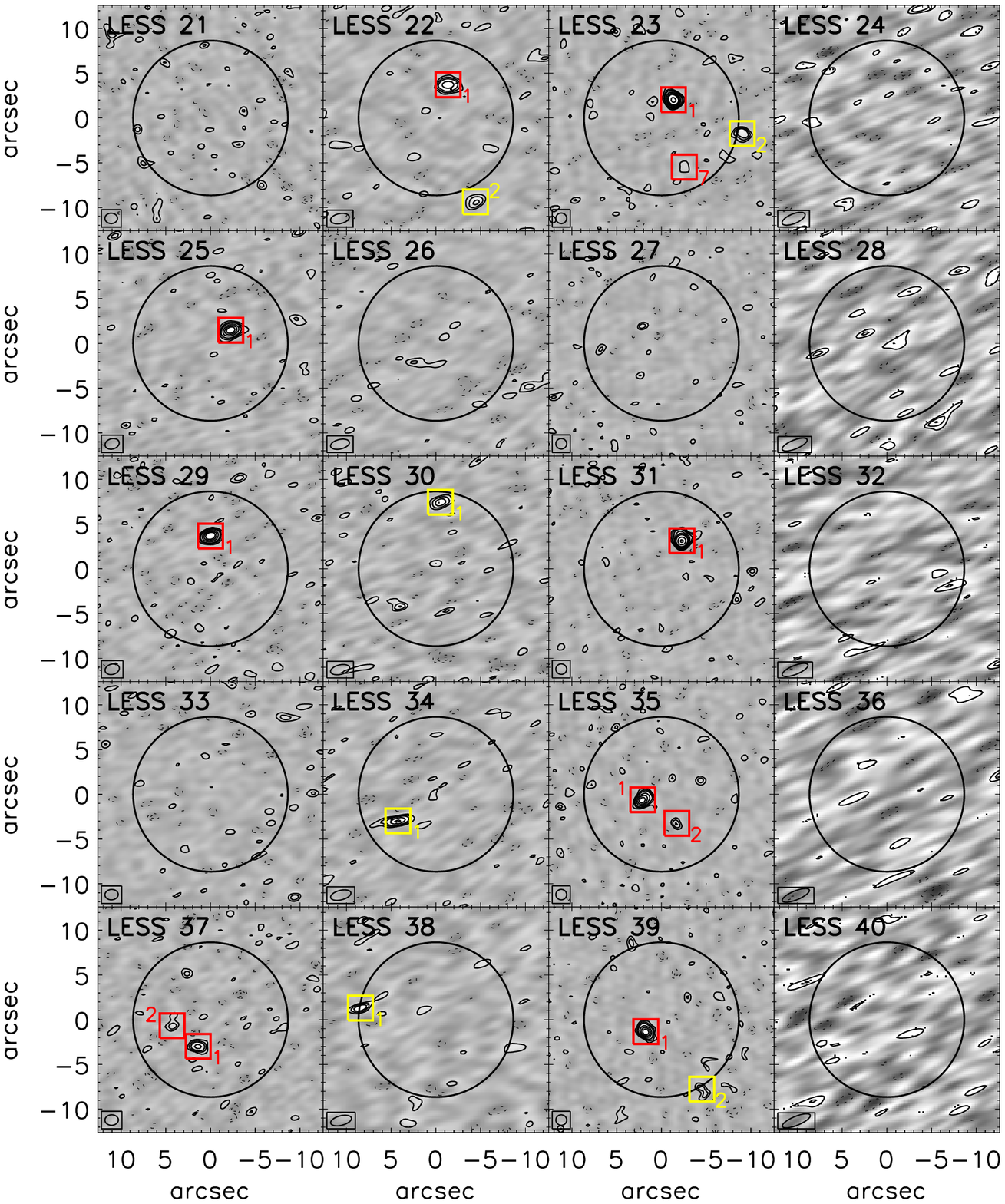}
\vspace{-20mm}
\begin{center}
Fig. \ref{fig:contourplots} (continued)
\end{center}
\end{figure*}

\begin{figure*}
\centering
\vspace{-20mm}
\includegraphics[scale=0.9]{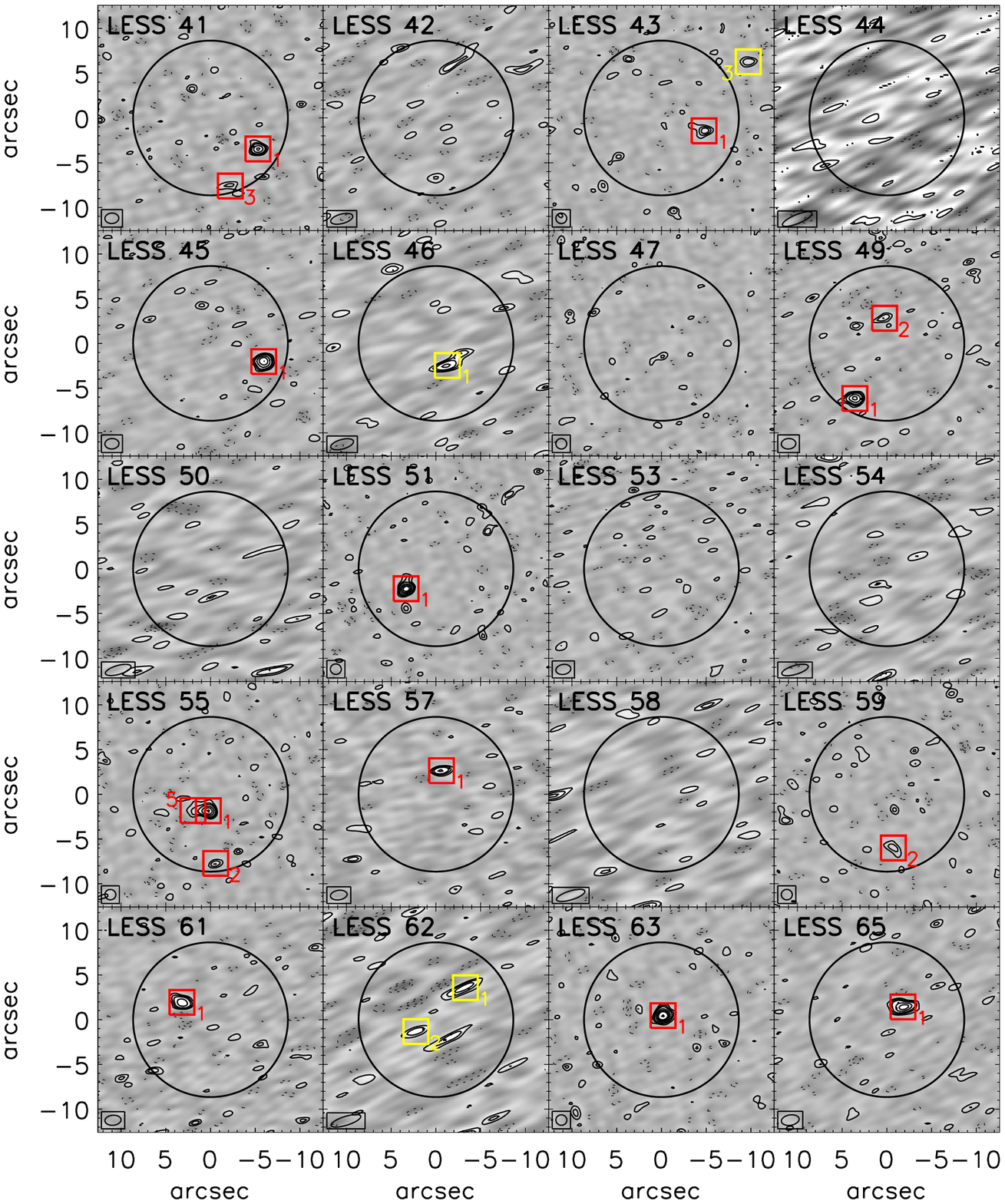}
\vspace{-20mm}
\begin{center}
Fig. \ref{fig:contourplots} (continued)
\end{center}
\end{figure*}

\begin{figure*}
\centering
\vspace{-20mm}
\includegraphics[scale=0.9]{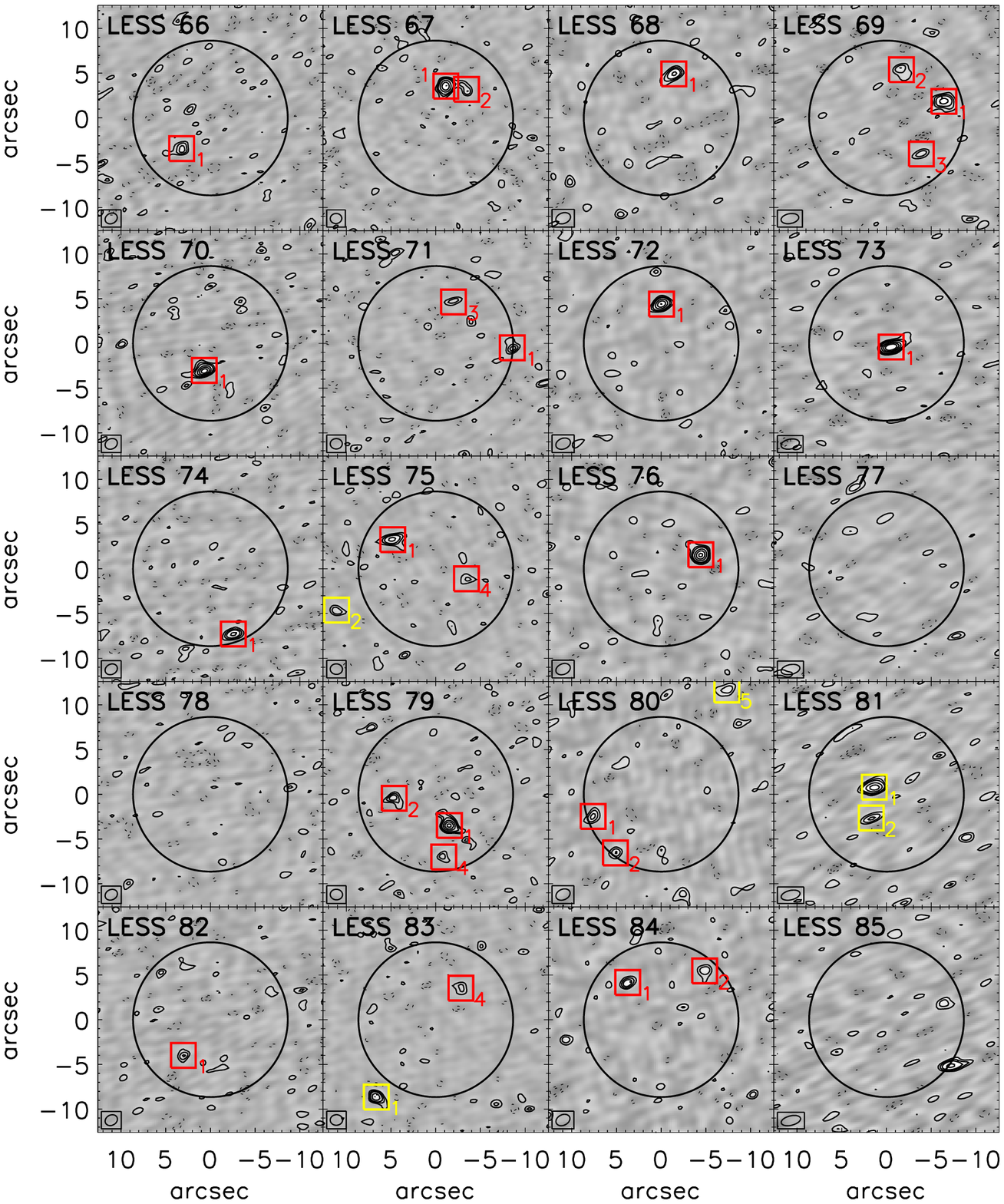}
\vspace{-20mm}
\begin{center}
Fig. \ref{fig:contourplots} (continued)
\end{center}
\end{figure*}

\begin{figure*}
\centering
\vspace{-20mm}
\includegraphics[scale=0.9]{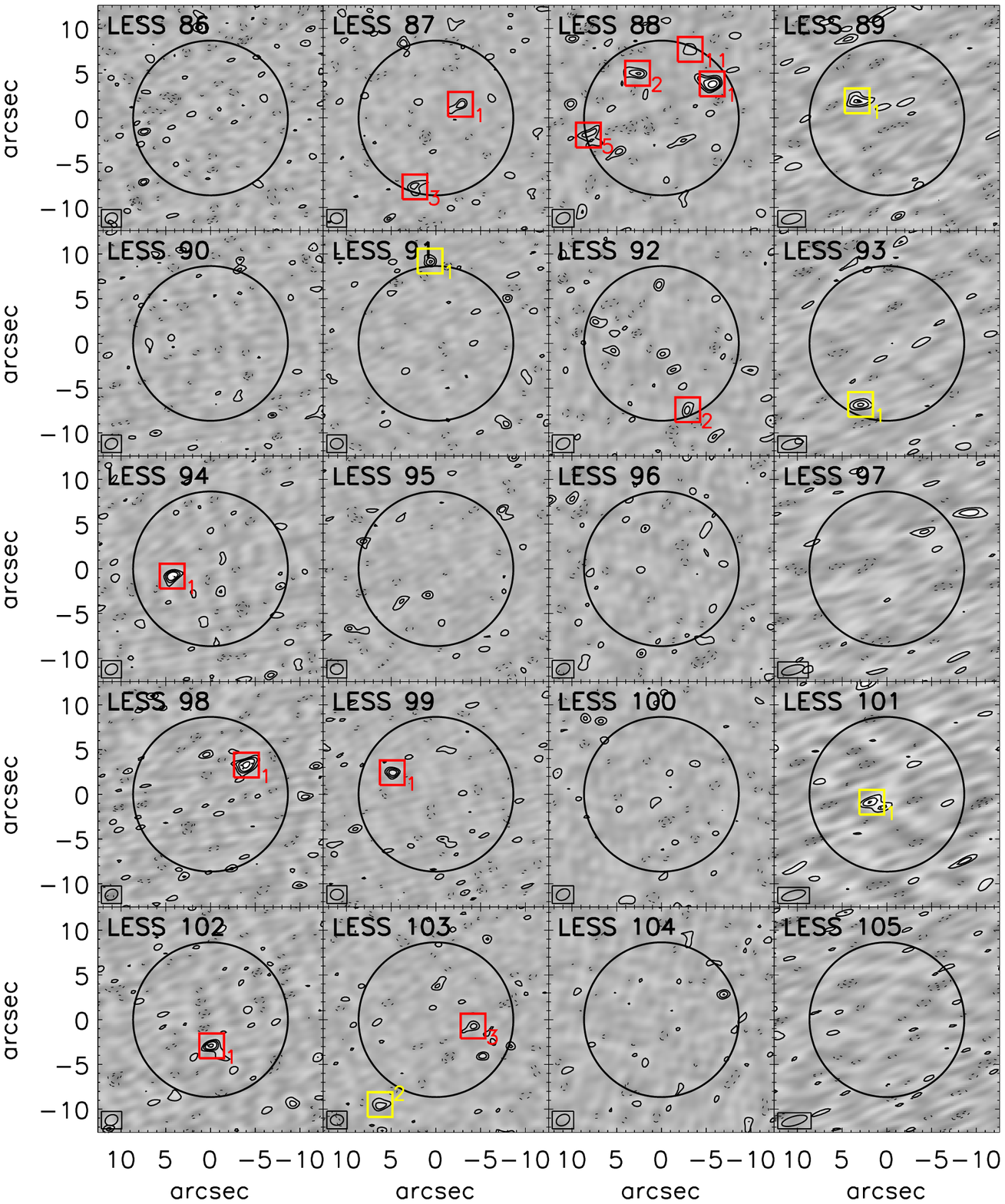}
\vspace{-20mm}
\begin{center}
Fig. \ref{fig:contourplots} (continued)
\end{center}
\end{figure*}

\begin{figure*}
\centering
\vspace{-20mm}
\includegraphics[scale=0.9]{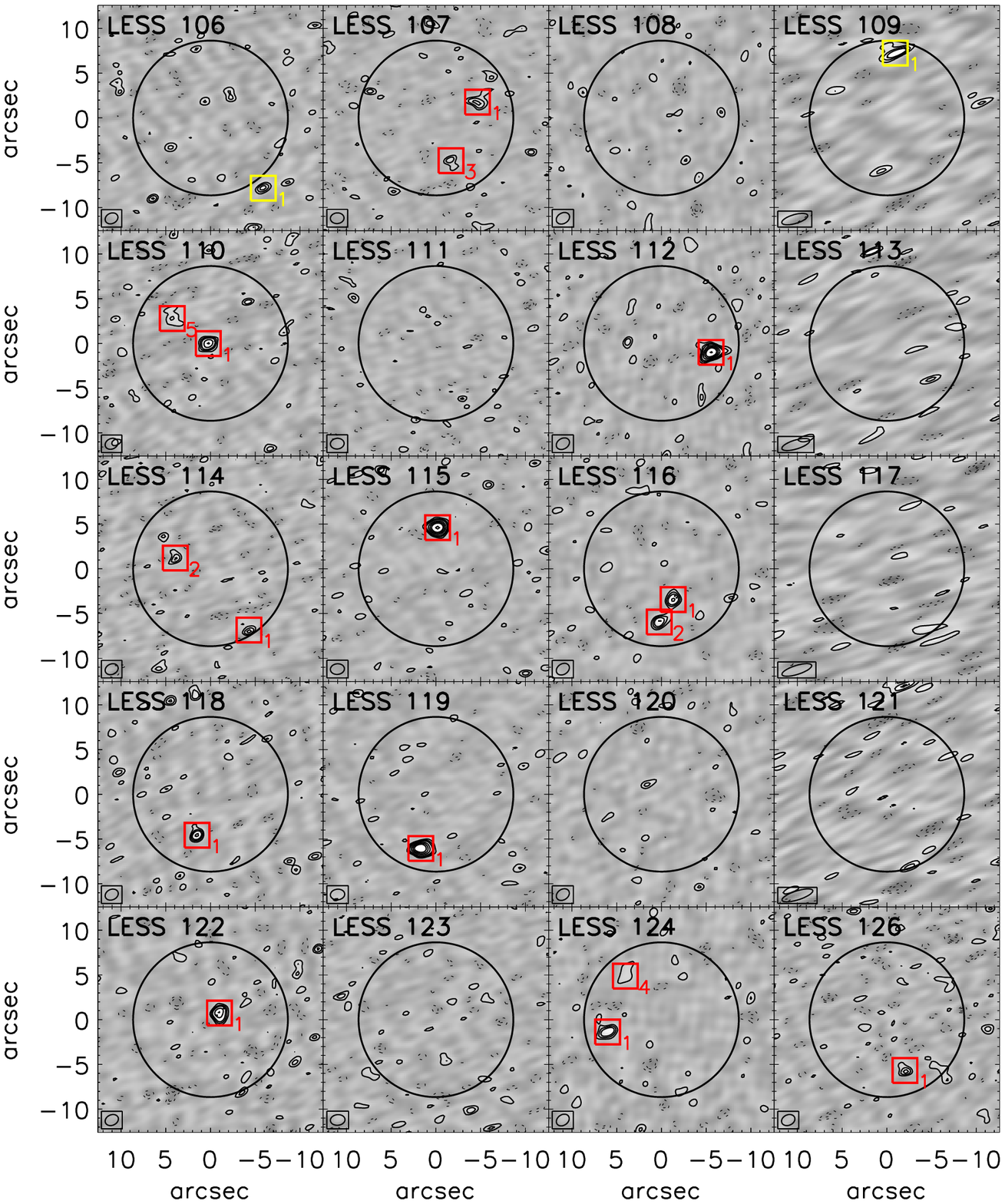}
\vspace{-20mm}
\begin{center}
Fig. \ref{fig:contourplots} (continued)
\end{center}
\end{figure*}


\begin{figure*}
\centering
\vspace{-20mm}
\includegraphics[scale=0.9]{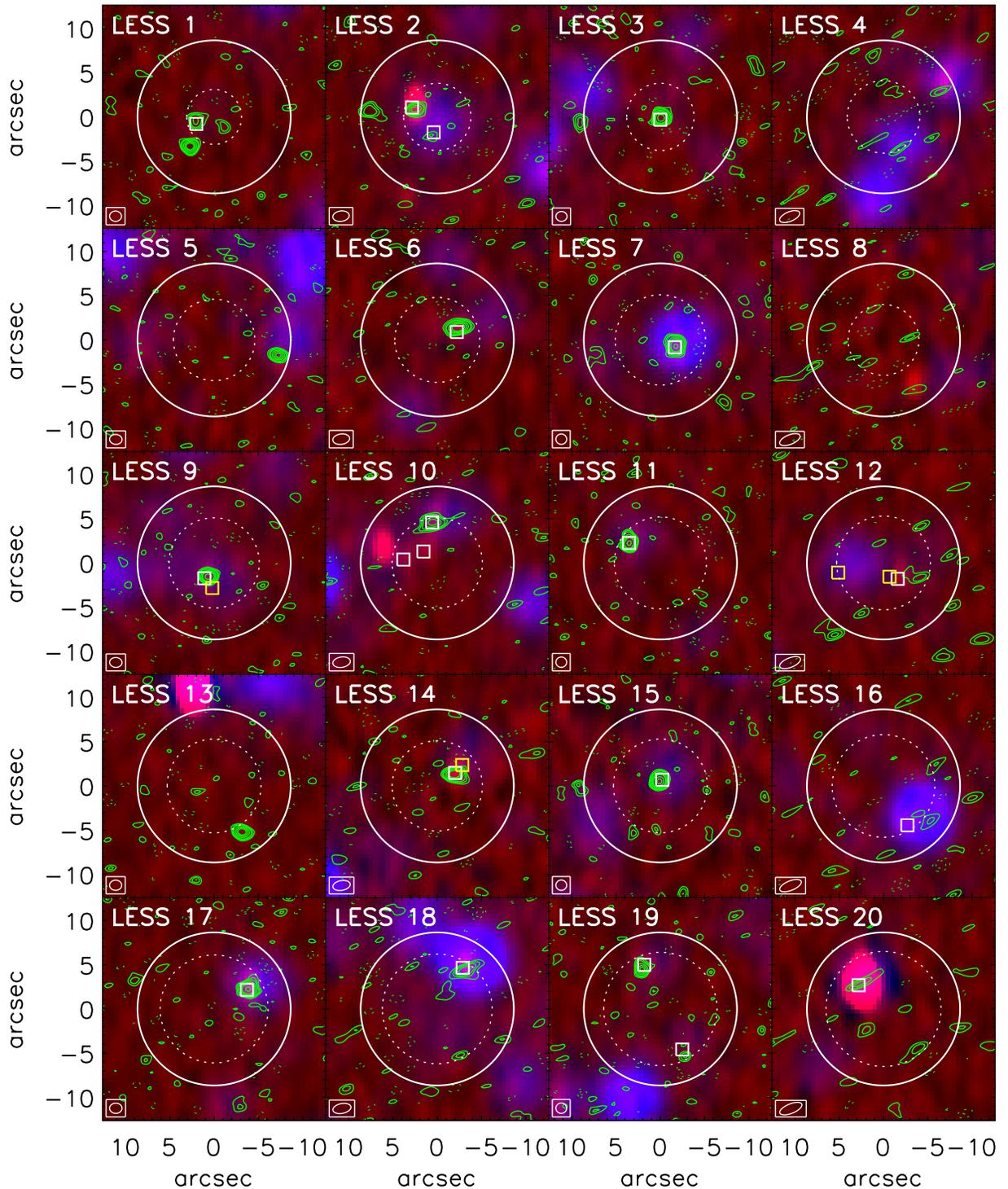}
\vspace{-20mm}
\caption{ALMA 870$\mu$m contours (green) on false--color multiwavelength images produced from the VLA 1.4 GHz (red) and MIPS 24$\mu$m (blue) data. 
The dashed circle indicates the search radius used by \citet{2011MNRAS.413.2314B} to statistically identify radio and mid--infrared counterparts to the LESS sources, and the small white/yellow squares indicate the positions of the predicted robust/tentative counterparts.
The synthesized beam of the ALMA data is shown in the bottom left corner of each map, and the large circle indicates the primary beam FWHM. 
ALMA contours are in steps of 1$\sigma$ starting at $\pm$2$\sigma$.
Note that LESS 52, 56, 64, and 125 were not observed with ALMA, and the quality of the ALMA maps for LESS 48 and 60 is so poor that we do not show them here.}
\label{fig:falsecolor}
\end{figure*}

\begin{figure*}
\centering
\vspace{-20mm}
\includegraphics[scale=0.9]{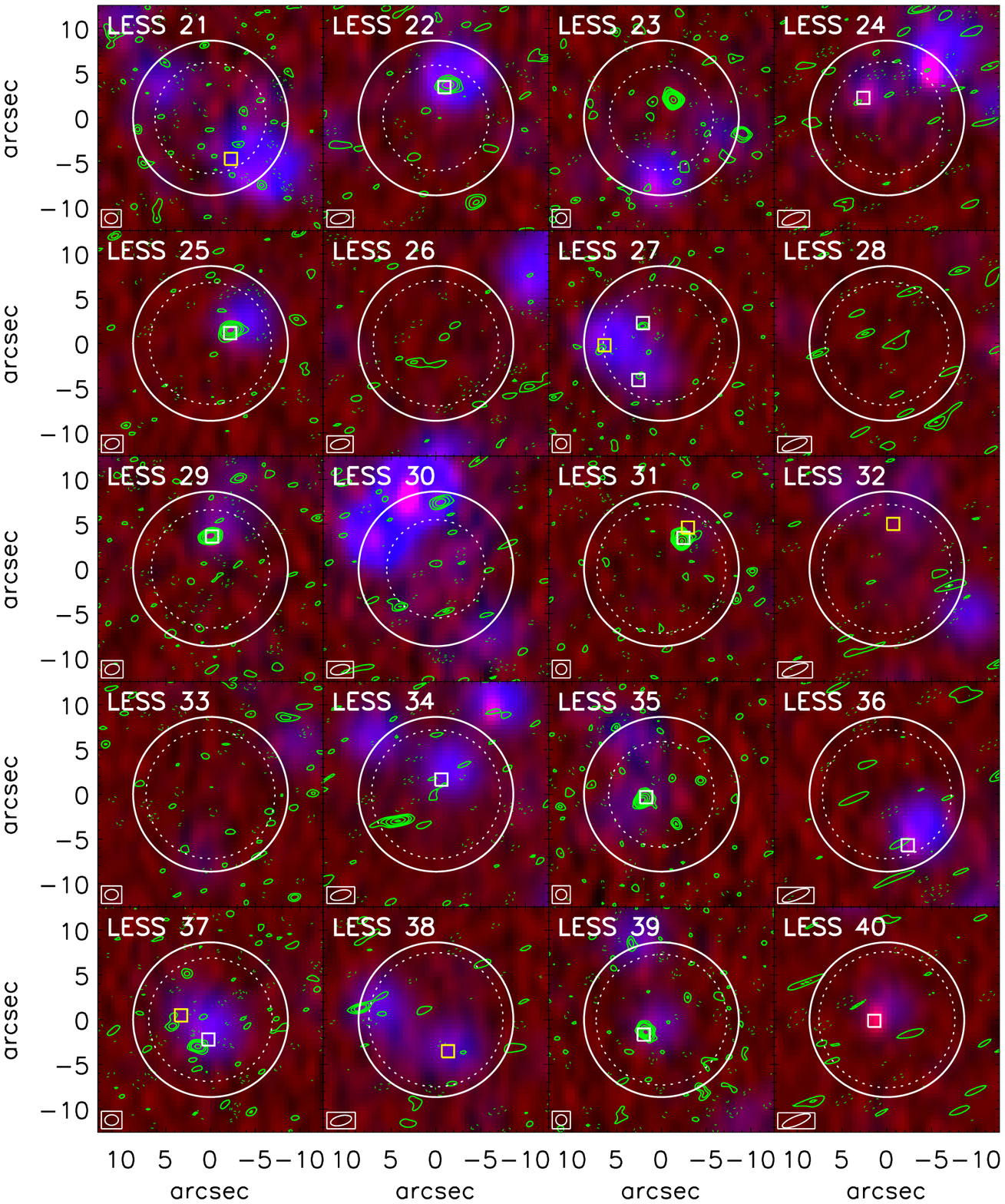}
\vspace{-20mm}
\begin{center}
Fig. \ref{fig:falsecolor} (continued)
\end{center}
\end{figure*}

\begin{figure*}
\centering
\vspace{-20mm}
\includegraphics[scale=0.9]{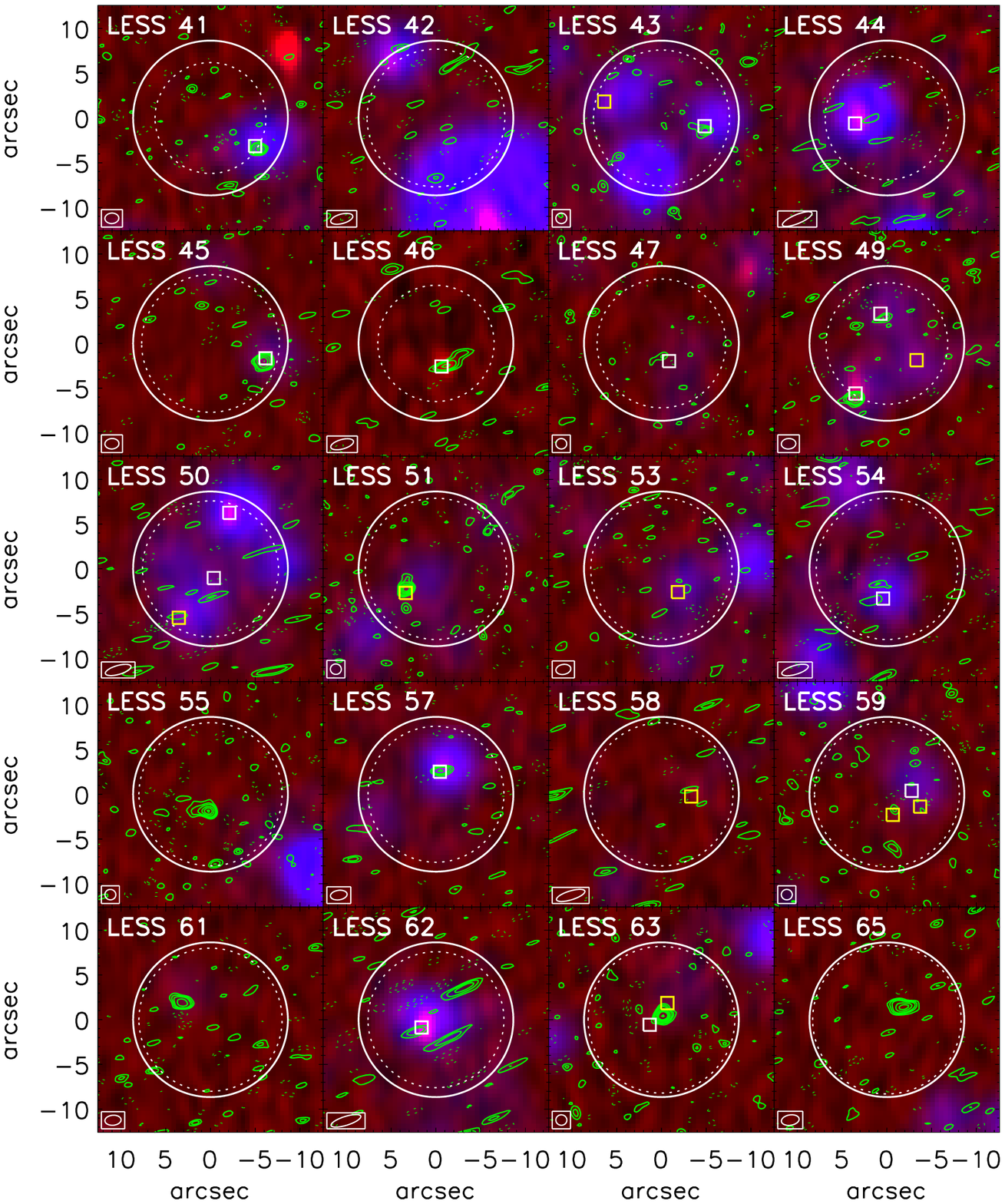}
\vspace{-20mm}
\begin{center}
Fig. \ref{fig:falsecolor} (continued)
\end{center}
\end{figure*}

\begin{figure*}
\centering
\vspace{-20mm}
\includegraphics[scale=0.9]{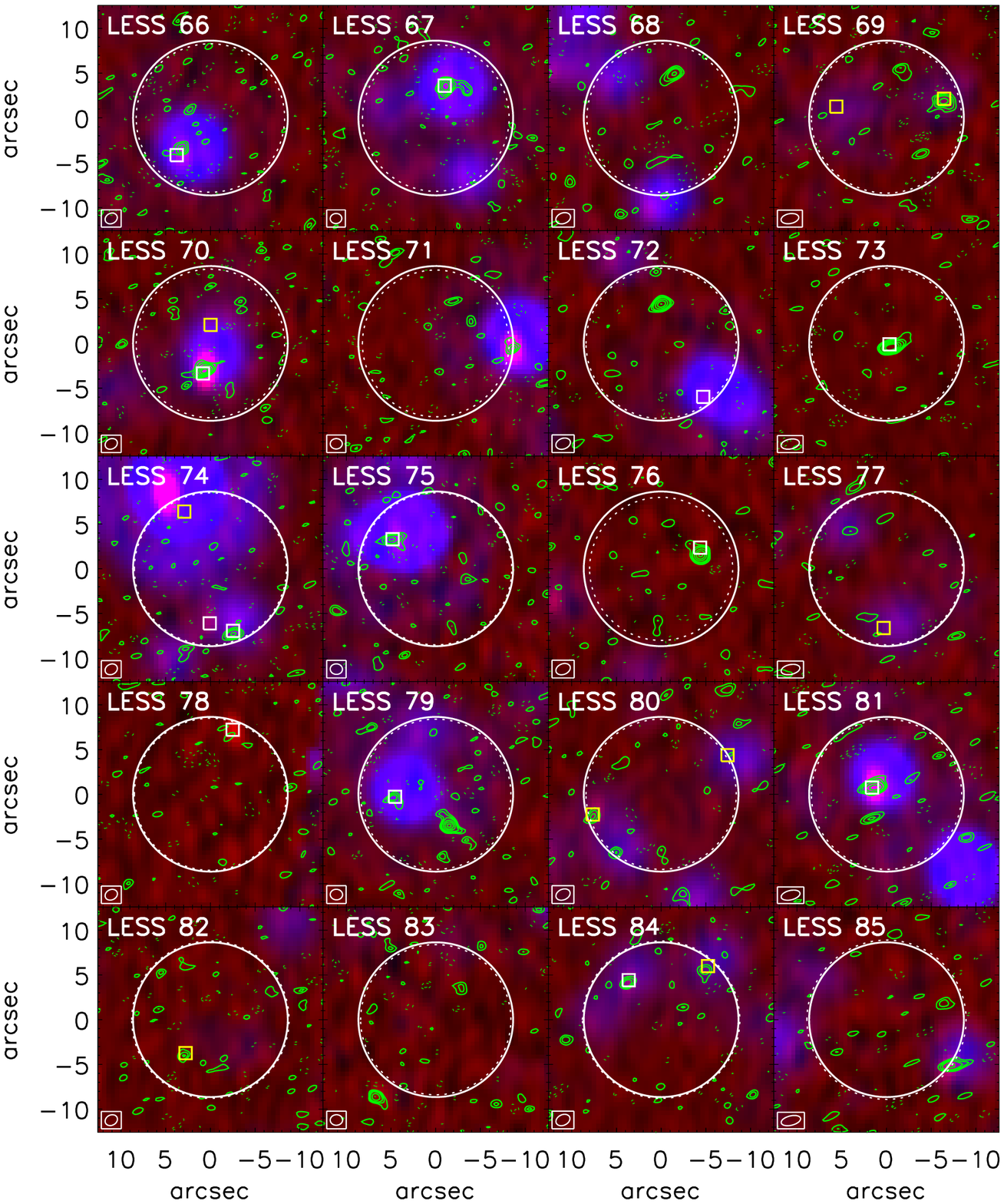}
\vspace{-20mm}
\begin{center}
Fig. \ref{fig:falsecolor} (continued)
\end{center}
\end{figure*}

\begin{figure*}
\centering
\vspace{-20mm}
\includegraphics[scale=0.9]{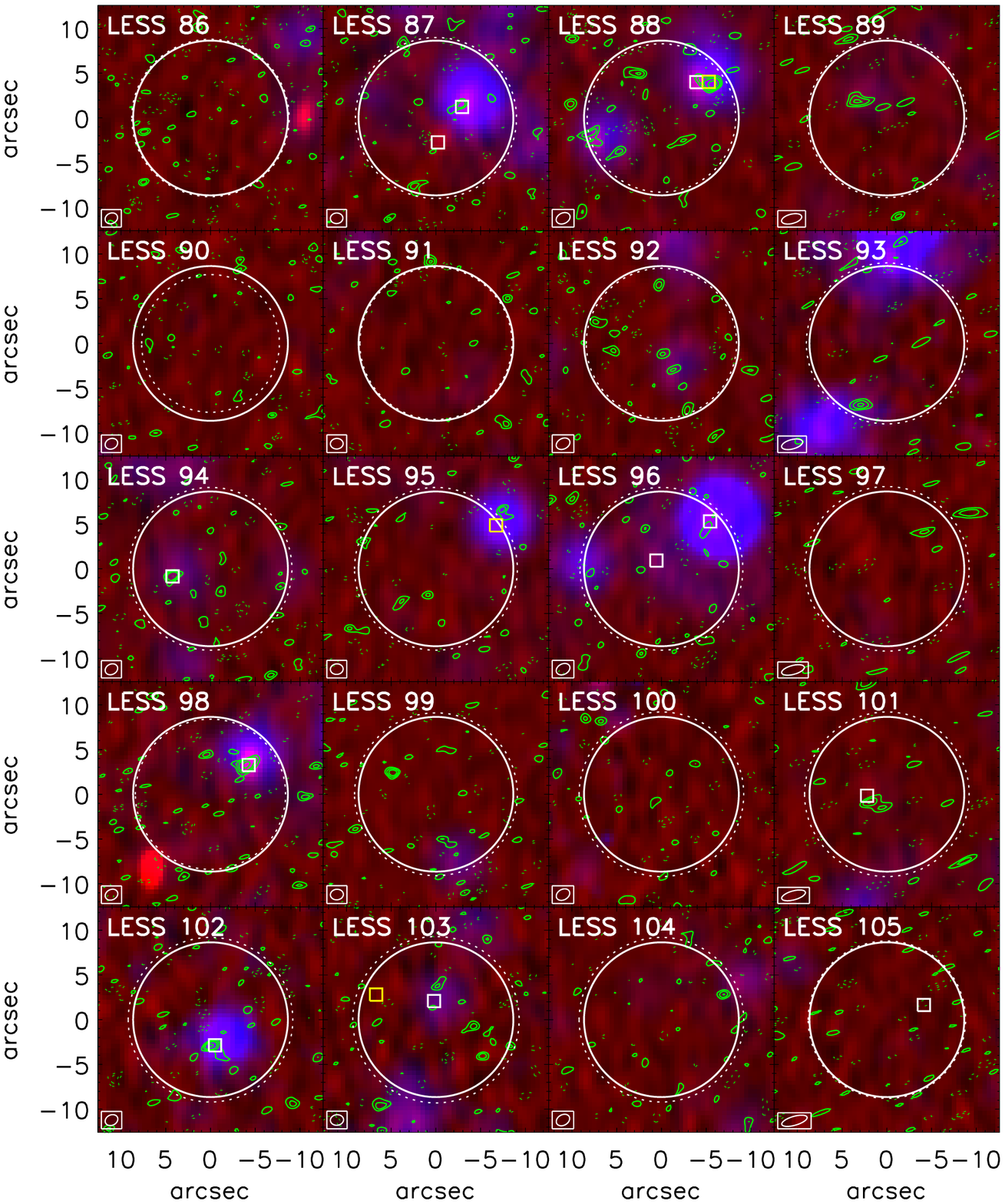}
\vspace{-20mm}
\begin{center}
Fig. \ref{fig:falsecolor} (continued)
\end{center}
\end{figure*}

\begin{figure*}
\centering
\vspace{-20mm}
\includegraphics[scale=0.9]{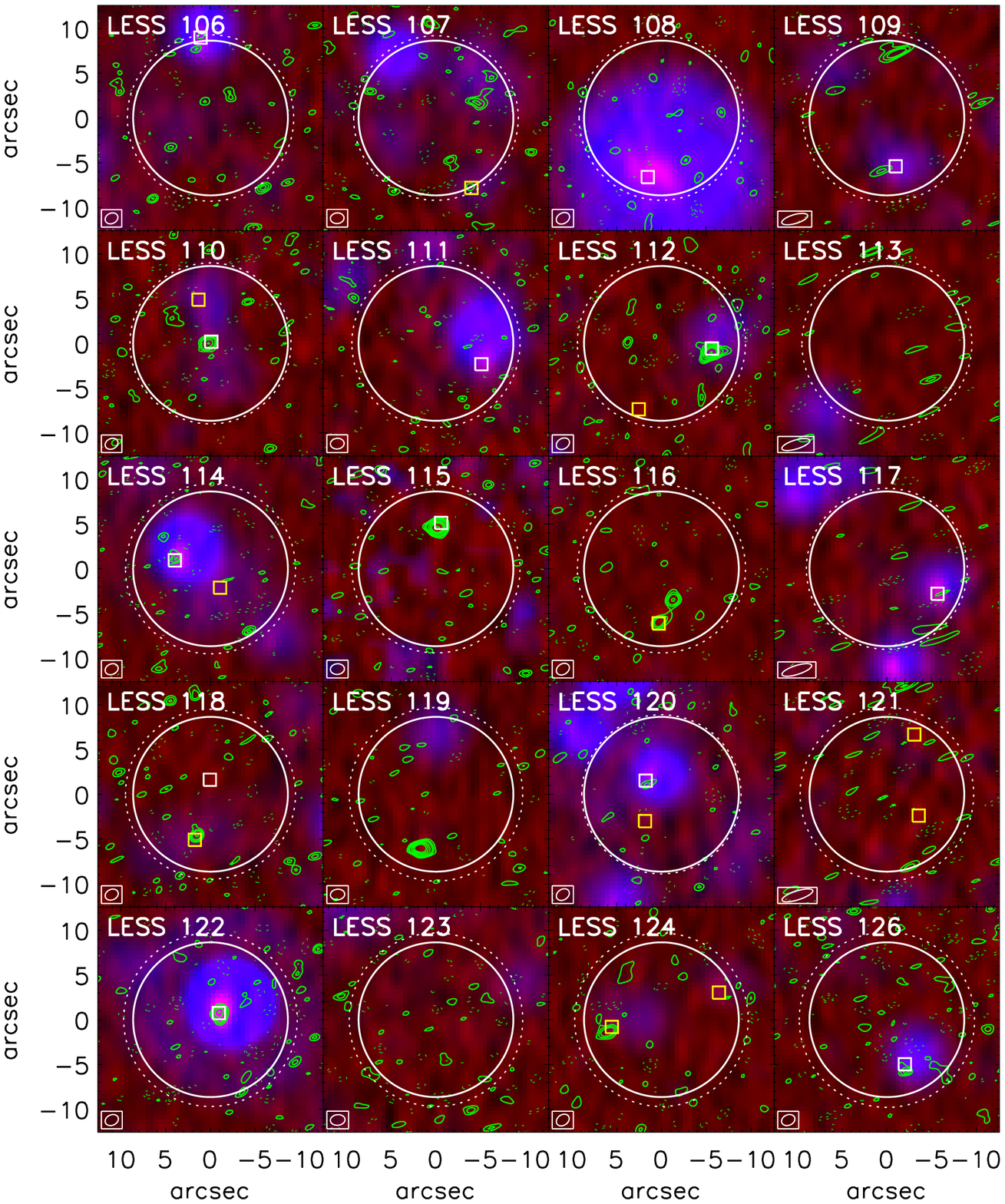}
\vspace{-20mm}
\begin{center}
Fig. \ref{fig:falsecolor} (continued)
\end{center}
\end{figure*}


\LongTables	

\clearpage

%
\begin{deluxetable*}{ l c c c c c c }
\tabletypesize{\small}
\tablewidth{0pt}
\tablecaption{ALESS Observations by LESS field \label{tab-2}}
\tablehead{
\colhead{LESS ID} & \colhead{Position} & \colhead{S$_{\rm LABOCA}$} & \colhead{S/N$_{\rm LABOCA}$} & \colhead{$\sigma_{\rm ALMA}$} & \colhead{ALMA Beam} & \colhead{SB$^{a}$}\\
 & (J2000) & \colhead{[mJy\,beam$^{-1}$]} & & \colhead{[mJy\,beam$^{-1}$]} & \colhead{[$^{\prime\prime}$]} &  
}
\startdata
 LESS 1    &   03:33:14.26   -27:56:11.2  & 14.5 $\pm$  1.2  & 12.5  & 0.41  & 1.34 $\times$ 1.16  &  SB1  \\
 LESS 2    &   03:33:02.50   -27:56:43.6  & 12.0 $\pm$  1.2  & 10.3  & 0.39  & 1.86 $\times$ 1.17  &  SB5  \\
 LESS 3    &   03:33:21.51   -27:55:20.2  & 11.7 $\pm$  1.2  & 10.1  & 0.40  & 1.26 $\times$ 1.11  &  SB3  \\
 LESS 4    &   03:31:36.01   -27:54:39.2  & 11.0 $\pm$  1.2  &  9.7  & 0.65  & 2.41 $\times$ 1.07  &  SB7  \\
 LESS 5    &   03:31:29.46   -27:59:07.3  & 10.0 $\pm$  1.2  &  8.5  & 0.41  & 1.41 $\times$ 1.12  &  SB1  \\
 LESS 6    &   03:32:57.14   -28:01:02.1  &  9.7 $\pm$  1.2  &  8.2  & 0.39  & 1.91 $\times$ 1.17  &  SB5  \\
 LESS 7    &   03:33:15.55   -27:45:23.6  &  9.2 $\pm$  1.2  &  7.9  & 0.31  & 1.22 $\times$ 1.17  &  SB3  \\
 LESS 8    &   03:32:05.07   -27:31:08.8  & 11.7 $\pm$  1.6  &  7.8  & 0.84  & 2.45 $\times$ 1.07  &  SB7  \\
 LESS 9    &   03:32:11.29   -27:52:10.4  &  9.2 $\pm$  1.2  &  7.7  & 0.46  & 1.40 $\times$ 1.11  &  SB1  \\
 LESS 10   &   03:32:19.02   -27:52:19.4  &  9.1 $\pm$  1.2  &  7.6  & 0.41  & 1.99 $\times$ 1.17  &  SB5  \\
 LESS 11   &   03:32:13.58   -27:56:02.5  &  9.1 $\pm$  1.2  &  7.6  & 0.35  & 1.22 $\times$ 1.17  &  SB3  \\
 LESS 12   &   03:32:48.12   -27:54:14.7  &  8.8 $\pm$  1.2  &  7.2  & 1.01  & 2.50 $\times$ 1.07  &  SB7  \\
 LESS 13   &   03:32:49.23   -27:42:46.6  &  8.8 $\pm$  1.2  &  7.2  & 0.42  & 1.36 $\times$ 1.15  &  SB1  \\
 LESS 14   &   03:31:52.64   -28:03:20.4  &  9.3 $\pm$  1.3  &  7.2  & 0.50  & 1.98 $\times$ 1.10  &  SB5  \\
 LESS 15   &   03:33:33.36   -27:59:30.1  &  8.9 $\pm$  1.3  &  7.0  & 0.36  & 1.22 $\times$ 1.16  &  SB3  \\
 LESS 16   &   03:32:18.89   -27:37:38.7  &  8.1 $\pm$  1.2  &  6.9  & 0.97  & 2.60 $\times$ 1.07  &  SB7  \\
 LESS 17   &   03:32:07.59   -27:51:23.0  &  7.6 $\pm$  1.3  &  6.4  & 0.38  & 1.40 $\times$ 1.16  &  SB1  \\
 LESS 18   &   03:32:05.12   -27:46:52.1  &  7.5 $\pm$  1.2  &  6.3  & 0.41  & 2.07 $\times$ 1.10  &  SB5  \\
 LESS 19   &   03:32:08.10   -27:58:18.7  &  7.3 $\pm$  1.2  &  6.2  & 0.33  & 1.21 $\times$ 1.17  &  SB3  \\
 LESS 20   &   03:33:16.56   -28:00:18.8  &  7.2 $\pm$  1.2  &  6.2  & 0.93  & 2.67 $\times$ 1.07  &  SB7  \\
 LESS 21   &   03:33:29.93   -27:34:41.7  &  7.6 $\pm$  1.3  &  6.1  & 0.36  & 1.41 $\times$ 1.16  &  SB1  \\
 LESS 22   &   03:31:47.02   -27:32:43.0  &  8.0 $\pm$  1.5  &  5.9  & 0.46  & 2.13 $\times$ 1.10  &  SB5  \\
 LESS 23   &   03:32:12.11   -28:05:08.5  &  8.2 $\pm$  1.5  &  5.9  & 0.35  & 1.21 $\times$ 1.17  &  SB3  \\
 LESS 24   &   03:33:36.79   -27:44:01.0  &  7.4 $\pm$  1.4  &  5.9  & 0.99  & 2.75 $\times$ 1.07  &  SB7  \\
 LESS 25   &   03:31:57.05   -27:59:40.8  &  6.7 $\pm$  1.3  &  5.8  & 0.44  & 1.62 $\times$ 1.15  &  SB1  \\
 LESS 26   &   03:31:36.90   -27:54:56.1  &  6.6 $\pm$  1.2  &  5.8  & 0.43  & 2.15 $\times$ 1.06  &  SB5  \\
 LESS 27   &   03:31:49.73   -27:34:32.7  &  7.2 $\pm$  1.4  &  5.8  & 0.33  & 1.21 $\times$ 1.17  &  SB3  \\
 LESS 28   &   03:33:02.92   -27:44:32.6  &  6.7 $\pm$  1.3  &  5.6  & 1.19  & 2.99 $\times$ 1.00  &  SB7  \\
 LESS 29   &   03:33:36.90   -27:58:13.0  &  7.1 $\pm$  1.4  &  5.6  & 0.38  & 1.63 $\times$ 1.14  &  SB1  \\
 LESS 30   &   03:33:44.37   -28:03:46.1  &  8.7 $\pm$  1.8  &  5.6  & 0.53  & 2.18 $\times$ 1.06  &  SB5  \\
 LESS 31   &   03:31:49.96   -27:57:43.9  &  6.3 $\pm$  1.3  &  5.5  & 0.32  & 1.21 $\times$ 1.17  &  SB3  \\
 LESS 32   &   03:32:43.57   -27:46:44.0  &  6.4 $\pm$  1.3  &  5.5  & 1.30  & 3.08 $\times$ 0.99  &  SB7  \\
 LESS 33   &   03:31:49.78   -27:53:32.9  &  6.4 $\pm$  1.3  &  5.5  & 0.35  & 1.49 $\times$ 1.16  &  SB1  \\
 LESS 34   &   03:32:17.64   -27:52:30.3  &  6.3 $\pm$  1.3  &  5.4  & 0.45  & 2.33 $\times$ 1.06  &  SB5  \\
 LESS 35   &   03:31:10.35   -27:37:14.8  &  8.1 $\pm$  1.8  &  5.4  & 0.32  & 1.21 $\times$ 1.17  &  SB3  \\
 LESS 36   &   03:31:49.15   -28:02:08.7  &  6.4 $\pm$  1.4  &  5.4  & 1.29  & 3.24 $\times$ 1.00  &  SB7  \\
 LESS 37   &   03:33:36.04   -27:53:47.6  &  6.7 $\pm$  1.5  &  5.3  & 0.37  & 1.50 $\times$ 1.17  &  SB1  \\
 LESS 38   &   03:33:10.20   -27:56:41.5  &  6.0 $\pm$  1.3  &  5.2  & 0.58  & 2.39 $\times$ 1.06  &  SB5  \\
 LESS 39   &   03:31:44.90   -27:34:35.4  &  6.2 $\pm$  1.4  &  5.2  & 0.32  & 1.21 $\times$ 1.17  &  SB3  \\
 LESS 40   &   03:32:46.74   -27:51:20.9  &  5.9 $\pm$  1.3  &  5.2  & 1.36  & 3.36 $\times$ 1.00  &  SB7  \\
 LESS 41   &   03:31:10.47   -27:52:33.2  &  7.6 $\pm$  1.9  &  5.2  & 0.42  & 1.55 $\times$ 1.17  &  SB1  \\
 LESS 42   &   03:32:31.02   -27:58:58.1  &  5.8 $\pm$  1.4  &  5.1  & 0.49  & 2.54 $\times$ 1.06  &  SB5  \\
 LESS 43   &   03:33:07.00   -27:48:01.0  &  5.9 $\pm$  1.4  &  5.1  & 0.34  & 1.22 $\times$ 1.17  &  SB3  \\
 LESS 44   &   03:31:30.96   -27:32:38.5  &  6.7 $\pm$  1.6  &  5.1  & 1.40  & 3.56 $\times$ 1.00  &  SB7  \\
 LESS 45   &   03:32:25.71   -27:52:28.5  &  5.8 $\pm$  1.4  &  5.1  & 0.38  & 1.57 $\times$ 1.17  &  SB1  \\
 LESS 46   &   03:33:36.80   -27:32:47.0  &  7.2 $\pm$  1.8  &  5.1  & 0.65  & 2.63 $\times$ 1.05  &  SB5  \\
 LESS 47   &   03:32:56.00   -27:33:17.7  &  6.3 $\pm$  1.5  &  5.1  & 0.35  & 1.22 $\times$ 1.16  &  SB3  \\
LESS 48	& 03:32:37.77   -27:30:02.0  & 6.8 $\pm$ 1.7	& 5.1	&  2.49	 &  3.77 $\times$ 0.80  &  SB7\\
 LESS 49   &   03:31:24.45   -27:50:40.9  &  5.9 $\pm$  1.4  &  5.1  & 0.43  & 1.62 $\times$ 1.17  &  SB1  \\
 LESS 50   &   03:31:41.15   -27:44:41.5  &  5.6 $\pm$  1.3  &  5.0  & 0.66  & 2.96 $\times$ 0.99  &  SB5  \\
 LESS 51   &   03:31:44.81   -27:44:25.1  &  5.6 $\pm$  1.3  &  5.0  & 0.33  & 1.22 $\times$ 1.17  &  SB3  \\
LESS 52 & 03:31:28.51   -27:56:01.3  & 5.6 $\pm$ 1.4	& 4.9	&  --    &  --                 &  --\\
 LESS 53   &   03:31:59.12   -27:54:35.5  &  5.6 $\pm$  1.4  &  4.9  & 0.42  & 1.68 $\times$ 1.12  &  SB1  \\
 LESS 54   &   03:32:43.61   -27:33:53.6  &  6.0 $\pm$  1.5  &  4.9  & 0.62  & 3.07 $\times$ 0.99  &  SB5  \\
 LESS 55   &   03:33:02.20   -27:40:33.6  &  5.5 $\pm$  1.4  &  4.9  & 0.35  & 1.22 $\times$ 1.16  &  SB3  \\
LESS 56 & 03:31:53.17   -27:39:36.1  & 5.4 $\pm$ 1.4	& 4.9	&  --    &  --		       &  --\\
 LESS 57   &   03:31:51.97   -27:53:29.7  &  5.5 $\pm$  1.4  &  4.9  & 0.57  & 1.88 $\times$ 1.06  &  SB1  \\
 LESS 58   &   03:32:25.79   -27:33:06.7  &  5.9 $\pm$  1.6  &  4.8  & 0.74  & 3.31 $\times$ 0.99  &  SB5  \\
 LESS 59   &   03:33:03.87   -27:44:12.2  &  5.3 $\pm$  1.4  &  4.8  & 0.31  & 1.22 $\times$ 1.16  &  SB3  \\
LESS 60 & 03:33:17.47   -27:51:21.5  & 5.2 $\pm$ 1.4	& 4.8	&  1.85  &  3.91 $\times$ 0.80 &  SB7\\
 LESS 61   &   03:32:45.63   -28:00:25.3  &  5.2 $\pm$  1.4  &  4.7  & 0.45  & 1.81 $\times$ 1.12  &  SB1  \\
 LESS 62   &   03:32:36.41   -27:34:52.5  &  5.4 $\pm$  1.5  &  4.7  & 0.68  & 3.47 $\times$ 0.99  &  SB5  \\
 LESS 63   &   03:33:08.46   -28:00:44.3  &  5.3 $\pm$  1.4  &  4.7  & 0.35  & 1.23 $\times$ 1.17  &  SB3  \\
LESS 64 & 03:32:01.00   -28:00:25.6  & 5.1 $\pm$ 1.4	& 4.7	&  --    &  --                 &  --\\
 LESS 65   &   03:32:52.40   -27:35:27.7  &  5.2 $\pm$  1.4  &  4.7  & 0.41  & 2.00 $\times$ 1.16  &  SB2  \\
 LESS 66   &   03:33:31.69   -27:54:06.1  &  5.3 $\pm$  1.5  &  4.7  & 0.39  & 1.41 $\times$ 1.14  &  SB6  \\
 LESS 67   &   03:32:43.28   -27:55:17.9  &  5.2 $\pm$  1.4  &  4.7  & 0.34  & 1.29 $\times$ 1.12  &  SB4  \\
 LESS 68   &   03:32:33.44   -27:39:18.5  &  5.1 $\pm$  1.4  &  4.7  & 0.44  & 1.70 $\times$ 1.19  &  SB8  \\
 LESS 69   &   03:31:34.26   -27:59:34.3  &  5.0 $\pm$  1.3  &  4.7  & 0.42  & 2.00 $\times$ 1.10  &  SB2  \\
 LESS 70   &   03:31:43.97   -27:38:32.5  &  5.0 $\pm$  1.4  &  4.6  & 0.41  & 1.42 $\times$ 1.14  &  SB6  \\
 LESS 71   &   03:33:06.29   -27:33:27.7  &  5.6 $\pm$  1.6  &  4.6  & 0.30  & 1.35 $\times$ 1.16  &  SB4  \\
 LESS 72   &   03:32:40.40   -27:38:02.5  &  5.0 $\pm$  1.4  &  4.6  & 0.42  & 1.68 $\times$ 1.18  &  SB8  \\
 LESS 73   &   03:32:29.33   -27:56:19.3  &  5.1 $\pm$  1.4  &  4.6  & 0.47  & 2.07 $\times$ 1.09  &  SB2  \\
 LESS 74   &   03:33:09.34   -27:48:09.9  &  5.1 $\pm$  1.4  &  4.6  & 0.40  & 1.42 $\times$ 1.14  &  SB6  \\
 LESS 75   &   03:31:26.83   -27:55:54.6  &  5.1 $\pm$  1.4  &  4.6  & 0.33  & 1.37 $\times$ 1.18  &  SB4  \\
 LESS 76   &   03:33:32.67   -27:59:57.2  &  5.1 $\pm$  1.5  &  4.5  & 0.47  & 1.67 $\times$ 1.18  &  SB8  \\
 LESS 77   &   03:31:57.23   -27:56:33.2  &  4.8 $\pm$  1.4  &  4.4  & 0.43  & 2.13 $\times$ 1.09  &  SB2  \\
 LESS 78   &   03:33:40.30   -27:39:56.9  &  5.1 $\pm$  1.7  &  4.4  & 0.35  & 1.43 $\times$ 1.14  &  SB6  \\
 LESS 79   &   03:32:21.25   -27:56:23.5  &  4.7 $\pm$  1.4  &  4.4  & 0.32  & 1.38 $\times$ 1.18  &  SB4  \\
 LESS 80   &   03:31:42.23   -27:48:34.4  &  4.6 $\pm$  1.4  &  4.4  & 0.47  & 1.65 $\times$ 1.18  &  SB8  \\
 LESS 81   &   03:31:27.45   -27:44:40.4  &  4.8 $\pm$  1.5  &  4.4  & 0.51  & 2.15 $\times$ 1.05  &  SB2  \\
 LESS 82   &   03:32:53.77   -27:38:10.9  &  4.5 $\pm$  1.4  &  4.4  & 0.37  & 1.43 $\times$ 1.14  &  SB6  \\
 LESS 83   &   03:33:08.92   -28:05:22.0  &  5.3 $\pm$  1.8  &  4.4  & 0.30  & 1.39 $\times$ 1.18  &  SB4  \\
 LESS 84   &   03:31:54.22   -27:51:09.8  &  4.6 $\pm$  1.4  &  4.3  & 0.47  & 1.64 $\times$ 1.18  &  SB8  \\
 LESS 85   &   03:31:10.28   -27:45:03.1  &  6.0 $\pm$  2.4  &  4.3  & 0.54  & 2.22 $\times$ 1.05  &  SB2  \\
 LESS 86   &   03:31:14.90   -27:48:44.3  &  5.1 $\pm$  1.8  &  4.3  & 0.40  & 1.43 $\times$ 1.16  &  SB6  \\
 LESS 87   &   03:32:51.09   -27:31:43.0  &  5.3 $\pm$  1.9  &  4.3  & 0.32  & 1.40 $\times$ 1.18  &  SB4  \\
 LESS 88   &   03:31:55.19   -27:53:45.3  &  4.5 $\pm$  1.4  &  4.3  & 0.37  & 1.63 $\times$ 1.17  &  SB8  \\
 LESS 89   &   03:32:48.44   -28:00:23.8  &  4.4 $\pm$  1.4  &  4.3  & 0.57  & 2.31 $\times$ 1.05  &  SB2  \\
 LESS 90   &   03:32:43.65   -27:35:54.1  &  4.5 $\pm$  1.5  &  4.2  & 0.40  & 1.44 $\times$ 1.16  &  SB6  \\
 LESS 91   &   03:31:35.25   -27:40:33.7  &  4.4 $\pm$  1.4  &  4.2  & 0.35  & 1.43 $\times$ 1.18  &  SB4  \\
 LESS 92   &   03:31:38.36   -27:43:36.0  &  4.3 $\pm$  1.4  &  4.2  & 0.38  & 1.62 $\times$ 1.17  &  SB8  \\
 LESS 93   &   03:31:10.84   -27:56:07.2  &  5.2 $\pm$  2.0  &  4.2  & 0.52  & 2.41 $\times$ 1.05  &  SB2  \\
 LESS 94   &   03:33:07.27   -27:58:05.0  &  4.4 $\pm$  1.4  &  4.2  & 0.44  & 1.45 $\times$ 1.15  &  SB6  \\
 LESS 95   &   03:32:41.74   -27:58:46.1  &  4.3 $\pm$  1.4  &  4.2  & 0.33  & 1.43 $\times$ 1.18  &  SB4  \\
 LESS 96   &   03:33:13.03   -27:55:56.8  &  4.3 $\pm$  1.4  &  4.2  & 0.40  & 1.61 $\times$ 1.17  &  SB8  \\
 LESS 97   &   03:33:13.65   -27:38:03.4  &  4.2 $\pm$  1.4  &  4.2  & 0.60  & 2.61 $\times$ 1.00  &  SB2  \\
 LESS 98   &   03:31:30.22   -27:57:26.0  &  4.2 $\pm$  1.4  &  4.1  & 0.47  & 1.47 $\times$ 1.15  &  SB6  \\
 LESS 99   &   03:32:51.45   -27:55:36.0  &  4.3 $\pm$  1.4  &  4.1  & 0.33  & 1.45 $\times$ 1.18  &  SB4  \\
 LESS 100  &   03:31:11.32   -28:00:06.2  &  4.8 $\pm$  1.9  &  4.1  & 0.35  & 1.60 $\times$ 1.17  &  SB8  \\
 LESS 101  &   03:31:51.47   -27:45:52.1  &  4.2 $\pm$  1.4  &  4.1  & 0.75  & 2.73 $\times$ 1.00  &  SB2  \\
 LESS 102  &   03:33:35.61   -27:40:20.1  &  4.3 $\pm$  1.5  &  4.1  & 0.46  & 1.47 $\times$ 1.15  &  SB6  \\
 LESS 103  &   03:33:25.35   -27:34:00.4  &  4.3 $\pm$  1.5  &  4.1  & 0.34  & 1.47 $\times$ 1.18  &  SB4  \\
 LESS 104  &   03:32:58.46   -27:38:03.0  &  4.0 $\pm$  1.4  &  4.1  & 0.38  & 1.59 $\times$ 1.17  &  SB8  \\
 LESS 105  &   03:31:15.78   -27:53:13.1  &  4.6 $\pm$  1.7  &  4.1  & 0.54  & 2.93 $\times$ 1.00  &  SB2  \\
 LESS 106  &   03:31:40.09   -27:56:31.4  &  4.0 $\pm$  1.4  &  4.0  & 0.45  & 1.48 $\times$ 1.16  &  SB6  \\
 LESS 107  &   03:31:30.85   -27:51:50.9  &  4.0 $\pm$  1.4  &  4.0  & 0.31  & 1.53 $\times$ 1.19  &  SB4  \\
 LESS 108  &   03:33:16.42   -27:50:33.1  &  4.0 $\pm$  1.4  &  4.0  & 0.38  & 1.58 $\times$ 1.18  &  SB8  \\
 LESS 109  &   03:33:28.08   -27:41:57.0  &  4.0 $\pm$  1.4  &  4.0  & 0.60  & 3.01 $\times$ 1.00  &  SB2  \\
 LESS 110  &   03:31:22.64   -27:54:17.2  &  4.1 $\pm$  1.5  &  4.0  & 0.47  & 1.49 $\times$ 1.16  &  SB6  \\
 LESS 111  &   03:33:25.58   -27:34:23.0  &  4.1 $\pm$  1.5  &  4.0  & 0.34  & 1.54 $\times$ 1.19  &  SB4  \\
 LESS 112  &   03:32:49.28   -27:31:12.3  &  4.6 $\pm$  2.0  &  4.0  & 0.37  & 1.57 $\times$ 1.18  &  SB8  \\
 LESS 113  &   03:32:36.42   -27:58:45.9  &  3.9 $\pm$  1.4  &  3.9  & 0.59  & 3.25 $\times$ 1.00  &  SB2  \\
 LESS 114  &   03:31:50.81   -27:44:38.5  &  3.9 $\pm$  1.4  &  3.9  & 0.43  & 1.51 $\times$ 1.16  &  SB6  \\
 LESS 115  &   03:33:49.71   -27:42:39.2  &  4.6 $\pm$  2.4  &  3.9  & 0.35  & 1.57 $\times$ 1.19  &  SB4  \\
 LESS 116  &   03:31:54.42   -27:45:25.5  &  3.8 $\pm$  1.4  &  3.8  & 0.41  & 1.56 $\times$ 1.17  &  SB8  \\
 LESS 117  &   03:31:28.02   -27:39:25.2  &  3.8 $\pm$  1.4  &  3.8  & 0.63  & 3.47 $\times$ 1.00  &  SB2  \\
 LESS 118  &   03:31:21.81   -27:49:36.8  &  3.8 $\pm$  1.5  &  3.8  & 0.44  & 1.52 $\times$ 1.16  &  SB6  \\
 LESS 119  &   03:32:56.51   -28:03:19.1  &  3.8 $\pm$  1.5  &  3.8  & 0.38  & 1.60 $\times$ 1.19  &  SB4  \\
 LESS 120  &   03:33:28.45   -27:56:55.9  &  3.7 $\pm$  1.5  &  3.8  & 0.41  & 1.56 $\times$ 1.17  &  SB8  \\
 LESS 121  &   03:33:33.32   -27:34:49.3  &  3.8 $\pm$  1.6  &  3.8  & 0.62  & 3.60 $\times$ 1.00  &  SB2  \\
 LESS 122  &   03:31:39.62   -27:41:20.4  &  3.6 $\pm$  1.5  &  3.8  & 0.41  & 1.53 $\times$ 1.17  &  SB6  \\
 LESS 123  &   03:33:30.88   -27:53:49.3  &  3.7 $\pm$  1.6  &  3.8  & 0.38  & 1.64 $\times$ 1.18  &  SB4  \\
 LESS 124  &   03:32:03.59   -27:36:05.0  &  3.5 $\pm$  1.4  &  3.7  & 0.40  & 1.55 $\times$ 1.17  &  SB8  \\
LESS 125 & 03:31:46.02   -27:46:21.2  & 3.6 $\pm$ 1.4	& 3.7	&  --    &  --                 &  --\\
 LESS 126  &   03:32:09.76   -27:41:02.0  &  3.6 $\pm$  1.4  &  3.7  & 0.39  & 1.55 $\times$ 1.17  &  SB6  
\enddata
\tablenotetext{a}{Scheduling block -- See Table 1 for more details.}
\end{deluxetable*}

\clearpage
\begin{deluxetable*}{ l c c c c c c c c }
\tabletypesize{\small}
\tablewidth{0pt}
\tablecaption{ALESS MAIN sample sources by LESS field \label{tab-3}}
\tablehead{
\colhead{LESS ID} & \colhead{ALESS ID} &\colhead{ALMA Position} & \colhead{$\delta$RA/$\delta$Dec} & \colhead{S$_{\rm pk}$} & \colhead{S$_{\rm int}$} & \colhead{S/N$_{\rm pk}$} & \colhead{S$_{\rm BEST,pbcorr}$} & \colhead{Biggs ID} \\
 & & (J2000) & [$^{\prime\prime}$] & [mJy] &  [mJy] &  & [mJy] 
}
\startdata
 LESS 1    & ALESS 001.1   &  03 33 14.46  -27 56 14.5  & 0.04/0.04  &  5.7  $\pm$  0.4  &  5.8  $\pm$  0.7  & 13.7  &  6.7  $\pm$  0.5 & --  \\
           & ALESS 001.2   &  03 33 14.41  -27 56 11.6  & 0.06/0.07  &  3.3  $\pm$  0.4  &  4.4  $\pm$  0.9  &  8.1  &  3.5  $\pm$  0.4 & \textbf{r} \\
           & ALESS 001.3   &  03 33 14.18  -27 56 12.3  & 0.12/0.12  &  1.8  $\pm$  0.4  &  2.5  $\pm$  0.9  &  4.4  &  1.9  $\pm$  0.4 & -- \\
 LESS 2    & ALESS 002.1   &  03 33 02.69  -27 56 42.8  & 0.06/0.09  &  3.6  $\pm$  0.4  &  3.3  $\pm$  0.6  &  9.1  &  3.8  $\pm$  0.4 & \textbf{r} \\
           & ALESS 002.2   &  03 33 03.07  -27 56 42.9  & 0.08/0.12  &  2.5  $\pm$  0.4  &  3.0  $\pm$  0.8  &  6.3  &  4.2  $\pm$  0.7 & -- \\
 LESS 3    & ALESS 003.1   &  03 33 21.50  -27 55 20.3  & 0.02/0.02  &  8.3  $\pm$  0.4  &  9.0  $\pm$  0.8  & 20.8  &  8.3  $\pm$  0.4 & \textbf{r} \\
 LESS 5    & ALESS 005.1   &  03 31 28.91  -27 59 09.0  & 0.04/0.05  &  4.6  $\pm$  0.4  &  4.8  $\pm$  0.7  & 11.4  &  7.8  $\pm$  0.7 & -- \\
 LESS 6    & ALESS 006.1   &  03 32 56.96  -28 01 00.7  & 0.04/0.05  &  5.6  $\pm$  0.4  &  5.9  $\pm$  0.7  & 14.4  &  6.0  $\pm$  0.4 & \textbf{r} \\
 LESS 7    & ALESS 007.1   &  03 33 15.42  -27 45 24.3  & 0.03/0.03  &  5.9  $\pm$  0.3  &  7.7  $\pm$  0.7  & 19.3  &  8.0  $\pm$  0.7 & \textbf{r} \\
 LESS 9    & ALESS 009.1   &  03 32 11.34  -27 52 11.9  & 0.03/0.03  &  8.5  $\pm$  0.5  &  8.4  $\pm$  0.8  & 18.6  &  8.8  $\pm$  0.5 & \textbf{r} \\
 LESS 10   & ALESS 010.1   &  03 32 19.06  -27 52 14.8  & 0.08/0.05  &  4.3  $\pm$  0.4  &  5.1  $\pm$  0.8  & 10.4  &  5.2  $\pm$  0.5 & \textbf{r} \\
 LESS 11   & ALESS 011.1   &  03 32 13.85  -27 56 00.3  & 0.03/0.03  &  6.2  $\pm$  0.3  &  6.7  $\pm$  0.6  & 17.9  &  7.3  $\pm$  0.4 & \textbf{r} \\
 LESS 13   & ALESS 013.1   &  03 32 48.99  -27 42 51.8  & 0.04/0.04  &  5.7  $\pm$  0.4  &  6.1  $\pm$  0.8  & 13.7  &  8.0  $\pm$  0.6 & -- \\
 LESS 14   & ALESS 014.1   &  03 31 52.49  -28 03 19.1  & 0.03/0.06  &  7.1  $\pm$  0.5  &  7.0  $\pm$  0.9  & 14.3  &  7.5  $\pm$  0.5 & \textbf{r} \\
 LESS 15   & ALESS 015.1   &  03 33 33.37  -27 59 29.6  & 0.02/0.02  &  9.0  $\pm$  0.4  &  9.7  $\pm$  0.7  & 24.6  &  9.0  $\pm$  0.4 & \textbf{r} \\
           & ALESS 015.3   &  03 33 33.59  -27 59 35.4  & 0.14/0.13  &  1.4  $\pm$  0.4  &  1.7  $\pm$  0.8  &  3.8  &  2.0  $\pm$  0.5 & -- \\
 LESS 17   & ALESS 017.1   &  03 32 07.30  -27 51 20.8  & 0.03/0.03  &  7.0  $\pm$  0.4  &  8.0  $\pm$  0.8  & 18.3  &  8.4  $\pm$  0.5 & \textbf{r} \\
 LESS 18   & ALESS 018.1   &  03 32 04.88  -27 46 47.7  & 0.08/0.09  &  3.3  $\pm$  0.4  &  4.1  $\pm$  0.8  &  8.1  &  4.4  $\pm$  0.5 & \textbf{r} \\
 LESS 19   & ALESS 019.1   &  03 32 08.26  -27 58 14.2  & 0.04/0.04  &  4.0  $\pm$  0.3  &  4.2  $\pm$  0.6  & 11.9  &  5.0  $\pm$  0.4 & \textbf{r} \\
           & ALESS 019.2   &  03 32 07.89  -27 58 24.1  & 0.12/0.12  &  1.4  $\pm$  0.3  &  1.5  $\pm$  0.6  &  4.2  &  2.0  $\pm$  0.5 & \textbf{r} \\
 LESS 22   & ALESS 022.1   &  03 31 46.92  -27 32 39.3  & 0.10/0.06  &  3.9  $\pm$  0.5  &  5.1  $\pm$  1.0  &  8.4  &  4.5  $\pm$  0.5 & \textbf{r} \\
 LESS 23   & ALESS 023.1   &  03 32 12.01  -28 05 06.5  & 0.03/0.03  &  6.4  $\pm$  0.3  &  7.1  $\pm$  0.7  & 18.4  &  6.7  $\pm$  0.4 & -- \\
           & ALESS 023.7   &  03 32 11.92  -28 05 14.0  & 0.14/0.14  &  1.3  $\pm$  0.3  &  2.1  $\pm$  0.9  &  3.6  &  1.8  $\pm$  0.5 & -- \\
 LESS 25   & ALESS 025.1   &  03 31 56.88  -27 59 39.3  & 0.05/0.04  &  5.8  $\pm$  0.4  &  6.5  $\pm$  0.8  & 13.2  &  6.2  $\pm$  0.5 & \textbf{r} \\
 LESS 29   & ALESS 029.1   &  03 33 36.90  -27 58 09.3  & 0.05/0.04  &  5.2  $\pm$  0.4  &  5.4  $\pm$  0.7  & 13.7  &  5.9  $\pm$  0.4 & \textbf{r} \\
 LESS 31   & ALESS 031.1   &  03 31 49.79  -27 57 40.8  & 0.02/0.02  &  7.1  $\pm$  0.3  &  7.4  $\pm$  0.6  & 22.0  &  8.1  $\pm$  0.4 & \textbf{r} \\
 LESS 35   & ALESS 035.1   &  03 31 10.51  -27 37 15.4  & 0.04/0.04  &  4.2  $\pm$  0.3  &  5.2  $\pm$  0.7  & 13.0  &  4.4  $\pm$  0.3 & \textbf{r} \\
           & ALESS 035.2   &  03 31 10.22  -27 37 18.1  & 0.13/0.13  &  1.2  $\pm$  0.3  &  1.3  $\pm$  0.6  &  3.9  &  1.4  $\pm$  0.4 & -- \\
 LESS 37   & ALESS 037.1   &  03 33 36.14  -27 53 50.6  & 0.08/0.08  &  2.6  $\pm$  0.4  &  3.1  $\pm$  0.7  &  7.1  &  2.9  $\pm$  0.4 & -- \\
           & ALESS 037.2   &  03 33 36.36  -27 53 48.3  & 0.15/0.16  &  1.4  $\pm$  0.4  &  1.6  $\pm$  0.7  &  3.7  &  1.6  $\pm$  0.4 & -- \\
 LESS 39   & ALESS 039.1   &  03 31 45.03  -27 34 36.7  & 0.04/0.04  &  4.1  $\pm$  0.3  &  4.7  $\pm$  0.6  & 12.9  &  4.3  $\pm$  0.3 & \textbf{r} \\
 LESS 41   & ALESS 041.1   &  03 31 10.07  -27 52 36.7  & 0.08/0.07  &  3.4  $\pm$  0.4  &  3.3  $\pm$  0.7  &  8.0  &  4.9  $\pm$  0.6 & \textbf{r} \\
           & ALESS 041.3   &  03 31 10.30  -27 52 40.8  & 0.18/0.15  &  1.5  $\pm$  0.4  &  2.8  $\pm$  1.2  &  3.6  &  2.7  $\pm$  0.8 & -- \\
 LESS 43   & ALESS 043.1   &  03 33 06.64  -27 48 02.4  & 0.09/0.09  &  1.8  $\pm$  0.3  &  2.3  $\pm$  0.7  &  5.4  &  2.3  $\pm$  0.4 & \textbf{r} \\
 LESS 45   & ALESS 045.1   &  03 32 25.26  -27 52 30.5  & 0.05/0.06  &  4.2  $\pm$  0.4  &  4.7  $\pm$  0.7  & 11.1  &  6.0  $\pm$  0.5 & \textbf{r} \\
 LESS 49   & ALESS 049.1   &  03 31 24.72  -27 50 47.1  & 0.06/0.08  &  3.7  $\pm$  0.4  &  3.9  $\pm$  0.8  &  8.8  &  6.0  $\pm$  0.7 & \textbf{r} \\
           & ALESS 049.2   &  03 31 24.47  -27 50 38.1  & 0.13/0.17  &  1.7  $\pm$  0.4  &  2.1  $\pm$  0.9  &  3.9  &  1.8  $\pm$  0.5 & \textbf{r} \\
 LESS 51   & ALESS 051.1   &  03 31 45.06  -27 44 27.3  & 0.04/0.04  &  4.0  $\pm$  0.3  &  4.3  $\pm$  0.6  & 12.1  &  4.7  $\pm$  0.4 & t \\
 LESS 55   & ALESS 055.1   &  03 33 02.22  -27 40 35.4  & 0.05/0.04  &  3.9  $\pm$  0.3  &  4.4  $\pm$  0.7  & 11.2  &  4.0  $\pm$  0.4 & -- \\
           & ALESS 055.2   &  03 33 02.16  -27 40 41.3  & 0.13/0.13  &  1.3  $\pm$  0.3  &  1.6  $\pm$  0.7  &  3.9  &  2.4  $\pm$  0.6 & -- \\
           & ALESS 055.5   &  03 33 02.35  -27 40 35.4  & 0.14/0.13  &  1.3  $\pm$  0.3  &  1.5  $\pm$  0.7  &  3.7  &  1.4  $\pm$  0.4 & -- \\
 LESS 57   & ALESS 057.1   &  03 31 51.92  -27 53 27.1  & 0.12/0.11  &  3.3  $\pm$  0.6  &  3.4  $\pm$  1.0  &  5.8  &  3.6  $\pm$  0.6 & \textbf{r} \\
 LESS 59   & ALESS 059.2   &  03 33 03.82  -27 44 18.2  & 0.12/0.11  &  1.4  $\pm$  0.3  &  2.3  $\pm$  0.8  &  4.4  &  1.9  $\pm$  0.4 & -- \\
 LESS 61   & ALESS 061.1   &  03 32 45.87  -28 00 23.4  & 0.07/0.09  &  3.8  $\pm$  0.5  &  4.4  $\pm$  0.9  &  8.3  &  4.3  $\pm$  0.5 & -- \\
 LESS 63   & ALESS 063.1   &  03 33 08.45  -28 00 43.8  & 0.03/0.03  &  5.6  $\pm$  0.3  &  6.0  $\pm$  0.6  & 16.1  &  5.6  $\pm$  0.3 & -- \\
 LESS 65   & ALESS 065.1   &  03 32 52.27  -27 35 26.3  & 0.08/0.05  &  4.0  $\pm$  0.4  &  4.9  $\pm$  0.8  &  9.7  &  4.2  $\pm$  0.4 & -- \\
 LESS 66   & ALESS 066.1   &  03 33 31.93  -27 54 09.5  & 0.11/0.10  &  2.0  $\pm$  0.4  &  2.4  $\pm$  0.8  &  5.2  &  2.5  $\pm$  0.5 & \textbf{r} \\
 LESS 67   & ALESS 067.1   &  03 32 43.20  -27 55 14.3  & 0.05/0.04  &  4.0  $\pm$  0.3  &  4.9  $\pm$  0.7  & 11.7  &  4.5  $\pm$  0.4 & \textbf{r} \\
           & ALESS 067.2   &  03 32 43.02  -27 55 14.7  & 0.13/0.12  &  1.4  $\pm$  0.3  &  1.7  $\pm$  0.7  &  4.2  &  1.7  $\pm$  0.4 & -- \\
 LESS 68   & ALESS 068.1   &  03 32 33.33  -27 39 13.6  & 0.08/0.11  &  2.9  $\pm$  0.4  &  3.2  $\pm$  0.8  &  6.6  &  3.7  $\pm$  0.6 & -- \\
 LESS 69   & ALESS 069.1   &  03 31 33.78  -27 59 32.4  & 0.06/0.11  &  3.2  $\pm$  0.4  &  4.3  $\pm$  0.9  &  7.7  &  4.9  $\pm$  0.6 & t \\
           & ALESS 069.2   &  03 31 34.13  -27 59 28.9  & 0.11/0.20  &  1.8  $\pm$  0.4  &  2.1  $\pm$  0.8  &  4.2  &  2.4  $\pm$  0.6 & -- \\
           & ALESS 069.3   &  03 31 33.97  -27 59 38.3  & 0.13/0.23  &  1.5  $\pm$  0.4  &  1.9  $\pm$  0.8  &  3.7  &  2.1  $\pm$  0.6 & -- \\
 LESS 70   & ALESS 070.1   &  03 31 44.02  -27 38 35.5  & 0.04/0.05  &  4.8  $\pm$  0.4  &  5.3  $\pm$  0.8  & 11.7  &  5.2  $\pm$  0.4 & \textbf{r} \\
 LESS 71   & ALESS 071.1   &  03 33 05.65  -27 33 28.2  & 0.11/0.11  &  1.4  $\pm$  0.3  &  1.9  $\pm$  0.7  &  4.8  &  2.9  $\pm$  0.6 & -- \\
           & ALESS 071.3   &  03 33 06.14  -27 33 23.1  & 0.15/0.15  &  1.1  $\pm$  0.3  &  1.5  $\pm$  0.7  &  3.6  &  1.4  $\pm$  0.4 & -- \\
 LESS 72   & ALESS 072.1   &  03 32 40.40  -27 37 58.1  & 0.05/0.07  &  4.1  $\pm$  0.4  &  4.1  $\pm$  0.7  &  9.9  &  4.9  $\pm$  0.5 & -- \\
 LESS 73   & ALESS 073.1   &  03 32 29.29  -27 56 19.7  & 0.05/0.06  &  6.1  $\pm$  0.5  &  6.2  $\pm$  0.8  & 12.9  &  6.1  $\pm$  0.5 & \textbf{r} \\
 LESS 74   & ALESS 074.1   &  03 33 09.15  -27 48 17.2  & 0.07/0.09  &  2.7  $\pm$  0.4  &  3.3  $\pm$  0.8  &  6.7  &  4.6  $\pm$  0.7 & \textbf{r} \\
 LESS 75   & ALESS 075.1   &  03 31 27.19  -27 55 51.3  & 0.08/0.07  &  2.3  $\pm$  0.3  &  3.3  $\pm$  0.7  &  7.1  &  3.2  $\pm$  0.4 & \textbf{r} \\
           & ALESS 075.4   &  03 31 26.57  -27 55 55.7  & 0.16/0.15  &  1.2  $\pm$  0.3  &  1.4  $\pm$  0.7  &  3.5  &  1.3  $\pm$  0.4 & -- \\
 LESS 76   & ALESS 076.1   &  03 33 32.34  -27 59 55.6  & 0.06/0.05  &  5.2  $\pm$  0.5  &  5.6  $\pm$  0.9  & 11.0  &  6.4  $\pm$  0.6 & \textbf{r} \\
 LESS 79   & ALESS 079.1   &  03 32 21.14  -27 56 27.0  & 0.05/0.04  &  3.6  $\pm$  0.3  &  3.7  $\pm$  0.6  & 11.2  &  4.1  $\pm$  0.4 & -- \\
           & ALESS 079.2   &  03 32 21.60  -27 56 24.0  & 0.12/0.10  &  1.6  $\pm$  0.3  &  1.9  $\pm$  0.6  &  5.0  &  2.0  $\pm$  0.4 & \textbf{r} \\
           & ALESS 079.4   &  03 32 21.18  -27 56 30.5  & 0.17/0.14  &  1.1  $\pm$  0.3  &  1.2  $\pm$  0.6  &  3.5  &  1.8  $\pm$  0.5 & -- \\
 LESS 80   & ALESS 080.1   &  03 31 42.80  -27 48 36.9  & 0.11/0.15  &  2.2  $\pm$  0.5  &  2.7  $\pm$  1.0  &  4.7  &  4.0  $\pm$  0.9 & t \\
           & ALESS 080.2   &  03 31 42.62  -27 48 41.0  & 0.13/0.17  &  1.9  $\pm$  0.5  &  1.8  $\pm$  0.8  &  3.9  &  3.5  $\pm$  0.9 & -- \\
 LESS 82   & ALESS 082.1   &  03 32 54.00  -27 38 14.9  & 0.13/0.14  &  1.5  $\pm$  0.4  &  1.7  $\pm$  0.7  &  4.1  &  1.9  $\pm$  0.5 & t \\
 LESS 83   & ALESS 083.4   &  03 33 08.71  -28 05 18.5  & 0.15/0.13  &  1.2  $\pm$  0.3  &  1.5  $\pm$  0.6  &  3.9  &  1.4  $\pm$  0.4 & -- \\
 LESS 84   & ALESS 084.1   &  03 31 54.50  -27 51 05.6  & 0.13/0.11  &  2.4  $\pm$  0.5  &  2.4  $\pm$  0.8  &  5.1  &  3.2  $\pm$  0.6 & \textbf{r} \\
           & ALESS 084.2   &  03 31 53.85  -27 51 04.3  & 0.15/0.13  &  2.0  $\pm$  0.5  &  2.1  $\pm$  0.8  &  4.2  &  3.2  $\pm$  0.8 & t \\
 LESS 87   & ALESS 087.1   &  03 32 50.88  -27 31 41.5  & 0.15/0.14  &  1.2  $\pm$  0.3  &  1.7  $\pm$  0.7  &  3.8  &  1.3  $\pm$  0.4 & \textbf{r} \\
           & ALESS 087.3   &  03 32 51.27  -27 31 50.7  & 0.14/0.13  &  1.3  $\pm$  0.3  &  2.0  $\pm$  0.8  &  4.1  &  2.4  $\pm$  0.6 & -- \\
 LESS 88   & ALESS 088.1   &  03 31 54.76  -27 53 41.5  & 0.08/0.06  &  3.0  $\pm$  0.4  &  3.3  $\pm$  0.7  &  8.0  &  4.6  $\pm$  0.6 & t \\
           & ALESS 088.2   &  03 31 55.39  -27 53 40.3  & 0.16/0.12  &  1.6  $\pm$  0.4  &  2.3  $\pm$  0.8  &  4.2  &  2.1  $\pm$  0.5 & -- \\
           & ALESS 088.5   &  03 31 55.81  -27 53 47.2  & 0.18/0.13  &  1.5  $\pm$  0.4  &  2.5  $\pm$  0.9  &  4.0  &  2.9  $\pm$  0.7 & -- \\
           & ALESS 088.11  &  03 31 54.95  -27 53 37.6  & 0.28/0.17  &  1.3  $\pm$  0.4  &  6.5  $\pm$  2.0  &  3.5  &  2.5  $\pm$  0.7 & -- \\
 LESS 92   & ALESS 092.2   &  03 31 38.14  -27 43 43.4  & 0.14/0.19  &  1.3  $\pm$  0.4  &  1.6  $\pm$  0.7  &  3.6  &  2.4  $\pm$  0.7 & -- \\
 LESS 94   & ALESS 094.1   &  03 33 07.59  -27 58 05.8  & 0.08/0.10  &  2.7  $\pm$  0.4  &  2.9  $\pm$  0.8  &  6.1  &  3.2  $\pm$  0.5 & \textbf{r} \\
 LESS 98   & ALESS 098.1   &  03 31 29.92  -27 57 22.7  & 0.06/0.08  &  3.7  $\pm$  0.5  &  5.0  $\pm$  1.0  &  8.0  &  4.8  $\pm$  0.6 & \textbf{r} \\
 LESS 99   & ALESS 099.1   &  03 32 51.82  -27 55 33.6  & 0.12/0.11  &  1.6  $\pm$  0.3  &  1.5  $\pm$  0.5  &  4.8  &  2.1  $\pm$  0.4 & -- \\
 LESS 102  & ALESS 102.1   &  03 33 35.60  -27 40 23.0  & 0.08/0.10  &  2.8  $\pm$  0.5  &  3.2  $\pm$  0.9  &  6.2  &  3.1  $\pm$  0.5 & \textbf{r} \\
 LESS 103  & ALESS 103.3   &  03 33 25.04  -27 34 01.1  & 0.18/0.14  &  1.2  $\pm$  0.3  &  1.4  $\pm$  0.7  &  3.5  &  1.4  $\pm$  0.4 & -- \\
 LESS 107  & ALESS 107.1   &  03 31 30.50  -27 51 49.1  & 0.13/0.11  &  1.5  $\pm$  0.3  &  2.1  $\pm$  0.7  &  4.8  &  1.9  $\pm$  0.4 & -- \\
           & ALESS 107.3   &  03 31 30.72  -27 51 55.7  & 0.17/0.15  &  1.2  $\pm$  0.3  &  1.3  $\pm$  0.6  &  3.7  &  1.5  $\pm$  0.4 & -- \\
 LESS 110  & ALESS 110.1   &  03 31 22.66  -27 54 17.2  & 0.07/0.06  &  4.1  $\pm$  0.5  &  4.6  $\pm$  0.9  &  8.7  &  4.1  $\pm$  0.5 & \textbf{r} \\
           & ALESS 110.5   &  03 31 22.96  -27 54 14.4  & 0.17/0.14  &  1.9  $\pm$  0.5  &  8.5  $\pm$  2.4  &  4.0  &  2.4  $\pm$  0.6 & -- \\
 LESS 112  & ALESS 112.1   &  03 32 48.86  -27 31 13.3  & 0.04/0.04  &  5.7  $\pm$  0.4  &  5.9  $\pm$  0.7  & 15.5  &  7.6  $\pm$  0.5 & \textbf{r} \\
 LESS 114  & ALESS 114.1   &  03 31 50.49  -27 44 45.3  & 0.15/0.15  &  1.6  $\pm$  0.4  &  1.6  $\pm$  0.7  &  3.8  &  3.0  $\pm$  0.8 & -- \\
           & ALESS 114.2   &  03 31 51.11  -27 44 37.3  & 0.14/0.15  &  1.7  $\pm$  0.4  &  1.8  $\pm$  0.8  &  4.0  &  2.0  $\pm$  0.5 & \textbf{r} \\
 LESS 115  & ALESS 115.1   &  03 33 49.70  -27 42 34.6  & 0.04/0.03  &  5.7  $\pm$  0.3  &  6.0  $\pm$  0.6  & 16.4  &  6.9  $\pm$  0.4 & \textbf{r} \\
 LESS 116  & ALESS 116.1   &  03 31 54.32  -27 45 28.9  & 0.08/0.10  &  2.7  $\pm$  0.4  &  2.9  $\pm$  0.7  &  6.6  &  3.1  $\pm$  0.5 & -- \\
           & ALESS 116.2   &  03 31 54.44  -27 45 31.4  & 0.09/0.11  &  2.5  $\pm$  0.4  &  2.7  $\pm$  0.8  &  6.0  &  3.4  $\pm$  0.6 & t \\
 LESS 118  & ALESS 118.1   &  03 31 21.92  -27 49 41.4  & 0.09/0.11  &  2.6  $\pm$  0.4  &  2.4  $\pm$  0.7  &  5.9  &  3.2  $\pm$  0.5 & t \\
 LESS 119  & ALESS 119.1   &  03 32 56.64  -28 03 25.2  & 0.04/0.04  &  5.7  $\pm$  0.4  &  6.3  $\pm$  0.7  & 15.2  &  8.3  $\pm$  0.5 & -- \\
 LESS 122  & ALESS 122.1   &  03 31 39.54  -27 41 19.7  & 0.06/0.07  &  3.6  $\pm$  0.4  &  4.3  $\pm$  0.8  &  8.8  &  3.7  $\pm$  0.4 & \textbf{r} \\
 LESS 124  & ALESS 124.1   &  03 32 04.04  -27 36 06.4  & 0.10/0.08  &  2.6  $\pm$  0.4  &  3.8  $\pm$  0.9  &  6.4  &  3.6  $\pm$  0.6 & t \\
           & ALESS 124.4   &  03 32 03.89  -27 36 00.1  & 0.18/0.16  &  1.5  $\pm$  0.4  &  8.4  $\pm$  2.3  &  3.9  &  2.2  $\pm$  0.6 & -- \\
 LESS 126  & ALESS 126.1   &  03 32 09.61  -27 41 07.7  & 0.12/0.16  &  1.6  $\pm$  0.4  &  1.6  $\pm$  0.7  &  4.1  &  2.2  $\pm$  0.5 & \textbf{r} 
\enddata
\tablecomments{See \S\ref{columns} for column definitions.}
\end{deluxetable*}

\clearpage
\begin{deluxetable*}{ l c c c c c c c c c}
\tabletypesize{\small}
\tablewidth{0pt}
\tablecaption{ALESS Supplementary sample \label{tab-4}}
\tablehead{
\colhead{LESS ID} & \colhead{ALESS ID} &\colhead{ALMA Position} & \colhead{$\delta$RA/$\delta$Dec}  & \colhead{S$_{\rm pk}$} & \colhead{S$_{\rm int}$} & \colhead{S/N$_{\rm pk}$} & \colhead{S$_{\rm BEST,pbcorr}$} & \colhead{Sample$^{a}$} & \colhead{Biggs ID} \\
 & & (J2000) & [$^{\prime\prime}$] & [mJy] &  [mJy] &  & [mJy] & & 
}
\startdata
 LESS 3    & ALESS 003.2   &  03 33 22.19  -27 55 20.9  & 0.09/0.10  &  2.3  $\pm$  0.4  &  4.2  $\pm$  1.1  &  5.7  &  4.8  $\pm$  0.9  &  2 & -- \\
           & ALESS 003.3   &  03 33 20.71  -27 55 14.0  & 0.12/0.12  &  1.7  $\pm$  0.4  &  2.4  $\pm$  0.9  &  4.3  &  7.0  $\pm$  1.6  &  2 & -- \\
           & ALESS 003.4   &  03 33 21.99  -27 55 09.8  & 0.13/0.12  &  1.6  $\pm$  0.4  &  2.4  $\pm$  0.9  &  4.0  &  6.4  $\pm$  1.6  &  2 & -- \\
 LESS 7    & ALESS 007.2   &  03 33 15.01  -27 45 30.6  & 0.11/0.11  &  1.4  $\pm$  0.3  &  2.3  $\pm$  0.8  &  4.5  &  3.5  $\pm$  0.8  &  2 & -- \\
 LESS 15   & ALESS 015.2   &  03 33 34.05  -27 59 30.2  & 0.11/0.10  &  1.8  $\pm$  0.4  &  2.2  $\pm$  0.8  &  4.8  &  3.8  $\pm$  0.8  &  2 & -- \\
           & ALESS 015.6   &  03 33 33.17  -27 59 42.2  & 0.13/0.13  &  1.5  $\pm$  0.4  &  1.8  $\pm$  0.7  &  4.1  &  6.1  $\pm$  1.5  &  2 & -- \\
 LESS 17   & ALESS 017.2   &  03 32 08.26  -27 51 19.7  & 0.14/0.12  &  1.6  $\pm$  0.4  &  2.1  $\pm$  0.8  &  4.2  &  3.7  $\pm$  0.9  &  2 & -- \\
           & ALESS 017.3   &  03 32 07.37  -27 51 33.9  & 0.15/0.22  &  1.6  $\pm$  0.4  &  4.0  $\pm$  1.3  &  4.1  &  5.1  $\pm$  1.2  &  2 & -- \\
 LESS 20   & ALESS 020.1   &  03 33 16.76  -28 00 16.0  & 0.24/0.19  &  3.9  $\pm$  0.9  &  7.0  $\pm$  2.2  &  4.2  &  4.5  $\pm$  1.1  &  3 &\textbf{r} \\
           & ALESS 020.2   &  03 33 16.27  -28 00 23.3  & 0.25/0.20  &  3.8  $\pm$  0.9  &  5.1  $\pm$  2.0  &  4.0  &  5.2  $\pm$  1.3  &  3 & -- \\
 LESS 22   & ALESS 022.2   &  03 31 46.69  -27 32 52.4  & 0.23/0.12  &  2.3  $\pm$  0.5  &  4.1  $\pm$  1.2  &  4.9  &  6.1  $\pm$  1.3  &  2 & -- \\
 LESS 23   & ALESS 023.2   &  03 32 11.43  -28 05 10.2  & 0.07/0.07  &  2.4  $\pm$  0.3  &  3.2  $\pm$  0.8  &  6.9  &  5.3  $\pm$  0.8  &  2 & -- \\
 LESS 30   & ALESS 030.1   &  03 33 44.33  -28 03 38.7  & 0.21/0.17  &  2.8  $\pm$  0.5  &  5.0  $\pm$  1.3  &  5.4  &  4.8  $\pm$  0.9  &  3 & -- \\
 LESS 34   & ALESS 034.1   &  03 32 17.96  -27 52 33.3  & 0.07/0.12  &  3.5  $\pm$  0.5  &  3.9  $\pm$  0.8  &  7.7  &  4.5  $\pm$  0.6  &  3 & -- \\
 LESS 38   & ALESS 038.1   &  03 33 10.84  -27 56 40.2  & 0.14/0.18  &  2.8  $\pm$  0.6  &  3.3  $\pm$  1.1  &  4.9  &  5.6  $\pm$  1.1  &  3 & -- \\
 LESS 39   & ALESS 039.2   &  03 31 44.56  -27 34 43.2  & 0.13/0.13  &  1.3  $\pm$  0.3  &  2.2  $\pm$  0.8  &  4.0  &  2.7  $\pm$  0.7  &  2 & -- \\
 LESS 43   & ALESS 043.3   &  03 33 06.27  -27 47 54.7  & 0.12/0.11  &  1.5  $\pm$  0.3  &  2.0  $\pm$  0.7  &  4.5  &  5.2  $\pm$  1.2  &  2 & -- \\
 LESS 46   & ALESS 046.1   &  03 33 36.70  -27 32 49.5  & 0.17/0.08  &  4.2  $\pm$  0.7  &  5.3  $\pm$  1.3  &  6.4  &  4.5  $\pm$  0.7  &  3 & \textbf{r} \\
 LESS 62   & ALESS 062.1   &  03 32 36.16  -27 34 48.9  & 0.13/0.29  &  3.4  $\pm$  0.7  &  6.3  $\pm$  1.7  &  5.0  &  4.3  $\pm$  0.8  &  3 & -- \\
           & ALESS 062.2   &  03 32 36.58  -27 34 53.8  & 0.13/0.36  &  2.7  $\pm$  0.7  &  3.8  $\pm$  1.4  &  4.0  &  2.9  $\pm$  0.7  &  3 & \textbf{r} \\
 LESS 75   & ALESS 075.2   &  03 31 27.67  -27 55 59.2  & 0.14/0.13  &  1.3  $\pm$  0.3  &  1.5  $\pm$  0.6  &  4.0  &  5.0  $\pm$  1.2  &  2 & -- \\
 LESS 80   & ALESS 080.5   &  03 31 41.68  -27 48 22.7  & 0.16/0.22  &  2.0  $\pm$  0.5  &  4.6  $\pm$  1.4  &  4.3  & 11.8  $\pm$  2.8  &  2 & -- \\
 LESS 81   & ALESS 081.1   &  03 31 27.55  -27 44 39.6  & 0.06/0.08  &  5.2  $\pm$  0.5  &  5.8  $\pm$  1.0  & 10.1  &  5.3  $\pm$  0.5  &  3 & \textbf{r} \\
           & ALESS 081.2   &  03 31 27.58  -27 44 43.1  & 0.13/0.20  &  2.2  $\pm$  0.5  &  2.8  $\pm$  1.1  &  4.2  &  2.4  $\pm$  0.6  &  3 & --\\
 LESS 83   & ALESS 083.1   &  03 33 09.42  -28 05 30.6  & 0.08/0.07  &  2.3  $\pm$  0.3  &  2.3  $\pm$  0.5  &  7.7  &  6.8  $\pm$  0.9  &  2 & -- \\
 LESS 89   & ALESS 089.1   &  03 32 48.69  -28 00 21.9  & 0.12/0.19  &  2.7  $\pm$  0.6  &  4.6  $\pm$  1.4  &  4.8  &  3.1  $\pm$  0.7  &  3 & -- \\
 LESS 91   & ALESS 091.1   &  03 31 35.30  -27 40 24.5  & 0.13/0.14  &  1.5  $\pm$  0.3  &  1.6  $\pm$  0.6  &  4.2  &  3.2  $\pm$  0.8  &  2 & -- \\
 LESS 93   & ALESS 093.1   &  03 31 11.06  -27 56 14.0  & 0.19/0.16  &  2.3  $\pm$  0.5  &  2.8  $\pm$  1.0  &  4.5  &  3.9  $\pm$  0.9  &  3 & --\\
 LESS 101  & ALESS 101.1   &  03 31 51.60  -27 45 53.0  & 0.27/0.10  &  3.3  $\pm$  0.8  &  8.2  $\pm$  2.3  &  4.4  &  3.4  $\pm$  0.8  &  3 & \textbf{r} \\
 LESS 103  & ALESS 103.2   &  03 33 25.82  -27 34 09.9  & 0.15/0.12  &  1.5  $\pm$  0.3  &  1.9  $\pm$  0.7  &  4.2  &  4.8  $\pm$  1.1  &  2 & -- \\
 LESS 106  & ALESS 106.1   &  03 31 39.64  -27 56 39.2  & 0.14/0.11  &  2.0  $\pm$  0.5  &  2.2  $\pm$  0.9  &  4.4  &  4.8  $\pm$  1.1  &  2 & -- \\
 LESS 109  & ALESS 109.1   &  03 33 28.01  -27 41 49.7  & 0.14/0.20  &  3.4  $\pm$  0.6  &  4.5  $\pm$  1.3  &  5.6  &  5.5  $\pm$  1.0  &  3 & -- 
\enddata
\tablenotetext{a}{Indicates why the SMG was selected as supplementary. See \S\ref{samples} for further details.}
\end{deluxetable*}

\end{document}